\documentclass[11pt,a4paper]{article}

\usepackage[
    a4paper,
    top=3cm,
    bottom=3cm,
    left=3.2cm,
    right=3.2cm
]{geometry}

\usepackage{amsmath,amssymb}

\usepackage[numbers,sort&compress]{natbib}

\usepackage{xcolor}

\usepackage{multirow}
\usepackage{longtable}
\usepackage{booktabs}

\usepackage{graphicx}
\usepackage{subcaption}
\usepackage{pdflscape}

\usepackage{hyperref}
\usepackage{pdfpages}

\usepackage{algorithm}      % entorno float 'algorithm'
\usepackage{algpseudocode}  % entorno 'algorithmic' (opcional pero útil)
\usepackage{float}          % para usar [H] (posición aquí)
\newenvironment{keywords}
{\begin{quotation}\noindent\small\textbf{Keywords:}\ }
{\end{quotation}}

\title{
    \vspace{-0.5cm}
    {\LARGE\bfseries
    CircaFMM: A Flexible Statistical Framework for\\[0.15em]
    Circadian Rhythmicity Detection in Transcriptomics}
}
\author{
    Joshua S. Abston$^{1}$,
    Yolanda Larriba$^{1,2,*}$,
    Itziar Fernández$^{1,3,*}$
}

\date{}

\begin{document}

\maketitle

\begin{center}
\small

$^{1}$Department of Statistics and Operations Research,
University of Valladolid, 47011 Valladolid, Spain

\vspace{0.25em}

$^{2}$Mathematics Research Institute of the University of Valladolid (IMUVA),
University of Valladolid, 47011 Valladolid, Spain

\vspace{0.25em}

$^{3}$Biomedical Research Networking Center in Bioengineering,
Biomaterials and Nanomedicine (CIBER-BBN), Valladolid, Spain

\vspace{0.6em}

$^{*}$Corresponding authors:
\texttt{yolanda.larriba@uva.es};
\texttt{itziar.fernandez@uva.es}

\end{center}

\begin{abstract}
\noindent\textbf{Motivation}: Detecting circadian rhythmicity in large-scale transcriptomic data is challenging when sampling is irregular, signals are weak, and expression profiles deviate from sinusoidal shapes. The fixed harmonic form of classical Cosinor-based methods may therefore be poorly suited to asymmetric and temporally complex rhythms. \\
\textbf{Results}: We present CircaFMM, a flexible framework for circadian rhythmicity detection based on the single-component Frequency Modulated Möbius (FMM) model. CircaFMM defines rhythmicity as time dependence induced by the Möbius phase parameter, so that the non-rhythmic null corresponds to a time-invariant mean profile. The resulting non-regular test combines time-density weighting, regularized estimation, and bootstrap calibration of a weighted likelihood-ratio-type statistic. Simulations showed controlled type-I error under irregular sampling and improved power over the classical Cosinor F-test for asymmetric waveforms. Applied to the GTEx-13T postmortem transcriptomic dataset, CircaFMM recovered canonical clock genes and circadian pathways while identifying additional rhythmic candidates with lower signal-to-noise-ratio, greater waveform asymmetry, and non-sinusoidal temporal profiles. \\
\textbf{Availability and implementation}: CircaFMM is available at \url{https://circafmm.shinyapps.io/circafmm_shinyapp/}, with source code at \url{https://github.com/FMMGroupVa/CircaFMM}.
\end{abstract}


\bigskip

\section{Introduction}

Circadian rhythms orchestrate a wide range of physiological and behavioral processes, including sleep–wake cycles, metabolism, cognition, and reproduction \citep{franken2024sleep,lok2025rhythms,tripathi2025ecology}. Their disruption has been associated with neurodegenerative, metabolic, cardiovascular, and psychiatric disorders, as well as cancer \citep{nassan2022circadian, ansu2023associations, tonon2024sleep, young2025circadian, zhang2026circadian}.
At the molecular level, circadian regulation is reflected in temporal patterns of gene expression across tissues and species, making transcriptomic data a central resource for studying the temporal organization of physiology and its disease-related disruption \citep{zhang2014circadian, hughes2017guidelines, mure2018diurnal, ruben2018database}. 

Circadian gene identification is commonly formulated as a hypothesis-testing problem that distinguishes rhythmic from non-rhythmic expression profiles. Many methods have been proposed for this purpose, including parametric, nonparametric, and likelihood-based approaches \citep{hughes2017guidelines,yoo2024identifying}. Among them, Cosinor-based methods remain a widely used reference because they are simple, interpretable, and rely on closed-form estimation under sinusoidal assumptions \citep{cornelissen2014cosinor}. In this framework, non-rhythmicity is encoded through zero amplitude, and rhythmicity is assessed by comparing with a sinusoidal alternative \citep{ding2021likelihood}. 
However, transcriptomic circadian studies often depart from the idealized conditions under which amplitude-based sinusoidal inference performs best.

One major difficulty is irregular temporal sampling. In many human transcriptomic studies, circadian variation is inferred from cross-sectional samples from different individuals at observed or proxy circadian times rather than from dense within-subject time series. This is especially relevant in postmortem studies, where time of death is often used as a proxy for circadian phase \citep{larriba2023circust, ananthasubramaniam2025rhythm}. Such designs introduce uncertainty due to recording error, inter-individual differences in circadian alignment, and pre-mortem factors such as sedation or hospitalization \citep{hammarlund2025cyclops}. They also lead to uneven temporal coverage, with some circadian windows densely sampled and others poorly represented \citep{kimura2023application,gorczyca2025weighted}. Under these conditions, isolated observations with unusual expression values can exert disproportionate influence, distorting waveform estimation and contributing to unstable rhythmicity calls under classical cosinor inference \citep{ahdesmaki2007robust, zong2023experimental}. 

A second challenge is the low signal-to-noise ratio (SNR). Transcriptomic rhythms are affected by biological heterogeneity, technical noise, and inter-individual variability, especially when temporal structure is reconstructed from independently sampled subjects \citep{de2019gene,baum2023data, larriba2023circust, ananthasubramaniam2025rhythm}. 
As a result, genuinely rhythmic genes may exhibit weak observed amplitudes despite retaining meaningful circadian organization, reducing the sensitivity of amplitude-based sinusoidal tests \citep{hollis2025reconstructed}.

A third difficulty is waveform shape: rhythmic profiles may show asymmetric peaks, unequal rising and falling phases, localized transitions, or other non-sinusoidal features \citep{thaben2014detecting,hutchison2015improved,hughes2017guidelines,rueda2019frequency}. When such profiles are forced into a sinusoidal model, rhythmicity may be underdetected or incompletely characterized. 

To address these challenges, we build on the single-component Frequency Modulated M\"obius model (FMM$_1$), a flexible parametric representation of oscillatory signals beyond simple cosine waveforms, whose theoretical and methodological properties have been developed in previous work \citep{rueda2019frequency,rueda2022unique,fernandez2025functional}.
In FMM$_1$, temporal variation is governed by a M\"obius phase function with a sharpness parameter $\omega$ controlling the departure from a regular sinusoidal profile through a nonlinear modulation of time. Values of $\omega$ close to one yield approximately sinusoidal waveforms, whereas smaller values allow more localized and asymmetric temporal patterns.

This parametrization provides a natural inferential formulation of rhythmicity. The boundary case $\omega = 0$ makes the M\"obius phase constant, eliminating phase-induced temporal dependence, whereas $\omega > 0$ induces temporal structure. Thus, rhythmicity can be assessed through phase-induced temporal structure, rather than as a nonzero amplitude in a fixed sinusoidal basis.

We propose CircaFMM, a flexible statistical framework for circadian rhythmicity detection in transcriptomic data, together with a likelihood-ratio test (LRT) comparing the time-invariant null model $\omega = 0$ with the rhythmic alternative $\omega > 0$. Because the null lies on the boundary of the parameter space, and temporal-structure parameters are not identifiable under the null, standard asymptotic LRT theory is not valid \citep{self1987asymptotic}. CircaFMM addresses this non-regular testing problem through inverse-density time weighting, regularized estimation, and bootstrap calibration of the test statistic. 
We evaluate CircaFMM through simulations and an application to the GTEx-13T human postmortem circadian transcriptomic dataset \citep{ruben2018database}, and provide a Shiny application implementing the methodology.

% -----------------------------------------------
% Metodos
\section{Materials and methods}

% ------------- Model FMM1 ------------------------
\subsection{FMM\texorpdfstring{$_1$}{1} model for circadian gene-expression profiles}
\label{sec:FMMmodel}

The FMM$_1$ model represents the expression profile of a gene observed at circadian time $t\in[0,2\pi)$ as
\begin{equation}\label{eq:FMM1}
X(t) = M + A \cos\!\big(\beta + \phi(t;\alpha,\omega)\big) + \varepsilon(t),
\qquad \varepsilon(t)\sim \mathcal{N}(0,\sigma^2)\nonumber
\end{equation}
where $X(t)$ denotes the observed gene expression, $M\in\mathbb{R}$ is the baseline expression level, $A\ge 0$ is the amplitude of the oscillatory component, $\beta\in[0,2\pi)$ is a shape parameter that modulates the symmetry and peak direction, $\phi(t;\alpha,\omega)$ is the M\"obius phase function, and $\sigma^2$ is a constant variance term. 
The systematic component is therefore
\[
\mu(t)=\mathbb{E}\{X(t)\}= M + A \cos\!\big(\beta + \phi(t;\alpha,\omega)\big),
\]
so that $X(t)=\mu(t)+\varepsilon(t)$.

The M\"obius phase is defined as
\begin{equation}\label{eq:FMMphase}
\phi(t; \alpha, \omega) = 2\arctan\!\left( \omega\tan\!\left( \frac{t - \alpha}{2} \right) \right),\nonumber
\end{equation}
with phase-location parameter $\alpha \in [0, 2\pi)$ and sharpness parameter $\omega \in [0, 1]$.
The parameter $\omega$ controls the departure from a regular sinusoidal profile. When $\omega=1$, the model reduces to a classical cosine with a circular phase shift. Smaller values of $\omega$ allow increasingly localized and asymmetric temporal patterns, thereby providing more flexible waveform shapes than a fixed harmonic representation.

A key feature of this parametrization is that the boundary case $\omega=0$ collapses the M\"obius phase to a constant, thereby removing phase-induced temporal variation from the mean profile. Conversely, for $A>0$, values $\omega>0$ induce genuine temporal dependence in $\mu(t)$. Thus, non-rhythmicity can be represented through the phase structure itself, rather than through zero amplitude in a fixed sinusoidal basis. Formal details supporting this phase-based test definition, including the constancy of the M\"obius phase at $\omega=0$, continuity of the composed mean profile, and the flat non-rhythmic regime, are provided in Supplementary Section~S1.

This phase-based representation motivates the CircaFMM rhythmicity test
\begin{equation}\label{eq:test_omega}
H_0:\ \omega=0
\qquad\text{vs}\qquad
H_1:\ \omega>0.
\end{equation}

Under this formulation, rhythmicity is assessed as the presence of temporal dependence induced by the M\"obius phase. Although a flat mean profile may also arise when $A=0$, the boundary condition $\omega=0$ collapses the entire phase-driven component to a time-invariant mean, irrespective of the remaining oscillatory parameters. In practice, evidence against $H_0$ requires that allowing $\omega>0$ produces a substantial improvement in fit relative to the flat null model, so a LRT provides a natural basis for inference.

The resulting testing problem is non-regular. First, the null value $\omega=0$ lies on the boundary of the parameter space. Second, under $H_0$, the phase is constant and the likelihood no longer depends on $\alpha$. Moreover, $M$, $A$, and $\beta$ are not separately identifiable beyond the scalar mean

\[
\mu_0 = M + A\cos(\beta).
\]

Consequently, the regularity conditions required for Wilks' theorem are not satisfied, and the usual $\chi^2$ reference distribution for likelihood-ratio statistics in regular nested models is not valid \citep{self1987asymptotic,brazzale2024likelihood}. This motivates the stabilized estimation strategy and bootstrap calibration procedure described in the following sections.

% ------------- Weighting ------------------------
\subsection{Weighting for irregular sampling}\label{sec:weights}

Circadian sampling designs often show uneven temporal coverage over $[0,2\pi)$, with densely sampled regions and sparse or poorly represented time windows. To account for this design feature, we use nonnegative time weights $w_i$ that depend uniquely on the observed sampling times  $\{t_i\}_{i=1}^n$.

Building on inverse-density weighting approaches developed for circular/trigonometric regression \citep{gorczyca2025weighted}, we estimate the empirical sampling-time density using a von~Mises kernel estimator as follows. We define the unnormalized weights as

\begin{equation*}
\tilde w_i =
\frac{1}
{\max\{\widehat f_{\hat\kappa}(t_i),\varepsilon_w\}},
\qquad i=1,\ldots,n,\nonumber
\end{equation*}

where $\widehat f_{\hat\kappa}$ denotes the kernel density estimate evaluated using the concentration parameter $\hat\kappa$ selected by leave-one-out log-likelihood cross-validation, and $\varepsilon_w>0$ is a small numerical floor preventing excessively large weights in very sparse regions. Then, weights are normalized as

\[
w_i =
n\,\frac{\tilde w_i}
{\sum_{j=1}^n \tilde w_j},
\qquad i=1,\ldots,n,
\]

so that \(\sum_{i=1}^n w_i=n\). 

Under this construction, observations in densely sampled intervals receive a smaller weight, whereas observations in under-sampled regions are upweighted. Because the weights are design-based, they are determined independently of the random variation in the observed expression values. Consequently, they reduce the leverage of sampling clusters and stabilize estimation and inference under irregular temporal designs. Further methodological details regarding the time-density weighting scheme and the selection of the concentration parameter are given in Supplementary Section~S2.

% ------------- Estimation ------------------------
\subsection{Regularized estimation in the small-\texorpdfstring{$\omega$}{omega} regime}
\label{sec:estimation}

Small $\omega$ values allow highly localized transitions in the phase-induced oscillatory component but can also produce unstable spike-like fits in finite samples.
Standard FMM fitting selects $(\alpha,\omega)$ over a least-squares grid, with optional local refinement \citep{rueda2019frequency,fernandez2022fmm}. Here, to reduce small-$\omega$ degeneracy, we use a stabilized grid-based selection followed by a weighted least-squares refit conditional on the selected pair.

For fixed $(\alpha,\omega)$, define the phase-induced basis functions
\[
C_{\alpha,\omega}(t_i)=\cos\{\phi(t_i;\alpha,\omega)\},
\qquad
S_{\alpha,\omega}(t_i)=\sin\{\phi(t_i;\alpha,\omega)\}.
\]

Using the reparametrization
\[
\delta = A\cos(\beta),\qquad \gamma=-A\sin(\beta),
\]
the FMM$_1$ mean profile is linear in $(M,\delta,\gamma)$ conditional on $(\alpha,\omega)$:
\[
X(t_i)= M + \delta\,C_{\alpha,\omega}(t_i) + \gamma\,S_{\alpha,\omega}(t_i) + \varepsilon(t_i).
\]
so these coefficients are estimated by weighted least-squares (WLS) using the design weights introduced in Section~\ref{sec:weights}.

To select $(\alpha,\omega)$, we rank candidate pairs over a grid $\mathcal{A}\times\Omega\subset[0,2\pi)\times(0,1]$ using a normalized weighted oscillatory alignment score,

\begin{equation}\label{eq:power_main}
P(\alpha,\omega)=
\frac{\langle \tilde x, C_{\alpha,\omega}\rangle_w^2+\langle \tilde x, S_{\alpha,\omega}\rangle_w^2}
{\langle C_{\alpha,\omega}, C_{\alpha,\omega}\rangle_w +\langle S_{\alpha,\omega}, S_{\alpha,\omega}\rangle_w}.\nonumber
\end{equation}

where $\langle a,b\rangle_w=\sum_{i=1}^n w_i a_i b_i$, $\tilde x_i=x_i-\bar x_w$,  $x_i=X(t_i)$ is the observed value at time $t_i$, and $\bar x_w$ denotes the weighted mean. This criterion measures the weighted alignment between the centered expression profile and the phase-induced oscillatory basis, providing a measure of oscillatory strength analogously to periodogram ranking in generalized Lomb--Scargle methods
\citep{zechmeister2009generalised}.

To further stabilize selection in the small-$\omega$ regime, we maximize the penalized criterion
\[
J(\alpha,\omega)=P(\alpha,\omega)-\lambda\,g(\omega)\,R(\alpha,\omega),
\qquad
g(\omega)=\Big(\frac{\omega_0}{\omega+\omega_0}\Big)^p, 
\]
where  $R(\alpha,\omega)$ is a curvature-type roughness functional of the phase-induced basis functions, computed either on the observed sampling times or on a dense equispaced grid when the sampling design contains highly irregular spacing. The modulation function $g(\omega)$ concentrates the penalty near the small-$\omega$ boundary, where spike-like profiles are most likely, while attenuating it for approximately sinusoidal or moderately asymmetric profiles. Here, $\omega_0\in(0,1)$ controls the extent of the penalized region, whereas $p\geq 1$ governs the sharpness of the penalty decay.
The selected pair $(\hat\alpha,\hat\omega)$ maximizes $J$, and the linear coefficients are then refitted by WLS conditional on this pair. 

The criterion includes a dataset-level regularization parameter $\lambda\ge 0$ controlling the strength of the roughness penalty. We select a single value $\lambda^\star$ by Monte Carlo calibration under a flat non-oscillatory null model, using a spike-sensitive diagnostic of fitted values and elbow selection. Because this calibration depends only on the sampling times, weights, grid, roughness mode and calibration settings, $\lambda^\star$ is reused for all genes and is not tuned to maximize rhythmic detections or downstream analyses.
Further details on the stabilized grid search, roughness functional, and local
refinement and $\lambda^\star$ calibration are provided in Supplementary Section~S3.

% ------------- LRT statistics ------------------------
\subsection{Weighted LRT-type statistic for sharpness-driven rhythmicity}
\label{sec:test_statistic}

To construct the CircaFMM test in \eqref{eq:test_omega}, we compare goodness-of-fit under the time-invariant null model and under the stabilized FMM$_1$ alternative. 
The comparison is based on the weighted residual sum of squares (WRSS), 
\[
\mathrm{WRSS}(\mu)
=
\sum_{i=1}^n w_i(x_i-\mu(t_i))^2,
\qquad
\sum_{i=1}^n w_i=n,
\]
where $\{w_i\}_{i=1}^n$ are the inverse-density time weights introduced in Section~\ref{sec:weights}.

Under $H_0:\omega=0$, the mean profile is constant, $\mu(t)\equiv \mu_0$, and the WLS estimator is the weighted mean $\hat\mu_0=\bar x_w$. We define
\[
\mathrm{WRSS}_0=\sum_{i=1}^n w_i(x_i-\hat\mu_0)^2.
\]

Under $H_1:\omega>0$, the FMM$_1$ model is fitting using the stabilized estimation procedure described in Section~\ref{sec:estimation}. Let $(\hat\alpha,\hat\omega)$ denote the selected nonlinear phase parameters, and let $\hat\mu_1(t_i;\hat\alpha,\hat\omega)$ be the fitted values after WLS refitting conditional on this selected pair. We define
\[
\mathrm{WRSS}_1(\hat\alpha,\hat\omega)
=
\sum_{i=1}^n w_i(x_i-\hat\mu_1(t_i;\hat\alpha,\hat\omega))^2.
\]

Following the form of the Gaussian likelihood-ratio statistic under weighted least squares, we define the CircaFMM weighted LRT-type statistic as

\begin{equation}\label{eq:LRT_WRSS}
\Lambda_w
=
n\log\!\left(\frac{\mathrm{WRSS}_0}{\mathrm{WRSS}_1(\hat\alpha,\hat\omega)}\right).
\end{equation}

Large values of $\Lambda_w$ indicate a substantial reduction in weighted residual variation after allowing phase-induced temporal structure, and therefore provide evidence against the non-rhythmic null model.

As discussed in Section~\ref{sec:FMMmodel}, $\Lambda_w$ does not have a standard asymptotic reference distribution because the null hypothesis is non-regular. In addition, the statistic depends on the sampling design, time-density weighting and data-driven selection of $(\hat\alpha,\hat\omega)$. We therefore calibrate $\Lambda_w$ using the bootstrap procedure described in Section~\ref{sec:boot}.

% ------------- Bootstrap inference ------------------------
\subsection{Bootstrap calibration of the weighted LRT-type statistic and sequential Monte Carlo p-values}
\label{sec:boot}

Under $H_0$, the LRT-type statistic $\Lambda_w$ has a design-dependent nonstandard distribution induced by the weighted and centered noise component.  A formal representation is provided in Supplementary Section~S4.1.  Since no closed-form reference distribution is available in this non-regular setting, we calibrated the CircaFMM weighted LRT-type statistic \eqref{eq:LRT_WRSS} by bootstrap. 

To accommodate unknown heteroskedasticity, we use a design-preserving wild bootstrap with Rademacher sign multipliers \citep{wu1986jackknife,liu1988bootstrap,davidson2008wild}.
Residuals are computed under the null model $\mu(t)\equiv\mu_0$, multiplied by independent Rademacher variables, optionally recentered under the same weighting scheme, and added back to the null fit to generate bootstrap samples. For each bootstrap sample, the full CircaFMM pipeline is repeated using the same time weights, $(\alpha,\omega)$ grid, roughness mode, dataset-level penalty $\lambda^\star$, and weighted refit under the alternative. This yields a bootstrap replicate statistic $\Lambda_w^\ast$. Further details on the design-preserving wild bootstrap  are provided in Section~S4.2 of the Supplementary Materials.

Monte Carlo $p$-values are computed from exceedances of $\Lambda_w^\ast$ over the observed statistic $\Lambda_{w,\mathrm{obs}}$. If $B$ bootstrap replicates are generated and 
\[
E_B
=
\sum_{b=1}^{B}
\mathbb{I}\{\Lambda_{w,b}^\ast \ge \Lambda_{w,\mathrm{obs}}\},
\]
the finite-sample corrected Monte Carlo $p$-value is
\begin{equation}
\label{eq:mc_pvalue_main}
\widehat p
=
\frac{E_B+1}{B+1}.\nonumber
\end{equation}

To reduce computation in transcriptome-wide testing, we use a two-stage sequential Monte Carlo strategy following \citet{besag1991sequential}. At each stage, $N$ denotes the maximum number of bootstrap replicates and $S$ the exceedance stopping limit. The first stage provides a low-cost screening step, whereas only genes with preliminary evidence against the null are promoted to a second stage with a larger bootstrap budget. In our experiments, we used \(N_1=200\), \(S_1=10\), promoted genes with at most five first-stage exceedances, and used \(N_2=50{,}000\), \(S_2=20\) in the refinement stage. This design balances early stopping for clearly non-significant genes with accurate estimation of small $p$-values for stronger signals. Additional details on the bootstrap scheme, sequential stopping rule, two-stage design, and runtime profiling are provided in Supplementary Sections~S4.3 and S4.4.

% -----------------------------------------------
% Simulaciones
\section{Simulation studies} \label{sec:simulations}

\subsection{Type-I error under $H_0$}
\label{sec:simH0}

We first evaluated type-I error control under the non-rhythmic null model $H_0:\omega=0$. We considered four sampling regimes: equispaced schedules, jittered schedules with small timing deviations, clustered designs with locally dense temporal regions, and gappy designs with a poorly sampled circadian window. These scenarios represent increasing departures from regular circadian sampling, with sample sizes $n\in\{24,48,96,192\}$. 
Under $H_0$, expression values were generated from null flat profiles with Gaussian noise ($\sigma=1$). For each sampling design and sample size, time-density weights and the dataset-level roughness parameter $\lambda^\star$ were computed once and then kept fixed across simulated genes. Full simulation settings are provided in Supplementary Section~S5.1.

For each scenario, empirical type-I error at nominal level $\alpha=0.05$ was estimated from $G=2000$ null genes and repeated over $B=10$ Monte Carlo runs. The $p$-values were obtained using the sequential wild-bootstrap calibration described in Section~\ref{sec:boot}. Across all sampling regimes and sample sizes, empirical type-I error remained close to the nominal level (Table~\ref{tab:type1}). Calibration was closest to nominal under equispaced and jittered schedules, with only mild upward deviations for clustered and gappy designs at the smallest sample sizes. These deviations decreased as $n$ increased, indicating adequate null calibration under both regular and irregular temporal sampling. The sequential two-stage bootstrap typically stopped early under \(H_0\), with a stable median number of replicates ($\widetilde N$)  of approximately 20 across all simulation settings.

\begin{table}[t]
\centering
\caption{Empirical type-I error under $H_0:\omega=0$ across sampling designs and sample sizes.
Results are reported as mean $\pm$ standard deviation across $B=10$ Monte Carlo repetitions ($G=2000$ null genes per repetition).
$\widetilde N$ is the median number of sequential bootstrap replicates used per gene, averaged over repetitions.}
\label{tab:type1}
\begin{tabular}{r l c c}
\hline
$n$ & design & $\widetilde N$ & type-I error \\
\hline
24  & equispaced & $19.40 \pm 0.52$ & $0.0506 \pm 0.0034$ \\
24  & jittered   & $19.35 \pm 0.47$ & $0.0508 \pm 0.0033$\\
24  & clustered  & $21.10 \pm 0.32$ & $0.0540 \pm 0.0045$ \\
24  & gappy      & $20.60 \pm 0.52$ & $0.0543 \pm 0.0040$ \\
48  & equispaced & $19.65 \pm 0.47$ & $0.0493 \pm 0.0047$ \\
48  & jittered   & $19.60 \pm 0.52$ & $0.0499 \pm 0.0063$\\
48  & clustered  & $20.80 \pm 0.79$ & $0.0554 \pm 0.0044$ \\
48  & gappy      & $20.30 \pm 0.48$ & $0.0531 \pm 0.0050$ \\
96  & equispaced & $19.45 \pm 0.50$ & $0.0497 \pm 0.0057$ \\
96  & jittered   & $19.70 \pm 0.67$ & $0.0494 \pm 0.0042$\\
96  & clustered  & $19.60 \pm 0.52$ & $0.0526 \pm 0.0066$ \\
96  & gappy      & $19.80 \pm 0.63$ & $0.0516 \pm 0.0097$ \\
192 & equispaced & $19.70 \pm 0.48$ & $0.0513 \pm 0.0033$ \\
192 & jittered   & $19.40 \pm 0.52$ & $0.0513 \pm 0.0053$\\
192 & clustered  & $19.40 \pm 0.52$ & $0.0518 \pm 0.0053$ \\
192 & gappy      & $19.70 \pm 0.48$ & $0.0509 \pm 0.0048$ \\
\hline
\end{tabular}
\end{table}

Additional robustness analyses under heavy-tailed errors and ablation variants are reported in Supplementary Section~S5.2, showing broadly similar calibration patterns and no deterioration of type-I error control.

\subsection{Power under $H_1$ and comparison with cosinor}
\label{sec:simH1}

We assessed the power of the CircaFMM test under oscillatory alternatives generated from the FMM$_1$ model and benchmarked it against the classical Cosinor F-test. We retained the same four sampling regimes and sample sizes as in Section~\ref{sec:simH0}, together with the time-density weighting, $\lambda^\star$ calibration, noise model, and bootstrap testing pipeline described in Materials and methods.

Under $H_1$, expression data were generated over amplitudes $A\in\{0.5,0.75,1\}$, which following \citet{ding2021likelihood} correspond to amplitude-based signal-to-noise (SNR) levels because $\sigma=1.$ For the  sharpness parameter we considered $\omega\in\{0.05,0.15,0.25,0.4,0.7,1\}$. The case $\omega=1$ corresponds to a sinusoidal waveform, whereas smaller values of $\omega$ produce increasingly localized and asymmetric profiles. Full simulation settings, including the distributions of the remaining FMM$_1$ parameters, noise model and Cosinor fitting procedure, are provided in Supplementary Section~S5.1.

For each $(n,\mathrm{design},A,\omega)$ configuration, $G=2000$ were simulated and the experiment was repeated over $B=10$ Monte Carlo repetitions. Power was estimated at nominal level $\alpha=0.05$ as the proportion of rejections and reported as mean $\pm$  standard deviation across repetitions. 

\begin{figure*}[!t]
  \centering
  \includegraphics[width=\textwidth,height=0.92\textheight,keepaspectratio]{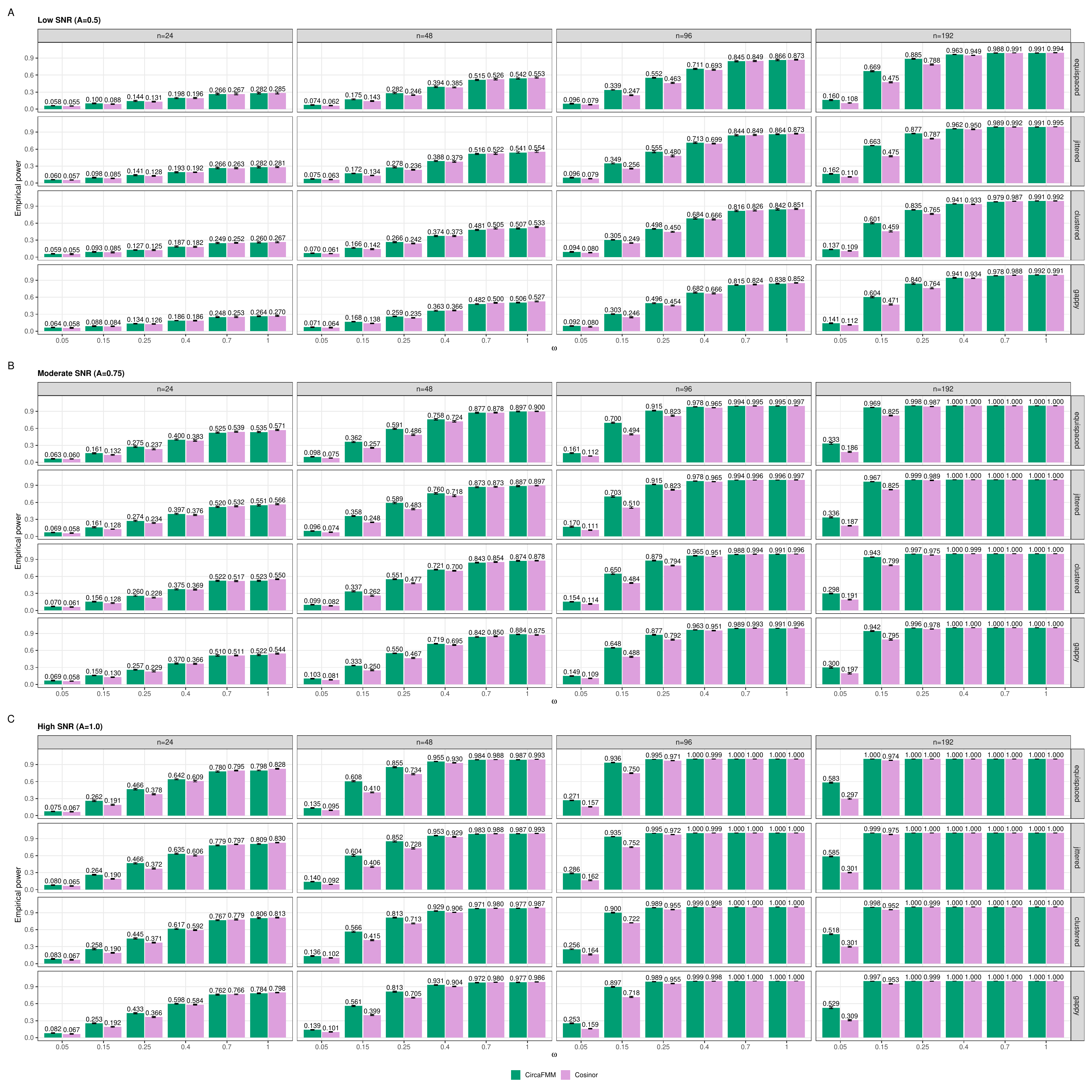}
  \caption{Power comparison between the proposed CircaFMM test and a classical cosinor F-test under oscillatory alternatives.
Panels (A)--(C) correspond to increasing signal-to-noise ratio levels, $\mathrm{SNR}=A/\sigma$ (here $\sigma=1$, so $\mathrm{SNR}=A$).
Within each panel, empirical power at nominal level $\alpha=0.05$ is shown as a function of the sharpness parameter $\omega$ across sampling-time designs (rows) and sample sizes $n$ (columns).
Bars report mean power over $B=10$ Monte Carlo repetitions with $G=2000$ genes per repetition; error bars show $\pm 1$ standard deviation.}
  \label{fig:power_snr}
\end{figure*}

Figure~\ref{fig:power_snr} shows that power increased with signal strength and sample size for both methods. For fixed SNR and $n$, detection was most difficult for small $\omega$, where the waveform contains localized temporal transitions. As expected, Cosinor was competitive in the sinusoidal regime ($\omega\approx 1$), where its model is correctly specified. In contrast, CircaFMM achieved higher power for smaller $\omega$ values that induce sharper and more asymmetric profiles, especially at moderate SNR and sample sizes where power was not yet saturated. These gains were also observed under clustered and gappy sampling schedules.

Additional power ablation analyses are reported in Supplementary Section~S5.3 and showed no systematic power loss due to time-density weighting or calibrated roughness penalization. A complementary analysis in Supplementary Section~S5.4 further illustrates the dependence of Cosinor sensitivity on variance-based SNR under a correctly specified sinusoidal model. Overall, the simulations indicate that CircaFMM remains competitive for sinusoidal rhythms while improving detection of structured non-sinusoidal profiles that are less well captured by a fixed Cosinor representation.

% -----------------------------------------------
% Real data application
\section{Application to the GTEx-13T circadian transcriptomic dataset}
\label{sec:realdata}

\subsection{GTEx-13T dataset and preprocessing}
\label{sec:gtex_preprocessing}

We applied CircaFMM to the GTEx-13T dataset, a subset of the GTEx \citep{gtex2020gtex} introduced by \citet{ruben2018database} for human postmortem circadian transcriptomic analyses.  The resource contains contains 4292 RNA-seq samples from 632 donors across 13 CYCLOPS-ordered \citep{anafi2017cyclops} tissues: subcutaneous adipose tissue (SUB), visceral adipose tissue (VIS), aorta (AOR), coronary artery (COR), tibial artery (TIB), sigmoid colon (COL), esophagus mucosa (ESO), heart atrial appendage (HEA), liver (LIV), lung (LUN), tibial nerve (NER), pituitary (PIT), and thyroid (THY).
After filtering, approximately 13000 highly expressed genes with Entrez identifiers were analyzed per tissue. To reduce the influence of isolated atypical observations under irregular temporal coverage, we applied a robust gene-wise outlier-attenuation procedure based on MAD local correction and LOESS smoothing. Preprocessing details and tissue-level summaries of temporal coverage and expression variability are provided in Supplementary Section~S6.1.

GTEx-13T provides a heterogeneous but biologically well-characterized benchmark for evaluating rhythmicity detection under realistic transcriptomic conditions, including uneven temporal coverage, tissue-specific rhythmic regulation, and variable signal strength. Tissue-level summaries of temporal coverage and expression variability are reported in Supplementary Section~S6.1.2. 

\subsection{CircaFMM rhythmicity detections across tissues}
\label{sec:gtex_detections}

Rhythmicity was analyzed independently for each GTEx-13T tissue using the CircaFMM rhythmicity test. Within each tissue, the time-density weights (\(w_i\)) and calibrated dataset-level roughness parameter (\(\lambda^\star\)) were computed once and reused across all genes. The resulting tissue-specific calibration settings and associated analyses are presented in Supplementary Sections~S2 and S3.

CircaFMM $p$-values were adjusted for multiple testing using the Benjamini--Hochberg procedure \citep{benjamini1995controlling}, and genes with FDR-adjusted $p$-values (\textit{adj.p}) below 0.01 were classified as rhythmic. CircaFMM identified substantial circadian activity across GTEx-13T tissues (Table~\ref{tab:rhythmics}) with detection rates varying markedly by tissue. The highest proportions of rhythmic genes were observed in VIS (57.2\%), TIB (41.8\%), and THY (39.5\%), whereas LIV and LUN showed the lowest proportions of CircaFMM detections, with 1.6\% and 8.9\%, respectively. This variability is consistent with the tissue-dependent nature of circadian transcriptomic regulation reported in large-scale mammalian studies, particularly the high rhythmic activity observed in adipose tissues and the comparatively reduced rhythmic signal in liver or lung tissues \citep{mure2018diurnal}.

Cross-tissue overlap among CircaFMM detections was moderate, reflecting the tissue-specific organization of circadian gene expression (Supplementary Figure~S7). This is consistent with previous evidence that circadian oscillations are widespread across the human transcriptome but often tissue-dependent, with nearly half of genes rhythmic in at least one tissue \citep{zhang2014circadian}. Nevertheless, CircaFMM identified 30 robust cross-tissue detections classified as rhythmic in at least 10 of the 13 tissues. This set included canonical clock-associated genes such as \textit{NR1D2}, \textit{PER2}, \textit{PER3}, \textit{CRY2}, \textit{DBP}, \textit{HLF}, \textit{TEF} and \textit{CIART}, with \textit{NR1D2} classified as rhythmic in all 13 tissues \citep{mure2018diurnal}.
Additional cross-tissue candidates included examples with non-sinusoidal temporal waveforms, such as \textit{SERPINA3} and \textit{TMED4}, which have been linked to circadian-related regulatory processes \citep{demezer2023serpina3,casanova2024circadian}. These examples are reported in Supplementary Section~S6.2.1 and Supplementary Figures~S8--S9.

% ----------------------------------
\subsection{Functional validation of CircaFMM rhythmicity rankings}
\label{sec:gtex_gsea}

\begin{figure}[!t]
  \centering
  \includegraphics[width=\textwidth]{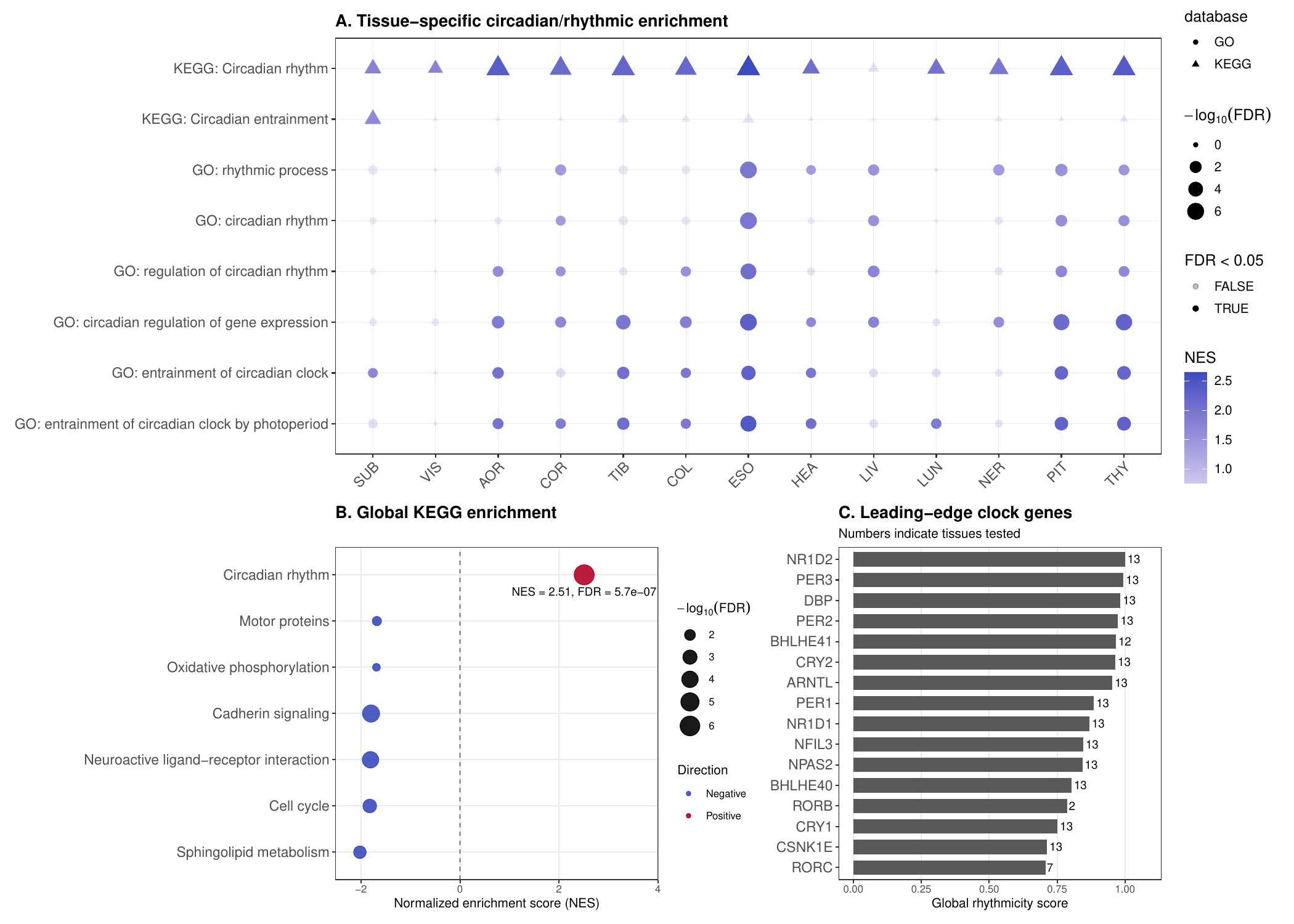}
  \caption{Functional characterization of tissue-specific and global CircaFMM rhythmicity rankings.
A) Tissue-specific preranked GSEA of predefined circadian/rhythmic GO Biological Process terms and KEGG pathways, using the observed CircaFMM likelihood-ratio statistic as ranking metric. Colour indicates the normalized enrichment score (NES), point size represents $-\log_{10}(\mathrm{FDR})$, and point shape denotes the database source.
B) Global KEGG enrichment using a tissue-normalized global rhythmicity score, defined as the mean within-tissue percentile rank of the observed CircaFMM statistic. KEGG \textit{Circadian rhythm} was the only significant pathway with positive NES.
C) Leading-edge genes driving the global KEGG \textit{Circadian rhythm} enrichment, ranked by global rhythmicity score. Numbers indicate the number of tissues in which each gene was evaluated.}
  \label{fig:GSEA}
\end{figure}

To assess the biological relevance of CircaFMM rhythmicity rankings, we performed preranked gene set enrichment analysis (GSEA) \citep{subramanian2005gene} using the observed CircaFMM likelihood-ratio statistic within each tissue. We focused on predefined circadian- and rhythm-related GO Biological Process terms \citep{gene2026gene} and KEGG pathways \citep{kanehisa2000kegg}. These gene sets showed predominantly positive normalized enrichment scores (NES) across tissues, indicating enrichment of circadian-related genes among the highest-ranked CircaFMM detections (Figure~\ref{fig:GSEA}A). The KEGG \textit{Circadian rhythm} pathway displayed the most consistent positive enrichment across tissues.

We also constructed a global rhythmicity score by averaging tissue-specific percentile ranks of the CircaFMM statistic across tissues. Using this global ranking, KEGG GSEA identified \textit{Circadian rhythm} as the only significantly enriched pathway with positive NES (NES = 2.51, FDR = $5.7 \times 10^{-7}$; Figure~\ref{fig:GSEA}B). The leading-edge genes included canonical molecular clock components such as \textit{NR1D2}, \textit{PER3}, \textit{DBP}, \textit{PER2}, \textit{CRY2}, \textit{ARNTL}, \textit{PER1}, \textit{NR1D1}, \textit{NPAS2}, and \textit{CRY1} (Figure~\ref{fig:GSEA}C).

Together, these results support the biological coherence of CircaFMM rhythmicity rankings at both tissue-specific and cross-tissue levels. 

% ------
\subsection{Comparison with Cosinor-based detections}
\label{sec:gtex_cosinor_comparison}

\begin{table*}[t]
\centering
\caption{CircaFMM rhythmicity detections and overlap with Cosinor-based calls across GTEx-13T tissues. The first row reports the percentage of analyzed genes classified as rhythmic by CircaFMM at FDR $<0.01$. Rows SR, FR and RR denote shared rhythmic genes detected by both methods, CircaFMM-only rhythmic genes, and Ruben-only rhythmic genes, respectively.
The total number of analyzed genes per tissue is shown in the last row.}
\label{tab:rhythmics}
\small
\setlength{\tabcolsep}{1.5pt}
\begin{tabular}{cccccccccccccc}

 & \multicolumn{13}{c}{Tissues} \\
\cline{2-14}
& SUB & VIS & AOR & COR & TIB & COL & ESO & HEA & LIV & LUN & NER & PIT & THY \\ 
\hline
\%CircaFMM & 26.70 & 57.16 & 30.34 & 37.72 & 41.80 & 32.00 & 15.11 & 34.80 & 1.60 & 8.88 & 36.47 & 24.10 & 39.49 \\ \hline
SR &  148 &1768  &115  &1304  &127  &244  &518  &748  &46  &97  &590& 150 &159\\
FR &  3371 & 5814 & 3861 & 3675 &5322  &3961  &1491 &3880 &160  &1086 &4230  &3039 &5034\\
RR & 1  &2  &0  & 112 & 1 & 19 & 148 & 113 & 133 &66  &1  &4 &8\\
\hline
Total &  13178& 13265 &13107& 13200& 13037& 13142 &13286& 13300 &12971& 13329 &13216& 13231 &13151\\\hline
\end{tabular}
\end{table*}

We compared CircaFMM detections with the Cosinor-based rhythmicity calls reported by \citet{ruben2018database} for the same GTEx-13T dataset. In that study, rhythmic genes were defined using multiple criteria, including FDR $<0.01$, relative amplitude $rAmp\ge 0.1$, and coefficient of determination $R^2\ge 0.1$. The impact of these criteria on rhythmic gene classification is illustrated in Figures~S9--S10, Table~S7, and discussed in detail in Supplementary Section~S6.2.2.
Genes were classified as shared rhythmic genes (SR), detected by both methods; Ruben-only rhythmic (RR), detected only by the Cosinor-based framework; or CircaFMM-only rhythmic  (FR). For phase comparisons, peak phases were aligned by setting the peak of \textit{ARNTL} to phase 0 in both approaches.

CircaFMM identified more rhythmic genes than the Cosinor-based framework in most tissues, with a median of 3861 rhythmic genes across tissues ($P_{25}=3039$, $P_{75}=4230$; Table~\ref{tab:rhythmics}). At the same time, it preserved a high recovery of previously reported rhythmic genes, with a median SR recovery proportion of 0.952 across tissues, whereas RR counts were low in most tissues. The main difference between methods was therefore the size of the FR group, suggesting that CircaFMM retains additional rhythmic profiles not captured by the Cosinor-based criteria.

The number of CircaFMM detections is broadly consistent in magnitude with recent larger GTEx analyses based on flexible circadian modelling frameworks \citep{palumaa2026local}, although direct comparisons are limited by differences in preprocessing, phase reconstruction, rhythmicity criteria, model class, and tissue coverage.

\begin{figure}
    \centering
    \includegraphics[width=1\linewidth]{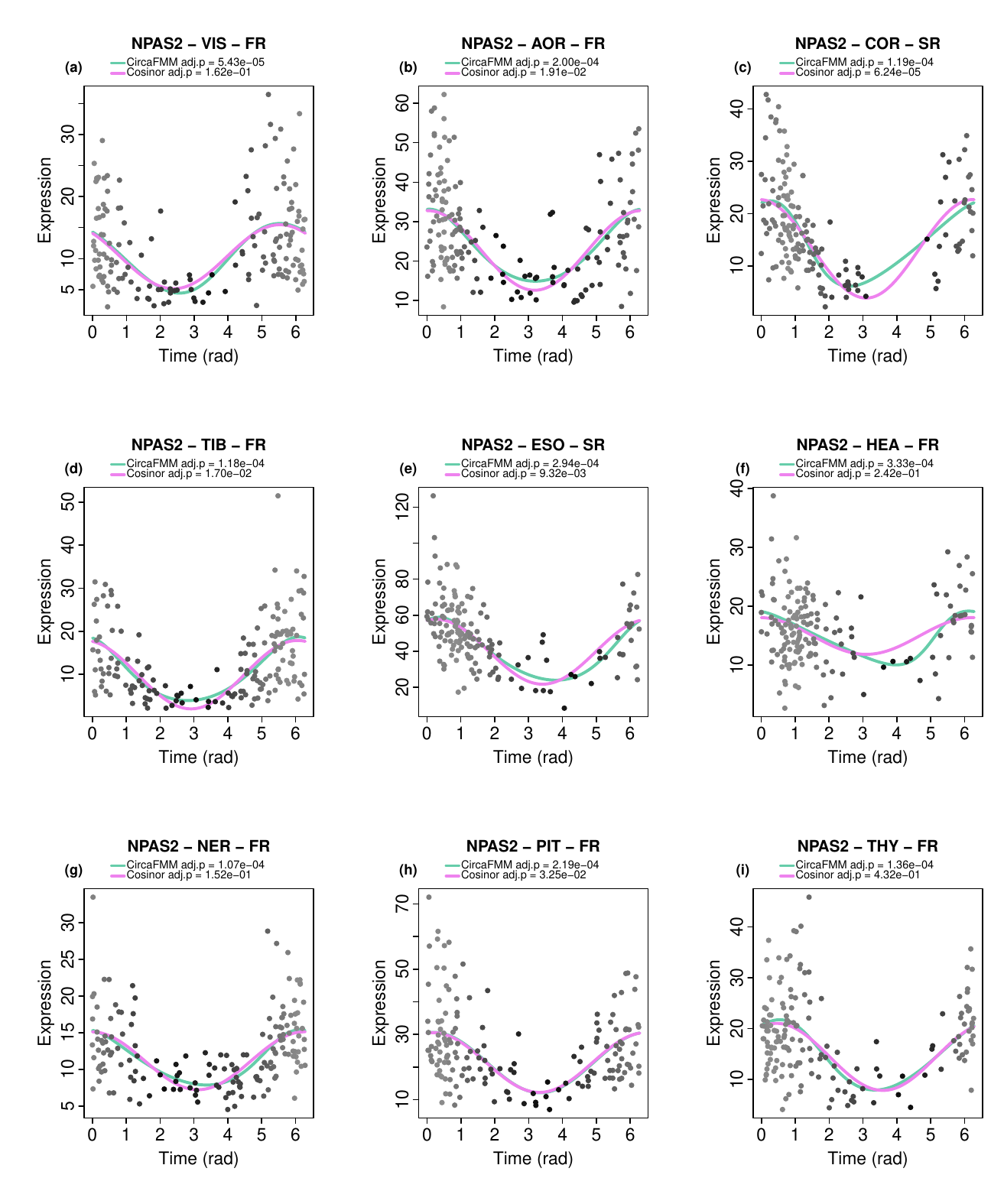}
    \caption{\textit{NPAS2} CircaFMM detections in VIS, AOR, COR, TIB, ESO, HEA, NER, PIT and THY. Points represent observed expression values, and grayscale intensity indicates observation weights $w_i$, with darker points corresponding to higher weights.}
    \label{fig:NPAS2}
\end{figure}

As a biologically relevant example, \textit{NPAS2}  was classified as rhythmic in nine tissues by CircaFMM, compared with five under Cosinor (Figure~\ref{fig:NPAS2}). This extends the set of ubiquitous rhythmic genes previously proposed by \citet{ruben2018database}, defined as genes detected as rhythmic in at least eight of the 13 tissues, and is consistent with the role of \textit{NPAS2} within the core circadian transcriptional network. 

\subsection{Characterization of CircaFMM-specific detections}
\label{sec:gtex_fr_characterization}

We next characterized the CircaFMM-specific rhythmic genes to assess whether they were enriched for features expected to challenge classical sinusoidal inference. We focused on two complementary properties: waveform asymmetry and signal strength. 

\subsubsection{Waveform asymmetry}

A key factor underlying the additional CircaFMM detections is that the M\"obius-based phase transformation defining the FMM framework naturally allows flexible asymmetric waveforms beyond classical sinusoidal oscillations \citep{rueda2019frequency}. To quantify these shape differences, we used the duration asymmetry index $AD_{\mathrm{Dur}}$, which compares the relative durations of the ascending and descending phases of the fitted FMM profile. This index ranges from $-1$ to $1$: values close to zero indicate approximately symmetric waveforms, whereas larger absolute values indicate stronger asymmetry. The sign identifies which phase is relatively longer. Details are provided in Supplementary Section~S6.3.1. 

We analyzed $AD_{\mathrm{Dur}}$ jointly with the selected sharpness parameter $\hat{\omega}$. SR genes were concentrated closer to the symmetric regime, with moderate-to-high $\hat{\omega}$ values and $AD_{\mathrm{Dur}}$ values close to zero, consistent with near-sinusoidal profiles. In contrast, FR genes showed a broader dispersion of $AD_{\mathrm{Dur}}$, particularly at smaller $\hat{\omega}$, where more asymmetric and structurally diverse waveforms can arise (Supplementary Table~S8). Consistently, $\vert AD_{\mathrm{Dur}}\vert$ was significantly higher for FR genes than for SR genes according to Wilcoxon rank-sum tests after FDR adjustment ($adj.p < 0.01$). This pattern is illustrated for NER in Figure~\ref{fig:asymNER}. Tissue-specific distributions for all tissues are shown in Supplementary Figures~S12--S13. 

\begin{figure} [h!]
    \centering
    \includegraphics[width=0.8\linewidth]{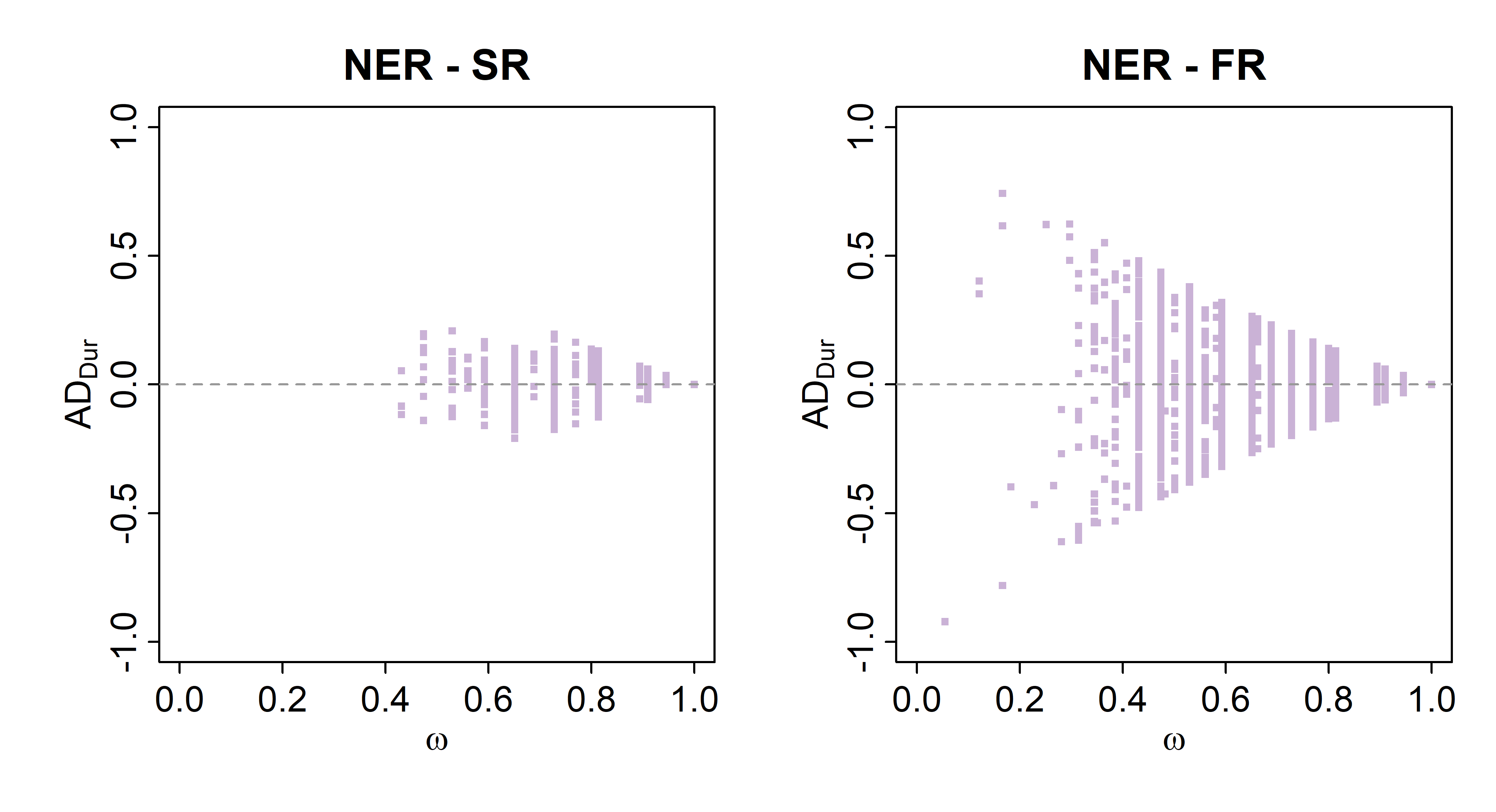}
    \caption{Relationship between waveform asymmetry and sharpness for SR
    and FR genes in NER. The duration asymmetry index $AD_{\mathrm{Dur}}$ is shown against the selected sharpness parameter $\hat{\omega}$. Left: shared rhythmic genes (SR). Right: CircaFMM-specific rhythmic genes (FR).}
    \label{fig:asymNER}
\end{figure}

\begin{figure} [!h]
    \centering
    \includegraphics[width=1\linewidth]{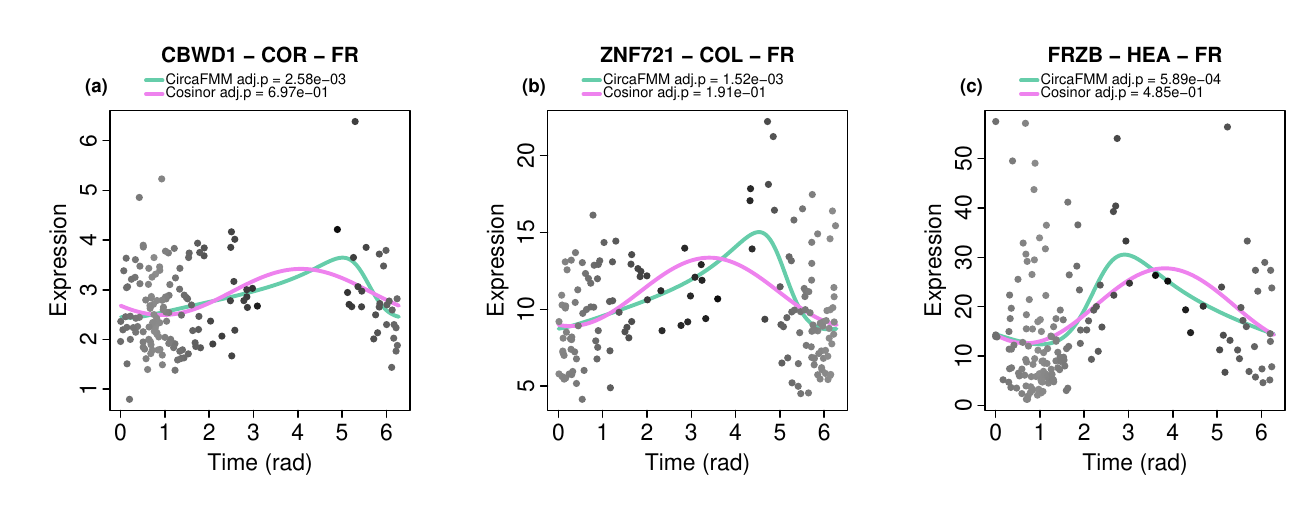}
    \caption{Representative CircaFMM-specific detections with asymmetric temporal profiles in COR, COL, and HEA. Points represent observed expression values, and grayscale intensity indicates observation weights $w_i$, with darker points corresponding to higher weights.}
    \label{fig:ASSYM_SNR}
\end{figure}

Representative asymmetric FR profiles are shown in Figure~\ref{fig:ASSYM_SNR} for COR, COL, and HEA, tissues with irregular temporal coverage according to Supplementary Table~S6. These examples illustrate clear non-sinusoidal rhythmic profiles recovered by CircaFMM but not retained by the Cosinor-based framework. Additional asymmetric FR examples are provided in Supplementary Figures~S14-S26 and Supplementary Table~S9.

\subsubsection{Signal strength}

We also evaluated whether CircaFMM-specific detections tended to occur in lower signal-strength regimes. For this purpose, we computed a gene-wise variance-based SNR using a nonparametric periodic-spline fit (Supplementary Section~S6.3.2). This definition is aligned with Supplementary Section~S5.4, where Cosinor sensitivity is shown to depend on explained oscillatory variance relative to residual variability.

\begin{figure}[ht]
\centering

\begin{minipage}{0.29\linewidth}
\centering
\includegraphics[width=\linewidth]{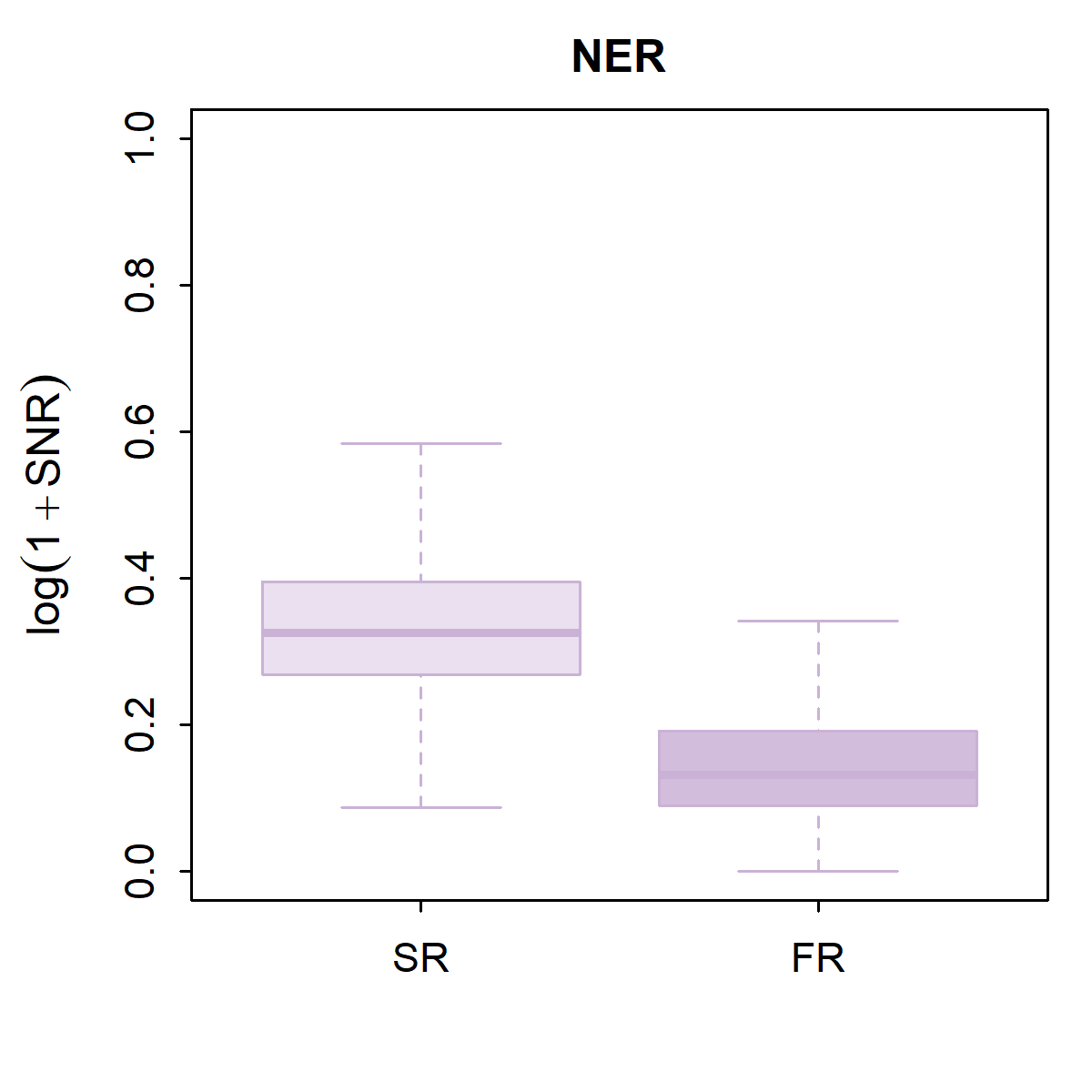}

\end{minipage}
\hfill
\begin{minipage}{0.29\linewidth}
\centering
\includegraphics[width=\linewidth]{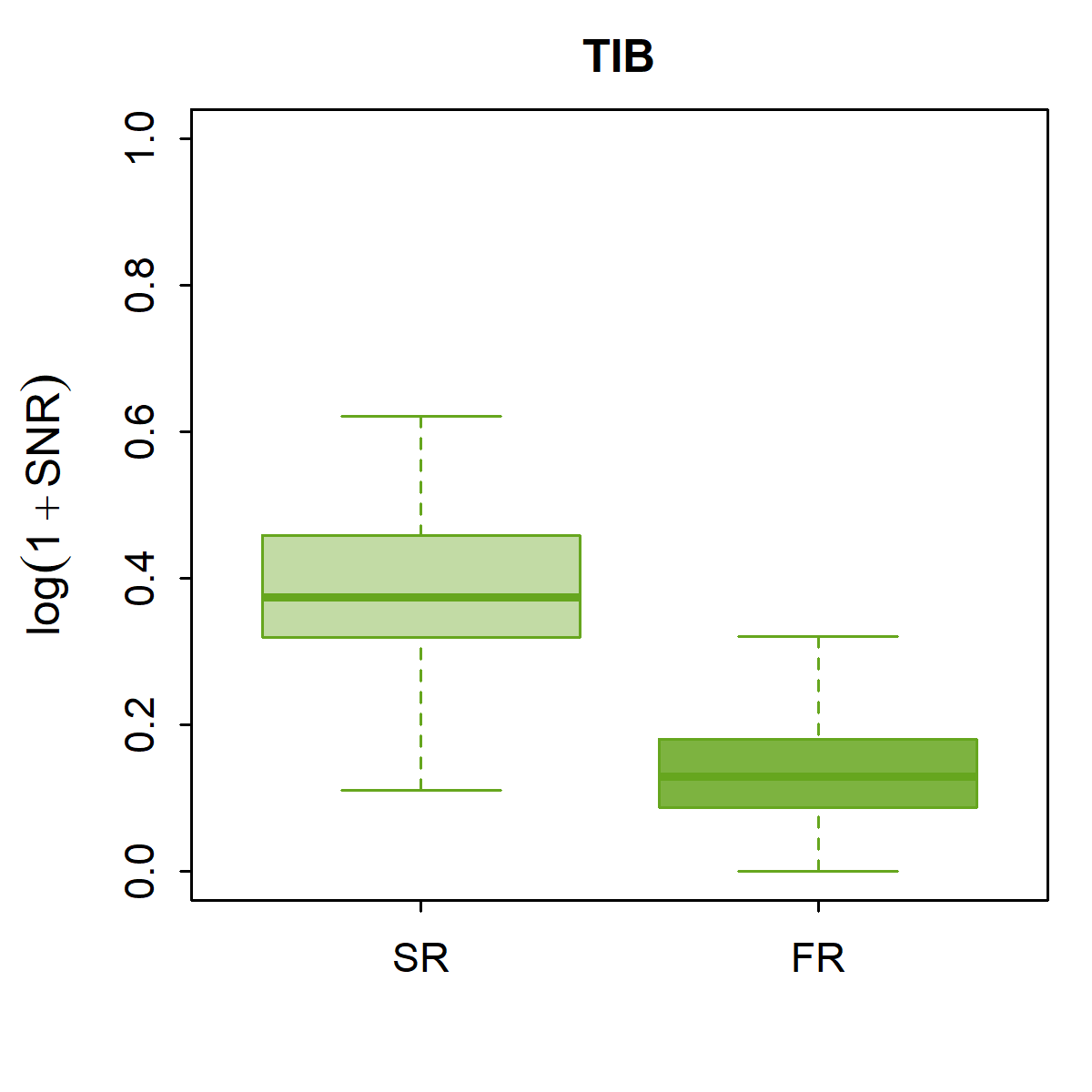}

\end{minipage}
\hfill
\begin{minipage}{0.29\linewidth}
\centering
\includegraphics[width=\linewidth]{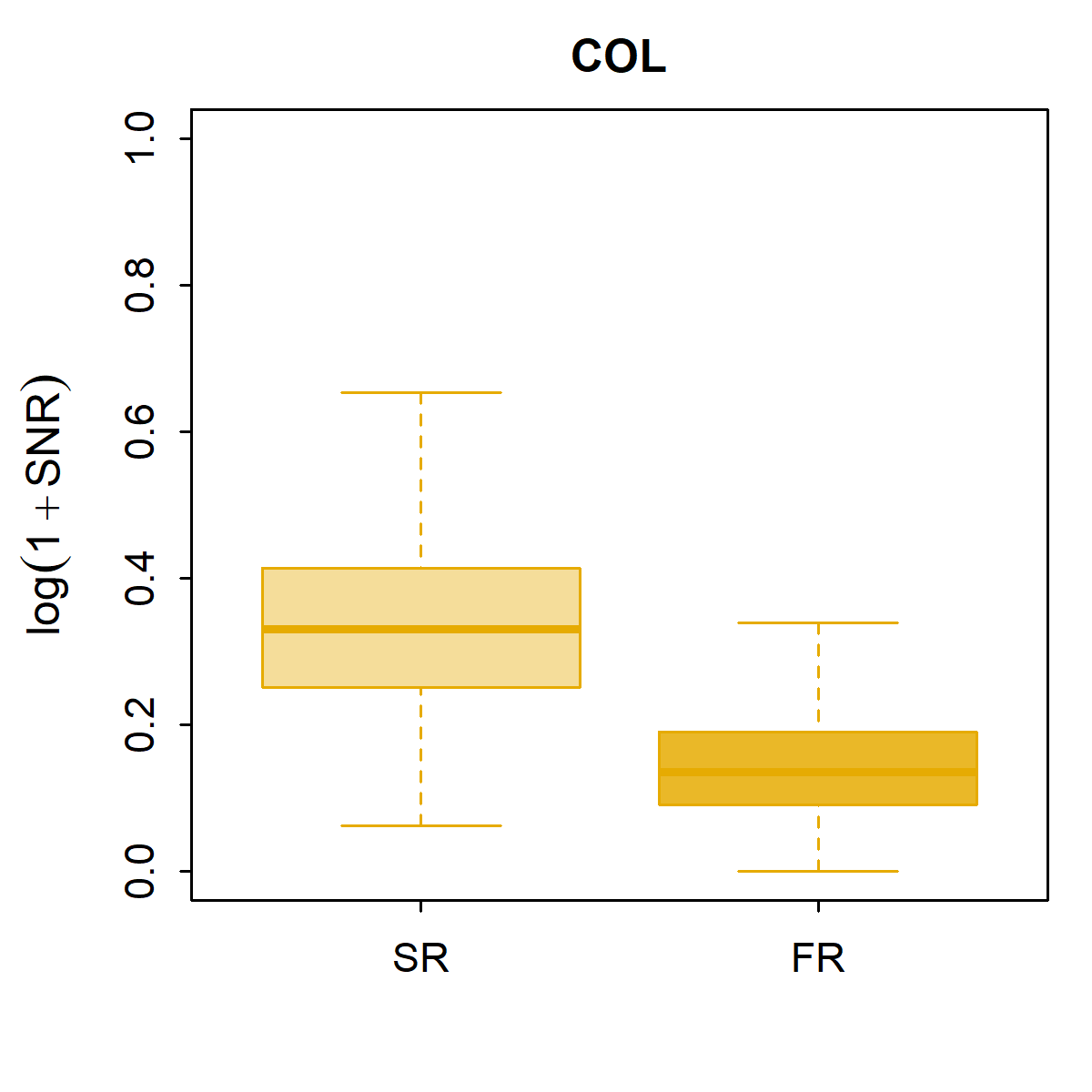}

\end{minipage}

\caption{Distribution of log-transformed variance-based SNR values for SR and FR genes in  representative low-SNR tissues. Boxplots compare shared rhythmic genes (SR) and CircaFMM-specific rhythmic genes (FR) in NER, TIB and COL tissues.}
\label{fig:SNR_NER}
\end{figure}

Across tissues, SR genes showed systematically higher SNR values than FR genes according to Wilcoxon rank-sum tests after FDR adjustment ($adj.p < 0.01$). Figure~\ref{fig:SNR_NER} illustrates this pattern in NER, TIB, and COL, three tissues with relatively low signal-strength regimes (Supplementary Table~S6). Results for all the tissues are shown in Supplementary Figure~S27.
Despite these lower SNR values, CircaFMM detected substantial rhythmicity in these tissues, all with rhythmicity rates above 30\% (Table~\ref{tab:rhythmics}). 
Representative low-SNR FR detections are shown in Figure~\ref{fig:SNR_NER_TIB}, with additional examples in Supplementary Figures~S28--S30 and Supplementary Tables~S10-S11.

\begin{figure}[ht]
    \centering
    \includegraphics[width=\linewidth]{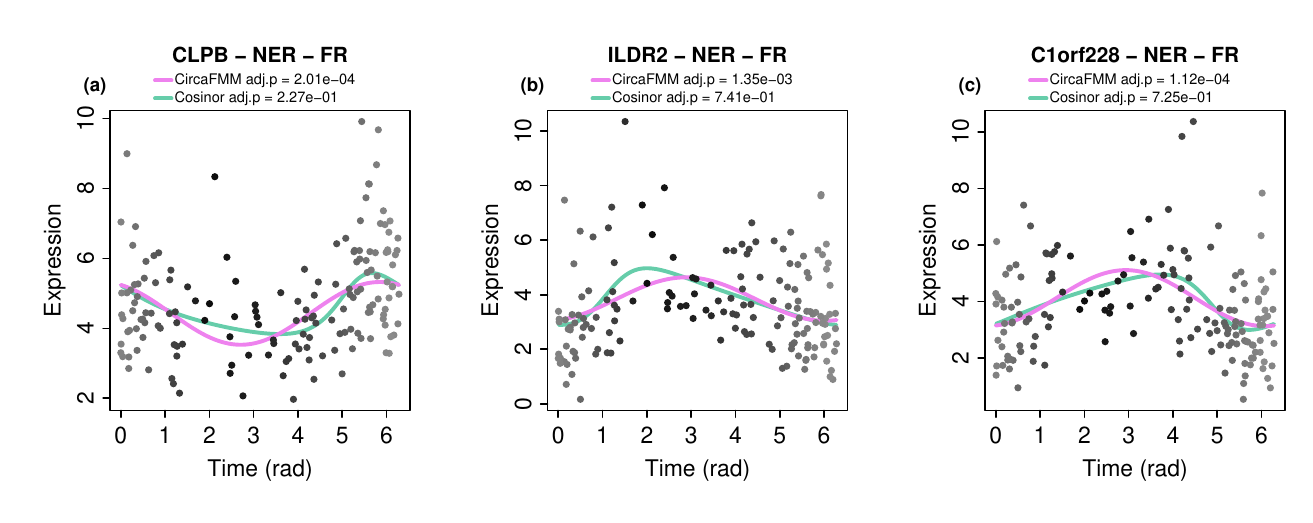}
    \caption{Representative CircaFMM-specific detections under low-SNR conditions in NER. Points represent observed expression values, and grayscale intensity indicates observation weights $w_i$, with darker points corresponding to higher weights.}
    \label{fig:SNR_NER_TIB}
\end{figure}

\subsection{Waveform heterogeneity captured by CircaFMM}
\label{sec:gtex_waveform_clustering}

We compared waveforms captured by CircaFMM and Cosinor using functional clustering analysis of the fitted expression profiles within each tissue. Clustering was performed separately for each method: Cosinor included SR and RR detections, whereas CircaFMM included SR and FR detections. Methodological details, complete cluster mean curves, cluster-wise summaries, and Cosinor--CircaFMM correspondence tables are provided in Supplementary Section~S6.4. 

Cosinor-based clustering revealed limited waveform diversity, with clusters largely restricted to symmetric sinusoidal profiles and typically ranging from two to five groups across tissues (Supplementary Figure~S31). In contrast, CircaFMM identified four to seven clusters per tissue, capturing approximately symmetric and asymmetric profiles, and waveforms with more localized temporal transitions (Supplementary Figure~S32 and Supplementary Table~S12). The correspondence between Cosinor and CircaFMM clusters was not generally one-to-one, indicating that the waveform families identified by CircaFMM are not merely refinements of sinusoidal clusters (Supplementary Table~S13).

\begin{figure}[t]
\centering

\begin{minipage}{\textwidth}
\centering
\includegraphics[width=\linewidth]{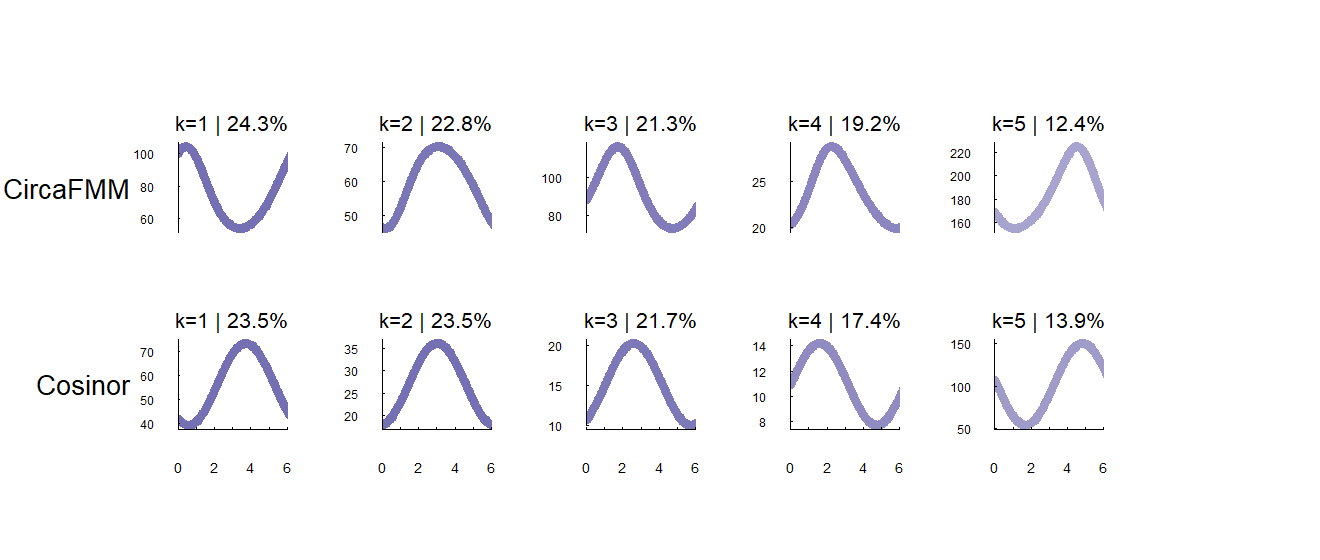}
%\caption*{(a) CircaFMM cluster structures.}
\end{minipage}

\vspace{-1cm}

\begin{minipage}{\textwidth}
\centering
\includegraphics[width=\linewidth]{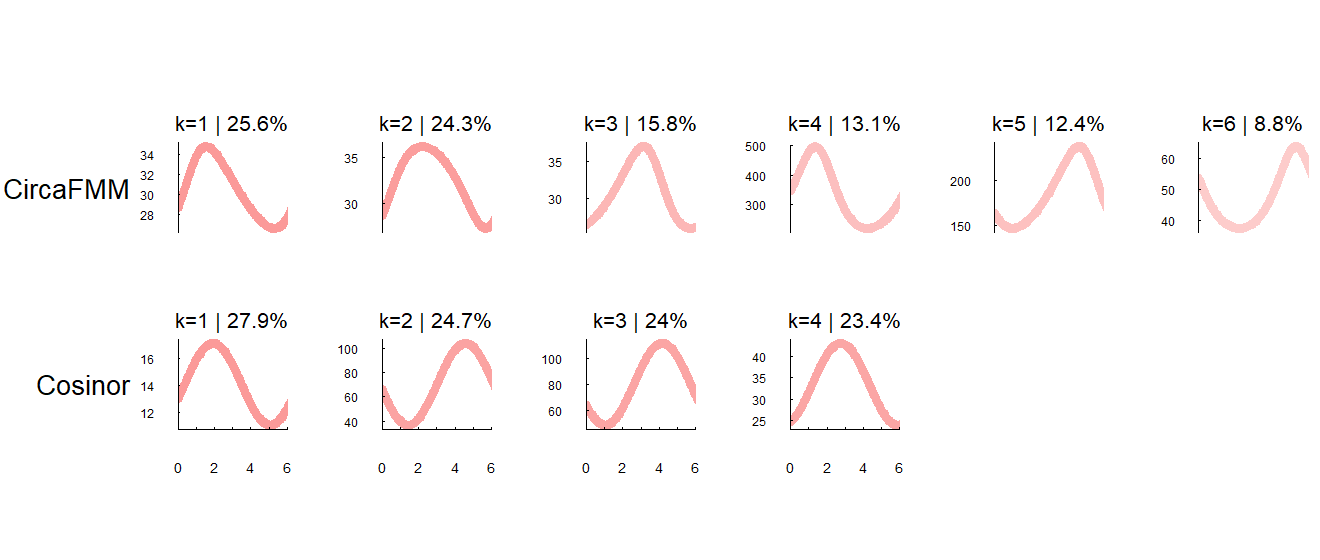}
%\caption*{(b) Cosinor cluster structures.}
\end{minipage}

\caption{Comparison of functional clustering structures identified from CircaFMM and Cosinor rhythmic detections. Top: AOR; Bottom:  PIT.  Curves represent the average fitted CircaFMM waveform for each cluster.  Clusters are sorted by prevalence within tissue, from most to least represented. Colour intensity is proportional to the percentage of genes assigned to the corresponding cluster, with darker shades indicating higher prevalence.}
\label{fig:cluster}
\end{figure}

Representative tissue examples are shown in Figure~\ref{fig:cluster}. In AOR, a tissue combining both symmetric and asymmetric waveform organizations, the dominant CircaFMM cluster ($k=1$, comprising 24.3\% of the  CircaFMM detections) was characterized by moderately asymmetric rhythmic profiles and showed limited correspondence with any Cosinor cluster. Similarly, in PIT, a tissue characterized by pronounced asymmetric waveform diversity, at least two CircaFMM clusters ($k=4$ and $k=6$) represented well-defined asymmetric waveforms without clear Cosinor counterparts. These results indicate that CircaFMM captures waveform heterogeneity that remains unresolved when rhythmicity is constrained to symmetric sinusoidal representations.

% --------------------------------------------
\subsection{CircaFMM circadian phase analysis}
\label{sec:gtex_phase_analysis}

\begin{figure}
    \centering
    \includegraphics[width=1\linewidth]{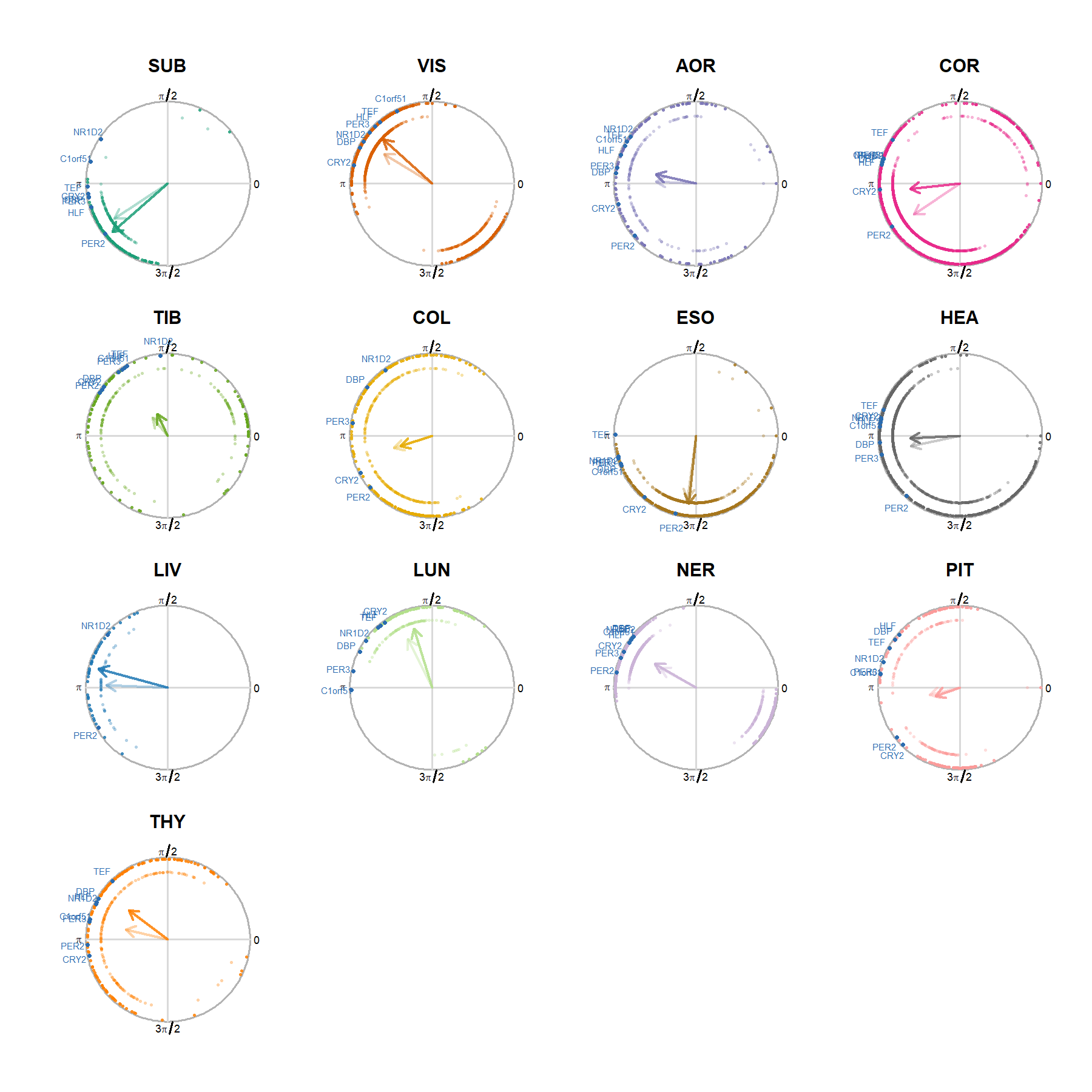}
    \caption{Circular distribution of peak expression times ($t^U$) for SR genes across tissues. Points represent gene-specific peak estimates obtained from CircaFMM (outer ring) and Cosinor (inner ring).  CircaFMM peak estimates corresponding to SR genes detected in at least 10 tissues are highlighted in blue. Arrows denote the mean resultant vector for each method, with direction indicating the circular mean peak time and length reflecting the concentration of phase estimates ($\overline{R}$), scaled relative to the corresponding ring radius.}
    \label{fig:peaksSR}
\end{figure}

We next examined whether this increased flexibility of CircaFMM preserved biologically meaningful circadian phase information, focusing on peak expression timing ($t^U$), a key landmark for interpreting the temporal organization of gene activity across tissues.
For SR genes, CircaFMM peak estimates showed structured tissue-specific distributions that closely resembled those obtained with Cosinor (Figure~\ref{fig:peaksSR}). Circular mean directions were nearly coincident in most tissues, indicating that CircaFMM preserved
the established phase organization of shared rhythmic genes.

Peak phases were generally concentrated within a dominant $6$-hour interval approximately spanning $[\pi/2,\,3\pi/2)$, or organized into two near-antiphasic groups, consistent with previous findings \citep{larriba2023circust}. 
Canonical SR clock genes detected by CircaFMM in at least 10 tissues also showed coherent phase organization, with peak phases shifted relative to \textit{ARNTL}, whose expression was centered near phase 0.  This agrees with the conserved temporal ordering of clock components reported in diurnal mammals \citep{zhang2014circadian,ruben2018database,mure2018diurnal}. 

Table~\ref{tab:peaks_agreement} summarizes the agreement between CircaFMM and Cosinor peak estimates. Circular correlations were high, exceeding 0.8 in 11 of 13 tissues and 0.95 in AOR, TIB, COL, PIT, and THY. 
Most tissues showed median absolute angular errors below 1 hour, indicating close gene-level agreement in peak timing.  Mean resultant lengths were slightly lower for CircaFMM than for Cosinor, consistent with greater waveform flexibility but preservation of the overall phase structure. Metric definitions and additional phase analyses are provided in Supplementary Section~S6.5.1.

\begin{table}[t]
\centering
\caption{Agreement between CircaFMM and Cosinor peak estimates across tissues for SR genes. Metrics include circular correlation ($\rho$), median absolute angular error in hours ($Me(|\Delta|)$), and concentration measures ($\overline{R}$) for both methods.}
\label{tab:peaks_agreement}
\begin{tabular}{lcccc}
\toprule
Tissue 
& $\rho$ 
& $Me(|\Delta|)$ (h) 
& $\overline{R}_{\mathrm{FMM}}$ 
& $\overline{R}_{\mathrm{Cosinor}}$ \\
\midrule

SUB & 0.858 & 0.682 & 0.914 & 0.951 \\
VIS & 0.698 & 0.551 & 0.819 & 0.835 \\
AOR & 0.981 & 0.551 & 0.511 & 0.597 \\
COR & 0.895 & 2.110 & 0.625 & 0.823 \\
TIB & 0.989 & 0.833 & 0.298 & 0.349 \\
COL & 0.973 & 0.937 & 0.413 & 0.592 \\
ESO & 0.774 & 0.975 & 0.849 & 0.947 \\
HEA & 0.935 & 0.796 & 0.617 & 0.744 \\
LIV & 0.946 & 0.875 & 0.891 & 0.914 \\
LUN & 0.854 & 0.584 & 0.767 & 0.809 \\
NER & 0.812 & 0.256 & 0.589 & 0.601 \\
PIT & 0.992 & 0.640 & 0.321 & 0.460 \\
THY & 0.968 & 1.230 & 0.612 & 0.643 \\

\bottomrule
\hline
\end{tabular}
\end{table}

For FR genes, peak-phase distributions were broader and spanned the full circular domain, reflecting more tissue-dependent temporal organization (Supplementary Figures~S33--S34). 
This broader distribution should not be interpreted as loss of temporal structure. Rather, it is consistent with the waveform heterogeneity described above and with CircaFMM-specific rhythmic candidates having more variable tissue-dependent timing.

To assess the reliability of FR phase estimates, we performed a parametric bootstrap analysis to derive $95\%$ percentile confidence intervals for peak $t^U$ and trough $t^L$ timings. Bootstrap standard errors were low and similar for both landmarks, averaging approximately 1.25 hours across tissues  (Table~\ref{tab:bootstrap_phase_uncertainty}). 
This indicates that, despite the non-sinusoidal nature of FR rhythms, CircaFMM achieved stable and balanced estimation of both peak and trough timings.

The proportion of genes with overlapping peak--trough 95\% confidence intervals was negligible in most tissues and exceeded 1\% only in HEA and THY. This limited overlap indicates that FR detections generally correspond to well-defined asymmetric waveforms with biologically interpretable circadian landmarks, rather than spiked or excessively asymmetric profiles that could artificially inflate rhythmicity calls. The slightly higher overlap in HEA and THY is consistent with their stronger asymmetry signatures and moderate-to-high sampling irregularity (Supplementary Figure~S32 and Supplementary Table~S6). Even in these settings, CircaFMM maintained stable and biologically coherent phase estimation.

Details of the bootstrap procedure are provided in Supplementary Section~S6.5.2. Complete bootstrap uncertainty estimates, together with the corresponding CircaFMM peak and trough phase estimates, are provided in Supplementary Data~1.
% La referencia al "data" o fichero?? en el supp hice una referencia a un fichero en la calibracion de lambda

\begin{table}[t]
\centering
\caption{Bootstrap-based uncertainty summaries for peak ($t^U$) and trough ($t^L$) phase estimates across tissues. The bootstrap standard errors $SE_{\mathrm{boot}}$ for $t^U$ and $t^L$ are summarized across genes within each tissue and reported as mean $\pm$ standard deviation, in hours (h). The overlap percentage denotes the proportion of genes whose 95\% bootstrap percentile confidence intervals for $t^U$ and $t^L$ overlap.}
\label{tab:bootstrap_phase_uncertainty}
\begin{tabular}{lccc}
\toprule
Tissue 
& $SE_{\mathrm{boot}}(t^U)$ (h)
& $SE_{\mathrm{boot}}(t^L)$ (h) 
& Overlap (\%) \\
\midrule

SUB & $1.293 \pm 0.445$ & $1.149 \pm 0.454$ & $0.356$ \\
VIS & $1.164 \pm 0.341$ & $1.183 \pm 0.382$ & $0.017$ \\
AOR & $1.194 \pm 0.307$ & $1.196 \pm 0.356$ & $0.000$ \\
COR & $1.275 \pm 0.630$ & $1.354 \pm 0.619$ & $0.054$ \\
TIB & $1.191 \pm 0.416$ & $1.316 \pm 0.428$ & $0.000$ \\
COL & $1.351 \pm 0.387$ & $1.268 \pm 0.433$ & $0.076$ \\
ESO & $1.409 \pm 0.445$ & $1.275 \pm 0.471$ & $0.268$ \\
HEA & $1.258 \pm 0.429$ & $1.483 \pm 0.553$ & $1.598$ \\
LIV & $1.028 \pm 0.238$ & $1.160 \pm 0.335$ & $0.000$ \\
LUN & $1.208 \pm 0.384$ & $1.460 \pm 0.420$ & $0.184$ \\
NER & $1.159 \pm 0.401$ & $1.204 \pm 0.512$ & $0.024$ \\
PIT & $1.285 \pm 0.364$ & $1.258 \pm 0.387$ & $0.000$ \\
THY & $1.245 \pm 0.467$ & $1.485 \pm 0.416$ & $1.192$ \\

\bottomrule
\end{tabular}
\end{table}

% -----------------------------------------------
% Discussion
\section{Discussion}

CircaFMM addresses a common limitation in circadian transcriptomic analysis: rhythmic profiles may be poorly described by a fixed sinusoidal model, especially under irregular sampling or weak signal strength. It combines FMM$_1$ waveform modelling \citep{rueda2019frequency}, a phase-based definition of rhythmicity, design-aware estimation and bootstrap-calibrated inference. Its main contribution is to test time dependence induced by the M\"obius phase, rather than nonzero amplitude in a fixed harmonic basis, while retaining precise estimates of waveform morphology and biologically interpretable circadian landmarks. The CircaFMM practical applicability is supported by a freely available application that enables users to analyse their own datasets through an intuitive interface and guided examples.

The simulation study supported the statistical behaviour of the CircaFMM test. Because the null is non-regular, bootstrap calibration was required instead of standard likelihood-ratio reference distributions. CircaFMM maintained type-I error close to nominal levels, improved power over Cosinor mainly for asymmetric or localized waveforms, and remained competitive for approximately sinusoidal profiles.

In GTEx-13T, anchoring the comparison to \citet{ruben2018database} allowed CircaFMM and Cosinor to be evaluated under the same CYCLOPS-derived ordering \citep{anafi2017cyclops} and across the same set of tissues.  CircaFMM recovered well-established circadian biology, while identifying additional CircaFMM-specific rhythmic candidates. These should not be interpreted as experimentally validated new circadian genes, but as statistically supported profiles with lower SNR, greater waveform asymmetry, and non-sinusoidal temporal organization. Stable peak/trough estimates support that they reflect rhythmic structure.

Several limitations should be considered. First, CircaFMM requires circadian times or sample-level phase estimates. This limitation is inherent to cross-sectional and postmortem transcriptomic studies, where unobserved internal circadian time, phase-reconstruction error, time-of-death uncertainty, and donor-, central-, or tissue-level phase differences may attenuate rhythmic signals or shift peak estimates. In GTEx-13T, rhythmicity calls are conditional on the CYCLOPS-derived ordering. But, circadian phase reconstruction remains an active methodological challenge, as reflected by recent developments such as CHIRAL, CIRCUST, updated CYCLOPS implementations, and deep-learning-based approaches \citep{talamanca2023sex,larriba2023circust,hammarlund2025cyclops,han2026dcpr,palumaa2026local}.

Second, observed temporal patterns in cross-sectional and postmortem studies such as GTEx-13T, may reflect not only circadian regulation but also inter-individual variability, postmortem effects, tissue composition, and unmeasured clinical or environmental factors \citep{ruben2018database,mure2018diurnal,wucher2023day}.

Finally, CircaFMM is computationally more demanding than methods with standard asymptotic reference distributions because inference relies on resampling-based calibration. Grid precomputation, reuse of design-dependent quantities, sequential Monte Carlo screening, and parallel computing substantially reduce this cost, but bootstrap calibration remains the main bottleneck for strongly rhythmic genes.

These limitations define the scope of interpretation rather than the utility of the approach: CircaFMM is designed for settings in which circadian phase is estimated, sampling is uneven, and rhythmic profiles may deviate from a simple cosine. Overall, CircaFMM provides a complementary inferential framework that extends statistical rhythmicity assessment beyond fixed sinusoidal assumptions while retaining interpretable circadian features and precise waveform characterization, making it particularly well-suited to postmortem circadian transcriptomic studies.

%\textcolor{red}{Future work will focus on extending the framework to multi-tissue and cell-type-resolved settings, where shared temporal structure may improve rhythmicity detection and characterization. Another promising direction is the integration of circadian phase reconstruction and rhythmicity inference within a unified framework, reducing dependence on a single estimated circadian ordering while improving statistical efficiency in large-scale transcriptomic studies.}

%%%%%%%%%%%%%%

\section*{Acknowledgments}
The authors gratefully acknowledge Dr. Marc de Ruben for providing additional information and helpful clarifications regarding the GTEx-13T dataset.

\section*{Funding}
This work was supported by the Spanish Ministry of Science, Innovation and Universities through project MOBIOS [PID2023-147839OB-I00] and by the Department of Education of the Junta de Castilla y León and FEDER Funds [CLU-2025-1-02- IMUVA]. J.A. was supported by a PhD fellowship from Fundación Ramón Areces, within the programme “Ayudas Fundación Ramón Areces para la realización de tesis doctorales en Ciencias de la Vida y de la Materia 2025”.

\section*{Availability and implementation}
CircaFMM is implemented in R. An interactive Shiny application, together with the source code, documentation, and example workflows, is available at \url{https://circafmm.shinyapps.io/circafmm_shinyapp/}.  The source code is freely available at \url{https://github.com/FMMGroupVa/CircaFMM}.

%USE THE BELOW OPTIONS IN CASE YOU NEED AUTHOR YEAR FORMAT.
% Bibliografía numérica
\bibliographystyle{unsrtnat}
\bibliography{reference}

\clearpage


\section*{Supplementary material}

# Supplementary Materials for "*CircaFMM: a flexible statistical framework for circadian rhythmicity detection in transcriptomics*"

Joshua S. Abston[1], Yolanda Larriba[1,2,*], Itziar Fernández[1,3,*]

[1]Departamento de Estadística e Investigación Operativa, Universidad de Valladolid, 47011,Valladolid, Spain [2]Mathematics Research Institute of the University of Valladolid (IMUVA), Universidad de Valladolid, 47011, Valladolid, Spain [3]Biomedical Research Networking Center in Bioengineering, Biomaterials and Nanomedicine (CIBER-BBN), Valladolid, Spain
* Corresponding authors: `yolanda.larriba@uva.es`; `itziar.fernandez@uva.es`

# Contents



# S1 Formal basis of the CircaFMM rhythmicity test

This section provides the formal results supporting the phase-based definition of non-rhythmicity used in the CircaFMM test. We first show that the Möbius phase is constant in time if and only if $\omega = 0$. We then justify that, although the phase function has a jump at $t = \alpha + \pi$ when $\omega > 0$, the composed oscillatory term $\cos(\beta + \phi(t; \alpha, \omega))$, and hence the mean profile $\mu(t)$, is well defined and continuous on the circle. Finally, we state the resulting flat non-rhythmic regime that motivates the null hypothesis $H_0 : \omega = 0$.

For completeness, we define the corresponding mean function

$$\mu(t) = M + A\cos\bigl(\beta + \phi(t;\alpha,\omega)\bigr).$$

**Proposition 1** (Constancy of the Möbius phase)**.** *Let* $t \in [0, 2\pi)$ *and define*

$$\phi(t;\alpha,\omega) = 2\arctan\left(\omega\,\tan\left(\frac{t-\alpha}{2}\right)\right), \qquad \alpha \in [0,2\pi),\ \omega \in [0,1].$$

*Then* $\phi(t;\alpha,\omega)$ *is constant in* $t$ *if and only if* $\omega = 0$*. In particular, when* $\omega = 0$*, then*

$$\phi(t;\alpha,0) \equiv 0, \qquad t \in [0,2\pi).$$

*Proof.* If $\omega = 0$, then for all $t \in [0, 2\pi)$,

$$\phi(t;\alpha,0) = 2\arctan(0) = 0,$$

and hence $\phi$ is constant.

Conversely, assume $\omega > 0$. At all points where the tangent term is finite, that is, for $t \neq \alpha + \pi$ modulo $2\pi$, the phase is differentiable and

$$\phi'(t;\alpha,\omega) = \frac{\omega}{\cos^2\bigl(\frac{t-\alpha}{2}\bigr) + \omega^2\sin^2\bigl(\frac{t-\alpha}{2}\bigr)} > 0.$$

Thus, on any open interval not containing the singular point $t = \alpha+\pi$ modulo $2\pi$, the phase is strictly increasing. Therefore, for $\omega > 0$, $\phi(t;\alpha,\omega)$ cannot be constant as a function of $t$. □

**Remark 1** (Jump of $\phi(t;\alpha,\omega)$ and continuity of $\mu(t)$)**.** *For* $\omega > 0$*, the tangent term*

$$\tan\left(\frac{t-\alpha}{2}\right)$$

*has a vertical asymptote at* $t = \alpha+\pi$ *modulo* $2\pi$*, which induces a jump discontinuity in the Möbius phase* $\phi(t;\alpha,\omega)$*. More precisely,*

$$\lim_{t\to(\alpha+\pi)^-} \phi(t;\alpha,\omega) = \pi, \qquad \lim_{t\to(\alpha+\pi)^+} \phi(t;\alpha,\omega) = -\pi.$$

*Since* $\cos(\cdot)$ *is* $2\pi$*-periodic and*

$$\cos(\beta+\pi) = \cos(\beta-\pi),$$

*the composed term*

$$t \mapsto \cos\bigl(\beta + \phi(t;\alpha,\omega)\bigr)$$

*has matching left and right limits at the jump point. Therefore, the oscillatory component, and hence* $\mu(t)$*, is well defined and continuous on the circle for all* $\omega \in [0,1]$*.*

**Corollary 1** (Flat non-rhythmic regime)**.** *The mean function* $\mu(t)$ *is constant in* $t$ *whenever* $A = 0$ *or* $\omega = 0$*. Conversely, if* $A > 0$*, then* $\mu(t)$ *is constant in* $t$ *if and only if* $\omega = 0$*.*

*Proof.* If $A = 0$, then $\mu(t) = M$, which is constant. If $\omega = 0$, Proposition 1 gives $\phi(t;\alpha,0) \equiv 0$, and therefore

$$\mu(t) = M + A\cos(\beta),$$

which is also constant.

Conversely, assume $A > 0$ and $\omega > 0$. Consider any open interval $I$ that does not contain the singular point $t = \alpha + \pi$ modulo $2\pi$. On $I$, the phase $\phi(t;\alpha,\omega)$ is differentiable and, by the proof of Proposition 1,

$$\phi'(t;\alpha,\omega) > 0, \qquad t \in I.$$

On this interval,

$$\mu'(t) = -A\sin\bigl(\beta + \phi(t;\alpha,\omega)\bigr)\,\phi'(t;\alpha,\omega).$$

If $\mu(t)$ were constant, then $\mu'(t) = 0$ for all $t \in I$. Since $A > 0$ and $\phi'(t;\alpha,\omega) > 0$ on $I$, this would imply

$$\sin\bigl(\beta + \phi(t;\alpha,\omega)\bigr) = 0, \qquad t \in I.$$

However, $\phi(t;\alpha,\omega)$ is strictly increasing on $I$, hence the image of $t \mapsto \beta + \phi(t;\alpha,\omega)$ is an interval containing infinitely values which cannot remain in the discrete set $\{k\pi : k \in \mathbb{Z}\}$ for all $t \in I$. This is a contradiction. Therefore, when $A > 0$ and $\omega > 0$, the mean function $\mu(t)$ cannot be constant. Hence, if $A > 0$ and $\mu(t)$ is constant in $t$, necessarily $\omega = 0$. □

# S2 Time-density weights on the circle

This section describes the procedure to compute time-density weights for irregular circadian sampling designs, characterized by densely sampled regions and sparsely represented time windows over $[0, 2\pi)$. We first introduce the general inverse-density weighting framework. We then estimate the sampling-time density using a standard von Mises kernel estimator and select its concentration parameter via leave-one-out cross-validation. Finally, we apply the procedure to the GTEx-13T dataset and evaluate the impact of the concentration parameter on the resulting weighting scheme.

## S2.1 Unnormalized and normalized inverse-density weights.

To account for uneven temporal coverage on $[0, 2\pi)$, we define nonnegative weights that depend only on the sampling times $\{t_i\}_{i=1}^n$, and not on gene-expression values. We first define unnormalized inverse-density weights $\tilde{w}_i$, and then normalize them to obtain the final weights $w_i$ satisfying $\sum_{i=1}^n w_i = n$.

Following design-aware weighting ideas for circular/trigonometric regression (Gorczyca and Sefas, 2025), we use an inverse-density construction to reduce the influence of densely sampled time windows and to approximate the contribution that would be obtained under more balanced temporal coverage. Specifically, we set

$$\tilde{w}_i = \frac{1}{\max\{\widehat{f}_{\hat{\kappa}}(t_i), \varepsilon_w\}}, \qquad i = 1, \dots, n,$$

where $\widehat{f}_{\hat{\kappa}}$ is a circular kernel density estimate of the empirical sampling-time density, with $\hat{\kappa}$ the concentration parameter selected by leave-one-out, and $\varepsilon_w > 0$ is a small numerical floor to avoid excessively large weights in very sparse regions. Final weights are obtained by normalization so that $\sum_{i=1}^n w_i = n$.

The final weights are then defined as

$$w_i = n\,\frac{\tilde{w}_i}{\sum_{m=1}^n \tilde{w}_m}, \qquad i = 1, \dots, n,$$

so that $\sum_{i=1}^n w_i = n$.

## S2.2 von Mises kernel density estimator

Let $t_1, \dots, t_n \in [0, 2\pi)$ denote sampling times. We estimate the circular sampling-time density by kernel density estimation (KDE) using a von Mises kernel, with concentration parameter $\kappa$. For $\kappa > 0$, the von Mises kernel is

$$K_\kappa(\Delta) = \frac{1}{2\pi I_0(\kappa)} \exp\big(\kappa \cos \Delta\big), \qquad \Delta \in (-\pi, \pi],$$

where $I_0$ is the modified Bessel function of order 0. Circular differences are computed via

$$\operatorname{angdiff}(\theta) = \operatorname{atan2}(\sin\theta, \cos\theta) \in (-\pi, \pi].$$

The KDE evaluated at $t$ is

$$\widehat{f}_\kappa(t) = \frac{1}{n} \sum_{j=1}^n K_\kappa\big(\operatorname{angdiff}(t - t_j)\big).$$

## S2.3 Choice of $\kappa$ by leave-one-out cross-validation

We select the concentration parameter $\kappa$ using leave-one-out log-likelihood cross-validation (LOO-CV) over a log-spaced candidate grid $\mathcal{K}$.

For each $\kappa \in \mathcal{K}$, let

$$\widehat{f}_{-i,\kappa}(t_i) = \frac{1}{n-1} \sum_{j \neq i} K_\kappa\big(\operatorname{angdiff}(t_i - t_j)\big)$$

denote the leave-one-out kernel density estimate evaluated at $t_i$.

The corresponding LOO-CV criterion is defined as

$$\mathrm{CV}_{\mathrm{LOO}}(\kappa) = \sum_{i=1}^n \log\Big(\max\{\widehat{f}_{-i,\kappa}(t_i), \varepsilon_{\mathrm{CV}}\}\Big),$$

where $\varepsilon_{\mathrm{CV}} > 0$ is a small numerical constant introduced to avoid underflow.

The selected concentration parameter is then

$$\hat{\kappa} = \arg\max_{\kappa \in \mathcal{K}} \mathrm{CV}_{\mathrm{LOO}}(\kappa).$$

The selected value $\hat{\kappa}$ is solely used to estimate the sampling-time density from which the inverse-density weights are computed.

For completeness, Algorithm S1 summarizes the steps used to select $\hat{\kappa}$ by LOO–CV and to construct the normalized inverse-density time-weights $\{w_i\}_{i=1}^n$.

**Algorithm S1** Inverse-density time-weights via von Mises KDE and LOO–CV.

**Inputs:** sampling times $\{t_i\}_{i=1}^n \subset [0, 2\pi)$, candidate set $\mathcal{K}$, floors $\varepsilon_{\mathrm{CV}} > 0$ and $\varepsilon_w > 0$
**for** $\kappa \in \mathcal{K}$ **do**
  **for** $i = 1, \ldots, n$ **do**
    $\widehat{f}_{-i,\kappa}(t_i) \leftarrow \frac{1}{n-1} \sum_{j \neq i} K_\kappa\big(\operatorname{angdiff}(t_i - t_j)\big)$
  **end for**
  $\mathrm{CV}_{\mathrm{LOO}}(\kappa) \leftarrow \sum_{i=1}^n \log\Big(\max\{\widehat{f}_{-i,\kappa}(t_i), \varepsilon_{\mathrm{LCV}}\}\Big)$
**end for**
$\hat{\kappa} \leftarrow \arg\max_{\kappa \in \mathcal{K}} \mathrm{CV}_{\mathrm{LOO}}(\kappa)$
**for** $i = 1, \ldots, n$ **do**
  $\widehat{f}_{\hat{\kappa}}(t_i) \leftarrow \frac{1}{n} \sum_{j=1}^n K_{\hat{\kappa}}\big(\operatorname{angdiff}(t_i - t_j)\big)$
  $\tilde{w}_i \leftarrow 1/\max\{\widehat{f}_{\hat{\kappa}}(t_i), \varepsilon_w\}$
**end for**
$w_i \leftarrow n\,\tilde{w}_i \Big/ \sum_{m=1}^n \tilde{w}_m \qquad (i = 1, \ldots, n)$
**Output:** normalized time-weights $\{w_i\}_{i=1}^n$

### S2.4 LOO–CV choice of $\kappa$ in GTEx-13T database

We applied the LOO–CV scheme to select the von Mises concentration $\widehat{\kappa}$ within each GTEx-13T tissue, using the candidate set

$$\mathcal{K} = \{\exp(u_\ell)\}_{\ell=1}^{10}, \qquad u_\ell \text{ equally spaced on } [\log 1, \log 64].$$

Because the concentration parameter $\kappa$ is not directly interpretable as an angular bandwidth, we also report the half-maximum angular radius $\Delta_{1/2}$, defined by

$$\frac{\exp\{\kappa \cos(\Delta_{1/2})\}}{\exp\{\kappa\}} = \frac{1}{2} \quad \Longrightarrow \quad \cos(\Delta_{1/2}) = 1 - \frac{\ln 2}{\kappa}, \qquad \Delta_{1/2} = \arccos\left(1 - \frac{\ln 2}{\kappa}\right),$$

which is measured in radians (rad), and well-defined only for $\kappa > \ln 2$. For readability, we also report $\Delta_{1/2}$ in degrees,

$$\Delta^{\circ}_{1/2} = \frac{180}{\pi}\,\Delta_{1/2}.$$

This quantity represents the angular distance from the kernel center at which its value decreases to one half of its maximum. Overall, larger values of $\widehat{\kappa}$ correspond to more localized smoothing, whereas smaller values correspond to broader smoothing.

Across the GTEx-13T tissues, LOO–CV selected a small set of recurrent concentration values (Table S1), indicating that similar levels of smoothing are adequate across tissues despite differences in sample size and expression profiles. Several tissues shared $\widehat{\kappa} = 6.35$, corresponding to a broader kernel with half-maximum radius $\Delta_{1/2} \approx 0.472$ rad ($\approx 27.0°$). A second group of tissues yielded $\widehat{\kappa} = 10.08$ ($\Delta_{1/2} \approx 0.373$ rad ($\approx 21.4°$), whereas SUB and COR selected a more localized kernel with $\widehat{\kappa} = 16.00$ ($\Delta_{1/2} \approx 0.295$ rad; $\approx 16.9°$). The broadest kernel was selected for TIB, with $\widehat{\kappa} = 4.00$ and $\Delta_{1/2} \approx 0.598$ rad ($\approx 34.2°$). Notably, COR and TIB correspond to the tissues with the lowest and highest proportion of temporal gaps, respectively, which is consistent with their selection of the most localized and broadest kernels (Table S6)

Figure S1 displays the LOO–CV curves across GTEx-13T tissues, together with the selected concentration parameter $\hat{\kappa}$ (vertical line).

## S3 Stabilizing optimization in nonregular regimes for the $\mathrm{FMM}_1$ model

This section provides the implementation details underlying the stabilized estimation procedure summarized in the main text. In particular, we describe the weighted least-squares structure conditional on the nonlinear phase parameters, the motivation for stabilizing the small-$\omega$ regime, the normalized oscillatory ranking criterion, the curvature-type roughness penalty, the optional local refinement step, and the Monte Carlo calibration of the regularization parameter $\lambda^\star$. Finally, we apply the procedure to the GTEx-13T dataset, performing a local sensitivity analysis around the calibrated value and investigating the impact of roughness regularization on sharpness-parameter estimation.

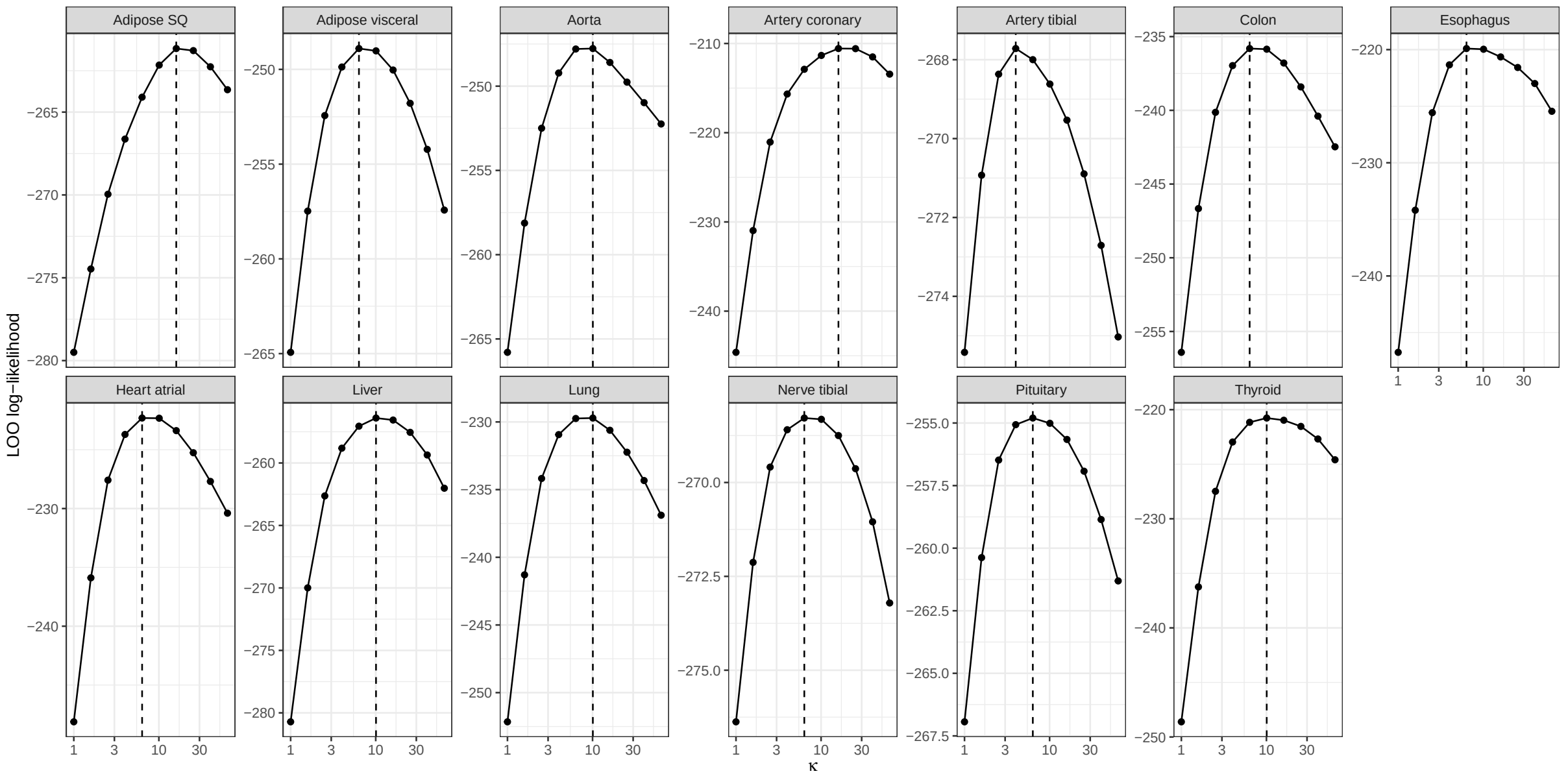


Figure S1: LOO–CV curves by tissue in the GTEx-13T dataset. The vertical line marks $\widehat{\kappa}$.

Table S1: Selected $\widehat{\kappa}$ and half-maximum angular radius $\Delta_{1/2}$ (radians and degrees) by tissue for the GTEx-13T dataset.

| Tissue | $n$ | $\widehat{\kappa}$ | $\Delta_{1/2}$ (rad) | $\Delta^{\circ}_{1/2}$ |
|---|---|---|---|---|
| Adipose subcutaneous (SUB) | 160 | 16.00 | 0.295 | 16.9 |
| Adipose visceral (VIS) | 160 | 6.35 | 0.472 | 27.0 |
| Aorta (AOR) | 160 | 10.08 | 0.373 | 21.4 |
| Coronary artery (COR) | 160 | 16.00 | 0.295 | 16.9 |
| Tibial artery (TIB) | 160 | 4.00 | 0.598 | 34.2 |
| Sigmoid colon (COL) | 160 | 6.35 | 0.472 | 27.0 |
| Esophagus mucosa (ESO) | 160 | 6.35 | 0.472 | 27.0 |
| Heart atrial appendage (HEA) | 160 | 6.35 | 0.472 | 27.0 |
| Liver (LIV) | 175 | 10.08 | 0.373 | 21.4 |
| Lung (LUN) | 160 | 10.08 | 0.373 | 21.4 |
| Tibial nerve (NER) | 160 | 6.35 | 0.472 | 27.0 |
| Pituitary (PIT) | 160 | 6.35 | 0.472 | 27.0 |
| Thyroid (THY) | 160 | 10.08 | 0.373 | 21.4 |

## S3.1 Model representation and weighted least-squares structure

Let $X(t_i)$ denote the expression level of a given gene measured at sampling time $t_i \in [0, 2\pi)$, for $i = 1, \ldots, n$, and let

$$X(t_i) = M + A\cos\big(\beta + \phi(t_i; \alpha, \omega)\big) + \varepsilon_i, \qquad \varepsilon_i \sim \mathcal{N}(0, \sigma^2),$$

be the single-component $\text{FMM}_1$ model.

For any fixed $(\alpha, \omega)$, define the phase-induced basis functions

$$C_{\alpha,\omega}(t_i) = \cos\big(\phi(t_i; \alpha, \omega)\big), \qquad S_{\alpha,\omega}(t_i) = \sin\big(\phi(t_i; \alpha, \omega)\big).$$

Using the reparametrization

$$\delta = A\cos(\beta), \qquad \gamma = -A\sin(\beta),$$

the model becomes linear in the coefficients $(M, \delta, \gamma)$ conditional on $(\alpha, \omega)$,

$$X(t_i) = M + \delta\, C_{\alpha,\omega}(t_i) + \gamma\, S_{\alpha,\omega}(t_i) + \varepsilon_i, \qquad \varepsilon_i \sim \mathcal{N}(0, \sigma^2).$$

Hence, for each $(\alpha, \omega)$, estimation of $(M, \delta, \gamma)$ reduces to a weighted least-squares problem.

Let $w_i \geq 0$ denote the time-density weights, normalized so that $\sum_{i=1}^{n} w_i = n$. We use the weighted inner product

$$\langle a, b \rangle_w = \sum_{i=1}^{n} w_i a_i b_i$$

and the associated weighted mean

$$\bar{x}_w = \frac{\langle x, \mathbf{1}\rangle_w}{\langle \mathbf{1}, \mathbf{1}\rangle_w}.$$

When assessing the oscillatory component induced by a candidate pair $(\alpha, \omega)$, we work with centered observations

$$\tilde{x}_i = x_i - \bar{x}_w,$$

so that the ranking step focuses on oscillatory alignment after removing the baseline contribution.

## S3.2 Motivation: small-$\omega$ degeneracy of SSE-based optimization

A key practical difficulty in fitting $\text{FMM}_1$ arises from the behavior of the Möbius phase for small values of $\omega$. The map

$$t \mapsto \tan\left(\frac{t-\alpha}{2}\right)$$

has a vertical asymptote at $t = \alpha + \pi$ modulo $2\pi$. Consequently, as $\omega \downarrow 0$, the phase function $\phi(t; \alpha, \omega)$ becomes nearly constant over most of the circle and changes very rapidly in a small neighborhood of $t \approx \alpha + \pi$.

This behavior is quantified by the derivative

$$\phi'(t; \alpha, \omega) = \frac{\omega}{\cos^2\left(\frac{t-\alpha}{2}\right) + \omega^2 \sin^2\left(\frac{t-\alpha}{2}\right)}, \qquad t \neq \alpha + \pi \pmod{2\pi}.$$

The denominator is minimized near $t = \alpha + \pi$, where the expression reaches values of order $1/\omega$. Thus, the maximum local rate of change diverges as $\omega \downarrow 0$. As a result, the oscillatory term

$$\cos\big(\beta + \phi(t; \alpha, \omega)\big)$$

vary extremely rapidly around $t \approx \alpha + \pi$ in small-$\omega$ regimes, yielding fitted profiles with highly localized transitions.

In the classical SSE/RSS-based fitting scheme followed by Rueda et al. (2019), $(\alpha, \omega)$ are selected on a coarse grid by minimizing a squared-error criterion. However, in finite samples, such a criterion can favor excessively localized transitions: placing a sharp transition near an isolated or influential observation may substantially reduce the squared residual at that point while incurring only a modest increase elsewhere. Since decreasing $\omega$ allows the transition to become increasingly concentrated, SSE-based optimization may exhibit a practical degeneracy in which the selected $\hat{\omega}$ is driven towards the lower boundary of its admissible range.

The same phenomenon may persist under weighted least squares, since the objective remains based on residual-error minimization. This motivates replacing direct SSE/RSS selection over $(\alpha, \omega)$ by a stabilized ranking criterion and a roughness penalty, as described below.

## S3.3 Normalized weighted oscillatory ranking criterion for selecting $(\alpha, \omega)$

To mitigate the small-$\omega$ degeneracy of squared-error selection, we replace RSS/SSE minimization over $(\alpha, \omega)$ by a normalized score based on the weighted alignment of the centered series with the two-dimensional oscillatory component induced by each candidate pair.

Conceptually, this mirrors the ranking of candidate oscillatory components in the generalized Lomb–Scargle framework, where each frequency defines a two-dimensional harmonic component through $\{\cos(ft), \sin(ft)\}$ and candidates are compared through normalized measures of fit to the centered series (Zechmeister and Kürster, 2009). In our setting, the harmonic regressors are replaced by the phase-induced basis functions $\{C_{\alpha,\omega}(t), S_{\alpha,\omega}(t)\}$, and the role of frequency is played by the pair $(\alpha, \omega)$.

We quantify the oscillatory component associated with $(\alpha, \omega)$ by the normalized quadratic functional

$$P(\alpha, \omega) = \frac{\langle \tilde{x}, C_{\alpha,\omega}\rangle_w^2 + \langle \tilde{x}, S_{\alpha,\omega}\rangle_w^2}{\langle C_{\alpha,\omega}, C_{\alpha,\omega}\rangle_w + \langle S_{\alpha,\omega}, S_{\alpha,\omega}\rangle_w}.$$

The numerator aggregates squared weighted inner products between the centered observations and the phase-induced basis functions, thereby quantifying the weighted projection of the centered expression profile onto the corresponding oscillatory subspace. The denominator normalizes by the total weighted energy of the basis functions. Consequently, $P(\alpha, \omega)$ measures the amount of oscillatory structure captured by the FMM basis for the candidate pair $(\alpha, \omega)$, providing a scale-normalized criterion for comparing candidate phase parameters across the search grid.

Unlike SSE, which is a pointwise quadratic loss, $P(\alpha, \omega)$ depends on global weighted inner products. As a consequence, it favors candidate oscillatory components that are coherently supported across the observed sampling times and is less prone to selecting extremely small $\omega$ values that accommodate isolated observations through highly localized transitions.

The score $P(\alpha, \omega)$ is used only to select the nonlinear phase parameters. Conditional on the selected pair, the linear coefficients are subsequently refitted by weighted least squares.

## S3.4 Roughness-regularized selection via a curvature-type penalty

The normalized projection score $P(\alpha, \omega)$ reduces the direct dependence on pointwise residual minimization, but it does not by itself penalize the roughness of the phase-induced basis functions. Therefore, additional stabilization is useful in the small-$\omega$ regime, where very localized transitions may still be selected in finite samples. We therefore regularize the selection of $(\alpha, \omega)$ by penalizing a curvature-type roughness functional computed from the phase-induced basis functions

$$\{C_{\alpha,\omega}(t), S_{\alpha,\omega}(t)\}.$$

### S3.4.1 Curvature roughness on the observed sampling times

Let $t_1 \leq \cdots \leq t_n$ be the circular ordering of the sampling times on $[0, 2\pi)$. Define circular gaps

$$\Delta t_i = t_{i+1} - t_i, \qquad i = 1, \ldots, n-1,$$

and

$$\Delta t_n = t_1 + 2\pi - t_n, \qquad t_{n+1} = t_1 + 2\pi.$$

For a fixed pair $(\alpha, \omega)$ and $i = 1, \ldots, n$, let

$$C_i = C_{\alpha,\omega}(t_i), \qquad S_i = S_{\alpha,\omega}(t_i),$$

and define first circular differences

$$\Delta C_i = C_{i+1} - C_i, \qquad \Delta S_i = S_{i+1} - S_i,$$

with the conventions $C_{n+1} = C_1$ and $S_{n+1} = S_1$. We then define discrete local slopes

$$\text{slope}_i^C = \frac{\Delta C_i}{\Delta t_i}, \qquad \text{slope}_i^S = \frac{\Delta S_i}{\Delta t_i}.$$

Changes in local slope are measured through circular first differences

$$\Delta\text{slope}_i^C = \text{slope}_{i+1}^C - \text{slope}_i^C, \qquad \Delta\text{slope}_i^S = \text{slope}_{i+1}^S - \text{slope}_i^S,$$

with circular indexing, i.e.

$$\text{slope}_{n+1}^C \equiv \text{slope}_1^C, \qquad \text{slope}_{n+1}^S \equiv \text{slope}_1^S.$$

The resulting curvature-type roughness functional is

$$R_{\text{obs}}(\alpha, \omega) = \sum_{i=1}^{n} \left(\Delta\text{slope}_i^C\right)^2 + \sum_{i=1}^{n} \left(\Delta\text{slope}_i^S\right)^2.$$

By construction, $R_{\text{obs}}(\alpha, \omega)$ depends only on the sampling times and the phase-induced basis functions. It is independent of the observed expression values and of the linear coefficients $(M, \delta, \gamma)$.

### S3.4.2 Optional dense-grid roughness

When the sampling design contains large gaps, the location of the Möbius transition near $t \approx \alpha + \pi$, may fall entirely within an unobserved interval. Conversely, highly irregular or nearly duplicated sampling times may make observed-time slopes sensitive to very small gaps. In such cases, $R_{\text{obs}}(\alpha, \omega)$ can be sensitive to the particular placement of observed times.

As a more stable alternative, we optionally compute roughness on a dense equispaced uniform grid

$$\mathcal{T}_{\text{d}} = \{u_j\}_{j=1}^m \subset [0, 2\pi), \qquad u_j = 2\pi(j-1)/m,$$

with constant spacing

$$\Delta u = 2\pi/m.$$

Defining

$$C_j^{\text{d}} = C_{\alpha,\omega}(u_j), \qquad S_j^{\text{d}} = S_{\alpha,\omega}(u_j),$$

and using circular indexing $u_{m+1} = u_1 + 2\pi$, we compute local slopes

$$\text{slope}_j^{C,\text{d}} = \frac{C_{j+1}^{\text{d}} - C_j^{\text{d}}}{\Delta u}, \qquad \text{slope}_j^{S,\text{d}} = \frac{S_{j+1}^{\text{d}} - S_j^{\text{d}}}{\Delta u},$$

and slope differences

$$\Delta\text{slope}_j^{C,\text{d}} = \text{slope}_{j+1}^{C,\text{d}} - \text{slope}_j^{C,\text{d}}, \qquad \Delta\text{slope}_j^{S,\text{d}} = \text{slope}_{j+1}^{S,\text{d}} - \text{slope}_j^{S,\text{d}}.$$

The dense-grid roughness is then defined as

$$R_{\text{dense}}(\alpha, \omega) = \sum_{j=1}^{m} \left(\Delta\text{slope}_j^{C,\text{d}}\right)^2 + \sum_{j=1}^{m} \left(\Delta\text{slope}_j^{S,\text{d}}\right)^2.$$

In practice, $m$ is fixed, for example $m = 512$, so $R_{\text{dense}}(\alpha, \omega)$ depends only on $(\alpha, \omega)$ and can be precomputed once over the $(\alpha, \omega)$ grid.

### S3.4.3 Penalized criterion

We modulate the roughness penalty to concentrate regularization in the small-$\omega$ regime by

$$g(\omega) = \left(\frac{\omega_0}{\omega + \omega_0}\right)^p, \qquad \omega_0 \in (0,1), \quad p \geq 1.$$

where $\omega_0 \in (0,1)$ determines the transition scale of the penalty, and $p \geq 1$ controls its rate of decay with $\omega_0 = 0.1$ and $p = 3$ in our implementation. The role of $g(\omega)$ is to focus regularization on very small values of $\omega$, where spike-like solutions are most likely to occur, while attenuating the penalty for moderate and large $\omega$.

Hence, let $R(\alpha, \omega)$ denote either $R_{\text{obs}}(\alpha, \omega)$ or $R_{\text{dense}}(\alpha, \omega)$. We select $(\hat{\alpha}, \hat{\omega})$ by maximizing

$$J(\alpha, \omega) = P(\alpha, \omega) - \lambda\, g(\omega)\, R(\alpha, \omega),$$

where $\lambda \geq 0$ is a dataset-level regularization parameter. Conditional on $(\hat{\alpha}, \hat{\omega})$, the remaining coefficients are estimated by weighted least squares.

### S3.4.4 Optional local refinement of $(\hat{\alpha}, \hat{\omega})$

After selecting a coarse maximizer $(\alpha^{(0)}, \omega^{(0)})$ on the $(\alpha, \omega)$ grid, we optionally apply a short local refinement step to reduce grid discretization effects. The refinement follows a simple pattern-search/coordinate-search strategy that explores a discrete neighborhood around the current pair, while keeping the same penalized objective $J(\alpha, \omega)$ used in the grid search (Lewis et al., 2000; Kolda et al., 2003).

Let $(\alpha^{(s)}, \omega^{(s)})$, be the current iterate at refinement step $s$, with step parameters $\Delta_\alpha^{(s)} > 0$ and $\eta^{(s)} > 1$, and $s = 1, \ldots, S$. By default $\Delta_\alpha^{(s)} = 2\pi/48$ and $\eta^{(s)} = 1.25$. At each step, we consider eight candidate neighbors: two $\alpha$-moves,

$$(\alpha^{(s)} - \Delta_\alpha^{(s)}, \omega^{(s)}), \qquad (\alpha^{(s)} + \Delta_\alpha^{(s)}, \omega^{(s)}),$$

two $\omega$-moves,

$$(\alpha^{(s)}, \omega^{(s)}\eta^{(s)}), \qquad (\alpha^{(s)}, \omega^{(s)}/\eta^{(s)}),$$

and four combined moves,

$$(\alpha^{(s)} - \Delta_\alpha^{(s)}, \omega^{(s)}\eta^{(s)}), \qquad (\alpha^{(s)} + \Delta_\alpha^{(s)}, \omega^{(s)}\eta^{(s)}),$$
$$(\alpha^{(s)} - \Delta_\alpha^{(s)}, \omega^{(s)}/\eta^{(s)}), \qquad (\alpha^{(s)} + \Delta_\alpha^{(s)}, \omega^{(s)}/\eta^{(s)}).$$

The phase-location parameter $\alpha$ is wrapped modulo $2\pi$, and $\omega$ is constrained to the admissible interval $[\omega_{\min}, \omega_{\max}]$ used in the search, with $\omega_{\min} = 0$ and $\omega_{\max} = 1$ by default.

For each admissible candidate, we compute the local score $J(\alpha, \omega)$. Rather than selecting a single maximizer solely by this score, we first identify the set of candidates whose scores lie within a small relative tolerance of the best local score. Specifically, if $J_{\max}^{(s)}$ is the largest finite candidate score, candidates are retained when

$$J(\alpha, \omega) \geq J_{\max}^{(s)} - \tau \max\{1, |J_{\max}^{(s)}|\},$$

with $\tau = 10^{-4}$ in our implementation.

Among these near-maximizing candidates, we apply a secondary tie-breaking criterion based on the penalized residual objective

$$\text{WRSS}(\alpha, \omega) + \lambda\, g(\omega)\, R(\alpha, \omega),$$

where $\text{WRSS}(\alpha, \omega)$ denotes the weighted residual sum of squares and is obtained from the WLS fit conditional on the candidate pair. The candidate with the smallest value of this secondary objective is selected. If this objective is tied up to numerical tolerance, the candidate with the larger $\omega$ is preferred, thereby avoiding unnecessary movement toward the small-$\omega$ boundary.

The selected candidate is accepted only if its score $J(\alpha, \omega)$ strictly improves the current score. Otherwise, the current pair is retained, and the neighborhood is shrunk according to

$$\Delta_\alpha^{(s+1)} = \Delta_\alpha^{(s)}/2, \qquad \eta^{(s+1)} = \sqrt{\eta^{(s)}}.$$

The final iterate is taken as $(\hat{\alpha}, \hat{\omega})$, and the linear coefficients are then re-estimated by WLS. Algorithm S2 summarizes the stages described for the local refinement of $(\hat{\alpha}, \hat{\omega})$. In practice, this refinement adds negligible computational cost relative to the initial grid scan and mainly serves to reduce grid discretization effects. We perform a small fixed number of refinement steps, with default $S = 3$.

**Algorithm S2** Optional local refinement of $(\hat{\alpha}, \hat{\omega})$

1: Initialize $(\alpha^{(0)}, \omega^{(0)})$ from the grid search.
2: **for** $s = 0, \dots, S-1$ **do**
3: Construct the 8-neighbor set around $(\alpha^{(s)}, \omega^{(s)})$ using $\Delta_\alpha^{(s)}$ and $\eta^{(s)}$.
4: Evaluate $J(\alpha, \omega)$ for all admissible neighbors.
5: Let $J_{\max}^{(s)}$ be the largest candidate score and retain the $(\alpha, \omega)$ candidates with

$$J(\alpha, \omega) \geq J_{\max}^{(s)} - \tau \max(1, |J_{\max}^{(s)}|).$$

6: From the pairs in 5, select those minimizing

$$\mathrm{WRSS}(\alpha, \omega) + \lambda g(\omega) R(\alpha, \omega),$$

breaking remaining ties in favor of the larger $\omega$.
7: **if** $J(\alpha^\star, \omega^\star) > J(\alpha^{(s)}, \omega^{(s)})$ **then**
8: $(\alpha^{(s+1)}, \omega^{(s+1)}) \leftarrow (\alpha^\star, \omega^\star)$.
9: **else**
10: $(\alpha^{(s+1)}, \omega^{(s+1)}) \leftarrow (\alpha^{(s)}, \omega^{(s)})$.
11: $\Delta_\alpha^{(s+1)} = \Delta_\alpha^{(s)}/2, \quad \eta^{(s+1)} = \sqrt{\eta^{(s)}}$.
12: **end if**
13: **end for**
14: Return $(\hat{\alpha}, \hat{\omega}) = (\alpha^{(S)}, \omega^{(S)})$.

## S3.5 Calibration of the roughness parameter $\lambda$

The penalized criterion depends on a regularization parameter $\lambda \geq 0$, which controls the strength of the curvature-type roughness penalty. To avoid gene-specific tuning and to stabilize behavior in the small-$\omega$ regime, we select a single dataset-level value $\lambda^\star$ by Monte Carlo calibration under a non-oscillatory null model.

For a fixed sampling design, weighting scheme, $(\alpha, \omega)$ grid, roughness mode, and calibration settings, $\lambda^\star$ is determined before gene-wise testing using only design-level information, rather than the observed gene-expression signals. from the observed gene-expression signals. In particular, it is not tuned to maximize the number of rhythmic detections or any downstream biological enrichment.

Calibration is performed under $H_0$, corresponding to flat profiles, by generating $B$ independent replicates on the observed sampling times,

$$x^{(b)}(t_i) = \varepsilon_i^{(b)}, \qquad \varepsilon_i^{(b)} \sim \mathcal{N}(0, \sigma^2), \qquad b = 1, \dots, B.$$

Since the selection criterion is computed on centered observations, it is sufficient to calibrate under zero-mean Gaussian noise. Moreover, because the calibration relies on a scale-free spike diagnostic, defined below, we fix $\sigma = 1$ without loss of generality. The same inverse-density time-weights $\{w_i\}_{i=1}^n$ used in the analysis are employed during calibration. For ease of notation, we write $\varepsilon_i$ in place of $\varepsilon(t_i)$ to denote the model error term at time point $t_i$.

Let $\Lambda = \{\lambda_l\}_{l=1}^L$ denote the candidate set. By default, $L = 24$ and $\lambda_l = 2^{h_l}$, where $h_1, \dots, h_{24}$ are equally spaced over the interval $[-1, 4]$. For a given $\lambda$ and replicate $b$, we fit the stabilized $\mathrm{FMM}_1$ model by maximizing $J(\alpha, \omega)$ over the $(\alpha, \omega)$ grid using the penalized criterion described in the main analysis. Roughness can be computed either on the observed sampling times, $R(\alpha, \omega)$, or using a dense-time approximation $R_{\mathrm{dense}}(\alpha, \omega)$ evaluated on an equispaced grid over $[0, 2\pi)$.

Let $\hat{\mu}_\lambda^{(b)}(t_i)$ denote the fitted values. To quantify spike-like behavior, we use the circular maximum-jump diagnostic

$$D(\hat{\mu}) = \max_{i=1,\dots,n} \left|\hat{\mu}(t_{i+1}) - \hat{\mu}(t_i)\right|, \qquad t_{n+1} = t_1 + 2\pi,$$

where $t_1 \leq \cdots \leq t_n$ denotes the circular ordering of the sampling times.

For each $\lambda$, we summarize the upper-tail spike magnitude by

$$Q(\lambda) = \mathrm{Quantile}_q\Big(\{D(\hat{\mu}_\lambda^{(b)})\}_{b=1}^B\Big),$$

with $q = 0.975$ in our experiments.

To reduce Monte Carlo noise when comparing different $\lambda$ values, we adopt common random numbers (CRN), a standard variance-reduction technique in stochastic simulation (Glasserman and Yao, 1992; Law, 2024). Specifically, the same set of Gaussian draws $\{x^{(b)}\}_{b=1}^B$ is reused for all $\lambda$, stabilizing the shape of the curve $\lambda \mapsto Q(\lambda)$.

Because high-quantile estimates may still exhibit residual Monte Carlo variability for finite $B$, we repeat the entire calibration $R$ times using different seeds, obtaining $\{Q_r(\lambda)\}_{r=1}^{R}$. We aggregate the curves pointwise using the median,

$$\tilde{Q}(\lambda) = \text{median}\big(\{Q_r(\lambda)\}_{r=1}^{R}\big),$$

and perform selection on $\tilde{Q}(\lambda)$. In our experiments we used $B = 1000$ and $R = 5$.

We select $\lambda^\star$ as the elbow point of the $(\log_{10}\lambda,\ \tilde{Q}(\lambda))$ curve using a distance-to-line criterion: $\lambda^\star$ maximizes the perpendicular distance to the chord joining the endpoints. Prior to the elbow computation, we apply a light running-median smoother, of window $\xi = 3$ by default, to $\tilde{Q}(\lambda)$ to mitigate minor non-monotone fluctuations due to Monte Carlo error without changing the large-scale trend.

Algorithm S3 summarizes the procedure.

---

**Algorithm S3** Monte Carlo calibration of $\lambda$ with CRN, repetitions, and elbow selection.

---

1: **Inputs:** sampling times $\{t_i\}_{i=1}^{n}$, weights $\{w_i\}_{i=1}^{n}$, $(\alpha, \omega)$ grid, candidate set $\Lambda = \{\lambda_1, \dots, \lambda_L\}$, quantile level $q$, number of null replicates $B$, number of repetitions $R$, roughness mode $R_{\text{obs}}$ or $R_{\text{dense}}$, and smoothing window $xi$.
2: Normalize weights so that $\sum_i w_i = n$, and set $\sigma = 1$.
3: **for** $r = 1, \dots, R$ **do**
4: Generate $B$ common-random-number null replicates

$$x_r^{(b)}(t_i) \sim \mathcal{N}(0,1), \qquad b = 1, \dots, B, \quad i = 1, \dots, n.$$

5: **for** $\lambda \in \Lambda$ **do**
6: **for** $b = 1, \dots, B$ **do**
7: Fit $\text{FMM}_1$ by maximizing $J(\alpha, \omega)$ on the $(\alpha, \omega)$ grid using the chosen roughness mode.
8: Obtain the fitted values $\hat{\mu}_{\lambda,r}^{(b)}(t_i)$.
9: Compute the circular maximum-jump diagnostic

$$D(\hat{\mu}_{\lambda,r}^{(b)}) = \max_i \left| \hat{\mu}_{\lambda,r}^{(b)}(t_{i+1}) - \hat{\mu}_{\lambda,r}^{(b)}(t_i) \right|.$$

10: **end for**
11: Compute

$$Q_r(\lambda) \leftarrow \text{Quantile}_q\left( \{D(\hat{\mu}_{\lambda,r}^{(b)})\}_{b=1}^{B} \right).$$

12: **end for**
13: **end for**
14: Aggregate repetitions pointwise:

$$\tilde{Q}(\lambda) \leftarrow \text{median}\left(\{Q_r(\lambda)\}_{r=1}^{R}\right), \qquad \lambda \in \Lambda.$$

15: Smooth $\tilde{Q}(\lambda)$ with a running median of window $xi$.
16: Select $\lambda^\star$ as the elbow point of $(\log_{10}\lambda, \tilde{Q}(\lambda))$, using the maximum distance-to-chord rule.
17: **Output:** dataset-level calibrated value $\lambda^\star$.

---

### S3.5.1 $\lambda^\star$ calibration and sensitivity analysis in the GTEx-13T dataset

**Calibration curves.** For each of the 13 tissues in the GTEx-13T dataset, we estimated the calibration curve $\lambda \mapsto \tilde{Q}(\lambda)$ described above and used it to determine a tissue-specific regularization parameter $\lambda^\star$. Since the GTEx-13T samples are not collected on a regular time grid, but rather ordered by CYCLOPS to infer circadian times, the resulting sampling schemes are irregular. We therefore used the dense-time approximation $R_{\text{dense}}(\alpha, \omega)$, evaluated on an equispaced grid over $[0, 2\pi)$. This choice makes the penalty less dependent on the particular spacing of the observed samples and keeps it focused on discouraging overly localized fitted patterns, especially spurious small-$\omega$ solutions under the null hypothesis.

Figure S2 displays $\tilde{Q}(\lambda)$ as a function of $\log_{10}(\lambda)$ for the 13 tissues. In all tissues, the curves show a broadly decreasing trend, indicating that increasing $\lambda$ reduces the upper-tail magnitude of spike-like fitted profiles under the null model. The distance-to-chord elbow rule was then used to select a compromise point beyond which further increases in regularization yielded comparatively smaller reductions in the spike diagnostic. The resulting tissue-specific values $\lambda^\star$, shown in each panel, are then used for all genes in the corresponding tissue.

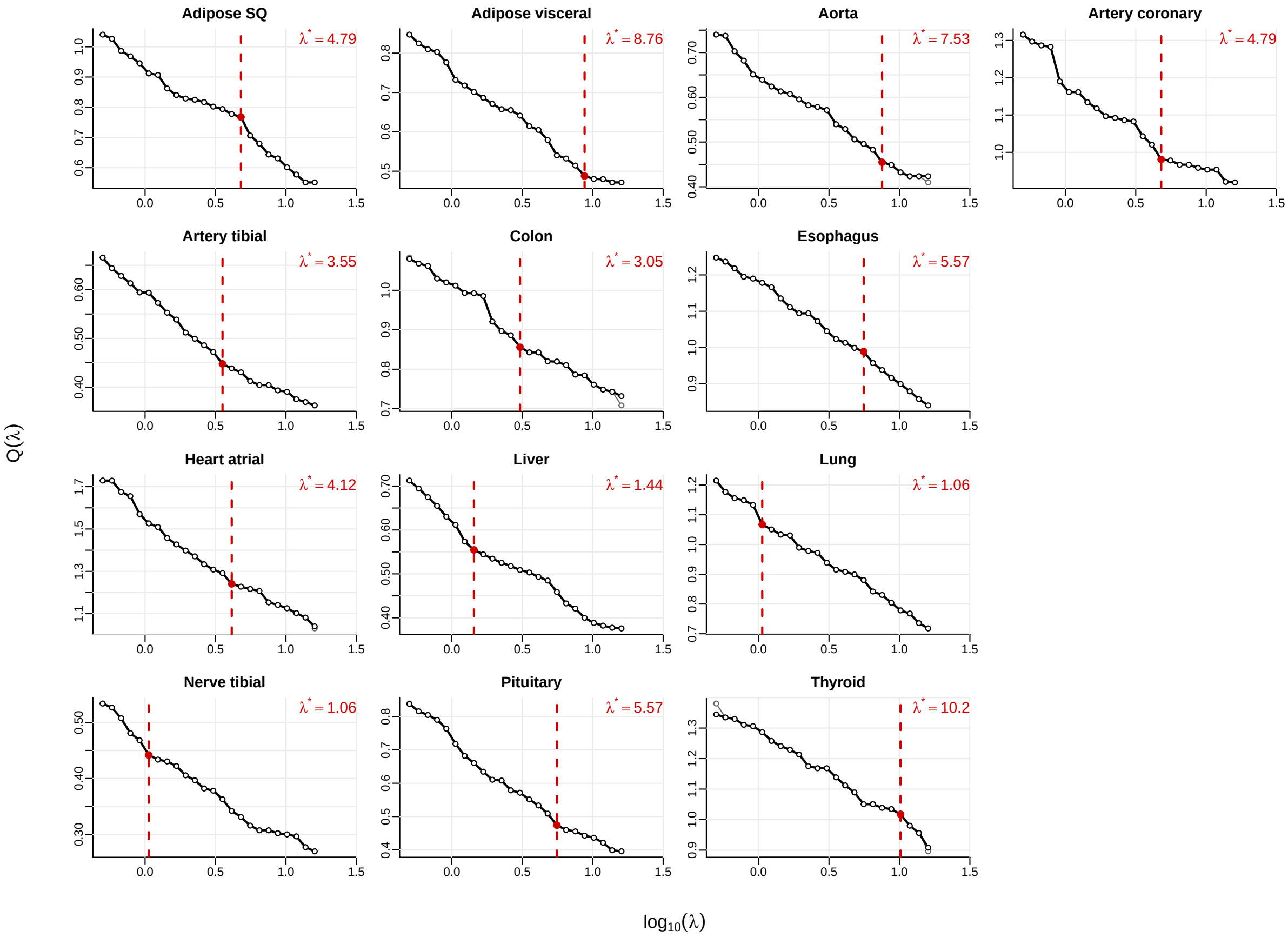


Figure S2: Monte Carlo calibration of the roughness penalty $\lambda$ under $H_0$ (flat profiles) for the 13 tissues in the GTEx-13T dataset. For each $\lambda$, we compute $\tilde{Q}(\lambda)$ as the median over $R = 5$ repetitions of the $q$-quantile ($q = 0.975$) of the spike diagnostic $D(\hat{\mu})$ across $B = 1000$ null replicates, using common random numbers across $\lambda$ within each repetition. Curves are displayed against $\log_{10} \lambda$. The dashed red vertical line indicates the selected tissue-specific $\lambda^\star$, obtained with the distance-to-chord elbow rule after light running-median smoothing. The selected $\lambda^\star$ is reported in each panel.

**Local sensitivity around $\lambda^\star$.** To assess local sensitivity around the calibrated value, we considered nine values of $\lambda$ uniformly spaced on the logarithmic scale from $\lambda^\star/4$ to $4\lambda^\star$. For each combination of tissue and $\lambda$, we summarized three gene-wise quantities computed at the selected pair $(\hat{\alpha}, \hat{\omega})$: the weighted coefficient of determination $R_w^2$, the selected sharpness parameter $\hat{\omega}$, and the penalized criterion $J(\hat{\alpha}, \hat{\omega})$. These quantities describe goodness of fit, stability of the selected frequency-modulation parameter in the regime targeted by the penalty, and the overall regularized ranking criterion, respectively.

Table S2: Summary of the local sensitivity analysis around the calibrated regularization parameter $\lambda^\star$ in the GTEx-13T dataset. For each tissue, the table reports the calibrated value $\lambda^\star$, the explored range of $\lambda$, the median values of the weighted coefficient of determination $R_w^2$ and selected sharpness parameter $\hat{\omega}$ obtained when fitting the model with $\lambda = \lambda^\star$, and the maximum absolute change in these medians across the nine explored values of $\lambda$ relative to $\lambda^\star$ ($\Delta_{\max}$). Detailed tissue–$\lambda$ summaries are provided in the Supplementary Data 2.

| **Tissue** | $\lambda^\star$ | $\lambda$ **range** | $R_w^2$ | | $\hat{\omega}$ | |
|---|---|---|---|---|---|---|
| | | | **Median** | $\Delta_{\max}$ | **Median** | $\Delta_{\max}$ |
| SUB | 4.793 | [1.198, 19.171] | 0.099 | 0.002 | 0.530 | 0.000 |
| VIS | 8.757 | [2.189, 35.028] | 0.156 | 0.001 | 0.728 | 0.000 |
| AOR | 7.532 | [1.883, 30.128] | 0.107 | 0.001 | 0.592 | 0.000 |
| COR | 4.793 | [1.198, 19.171] | 0.141 | 0.002 | 0.530 | 0.000 |
| TIB | 3.546 | [0.886, 14.183] | 0.093 | 0.000 | 0.592 | 0.000 |
| COL | 3.050 | [0.762, 12.199] | 0.129 | 0.003 | 0.592 | 0.000 |
| ESO | 5.572 | [1.393, 22.289] | 0.130 | 0.007 | 0.501 | 0.029 |
| HEA | 4.122 | [1.031, 16.490] | 0.183 | 0.004 | 0.474 | 0.000 |
| LIV | 1.436 | [0.359, 5.743] | 0.072 | 0.004 | 0.386 | 0.045 |
| LUN | 1.062 | [0.266, 4.249] | 0.109 | 0.004 | 0.431 | 0.000 |
| NER | 1.062 | [0.266, 4.249] | 0.083 | 0.001 | 0.560 | 0.022 |
| PIT | 5.572 | [1.393, 22.289] | 0.094 | 0.002 | 0.530 | 0.000 |
| THY | 10.181 | [2.545, 40.724] | 0.167 | 0.002 | 0.592 | 0.000 |

The sensitivity analysis showed weak variation in these quantities around a neighborhood of $\lambda^\star$. Across tissues, the maximum absolute change in the median $R_w^2$ was at most 0.007, and the median $\hat{\omega}$ was either unchanged or showed only small shifts, with the largest change equal to 0.045. These results indicate that the calibrated $\lambda^\star$ lies in a stable region of the fit–regularization trade-off. A compact summary is provided in Table S2, and detailed tissue–$\lambda$ summaries are provided in Supplementary Data 2.

**Effect of regularization relative to $\lambda = 0$.** To evaluate the effect of regularization in the small-$\omega$ region, where the penalty is designed to be most influential, we summarized the proportion of genes whose selected $\omega$ changed after regularization, both globally and among genes within the subset $\hat{\omega}_{\lambda=0} < 0.25$. For genes that changed within this small-$\omega$ subset, we further report the mean, median, and standard deviation of $\Delta\hat{\omega} = \hat{\omega}_{\lambda=\lambda^\star} - \hat{\omega}_{\lambda=0}$, in order to quantify the magnitude of the shift induced by the penalty.

Table S3 shows that the effect of regularization on the selected $\hat{\omega}$ is moderate at the global level, with the proportion of changed genes ranging from 8.9% to 26.3% across tissues. However, this effect becomes much more pronounced within the subset of genes with $\hat{\omega}_{\lambda=0} < 0.25$, where between 61.4% and 84.5% of genes change after regularization. Moreover, the reported values of $\Delta\hat{\omega}$ are consistently positive across tissues, indicating that the penalty tends to shift small unregularized estimates toward larger values.

Figure S3 illustrates this pattern by showing, for each tissue, examples of both stable genes and genes with large regularization-induced shifts. Across tissues, many genes remain close to the diagonal, indicating that their selected sharpness estimates are largely unchanged by regularization. In contrast, genes affected by the penalty typically correspond to very small unregularized $\hat{\omega}$ values. The representative examples show that regularization can prevent highly peaked fitted profiles while leaving already stable fits essentially unchanged.

# S4 Bootstrap calibration and sequential Monte Carlo inference

Because the null distribution of the CircaFMM weighted LRT-type statistic is generally unavailable in closed form, inference relies on bootstrap calibration. This section first formalizes the design-dependent structure of the statistic under the null hypothesis and then describes the design-preserving wild bootstrap used to approximate its reference distribution. Finally, we present the sequential Monte Carlo implementation adopted to make bootstrap calibration computationally tractable in large-scale transcriptomic applications and evaluate its computational performance in the GTEx-13T analysis.

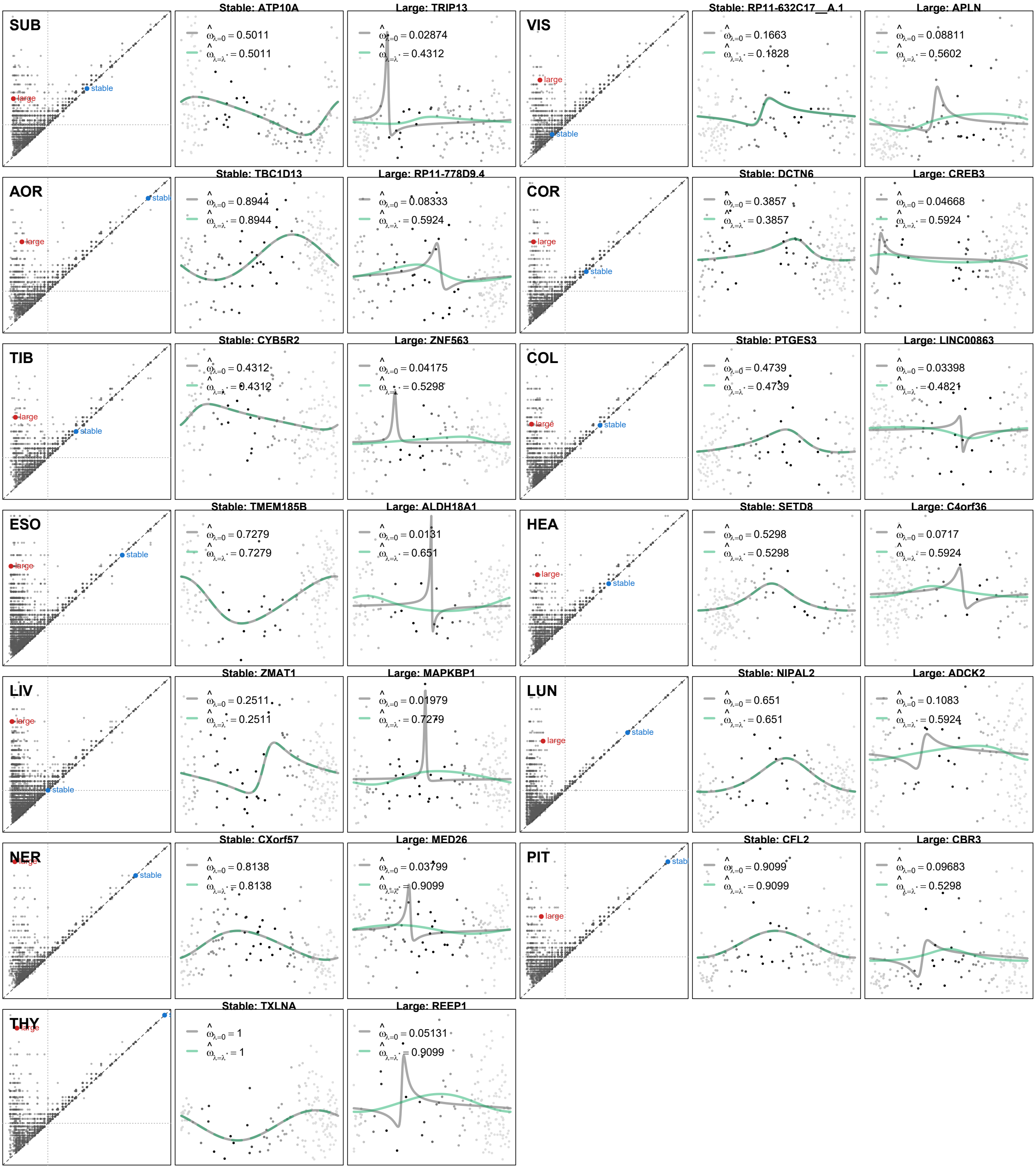


Figure S3: Effect of roughness regularization on selected $\hat{\omega}$ values across the 13 GTEx tissues. For each tissue, the first panel compares the selected $\hat{\omega}$ values obtained with the unregularized fit ($\lambda = 0$) and with the calibrated fit ($\lambda = \lambda^{\star}$). The remaining panels show two representative genes: one with stable estimates under regularization and one with a large regularization-induced change.

Table S3: Effect of roughness regularization on the selected sharpness parameter $\hat{\omega}$ across GTEx-13T tissues. For each tissue, the table reports: the total number of genes ($N$); the number and percentage of genes whose selected $\hat{\omega}$ differs between the unregularized ($\lambda = 0$) and regularized ($\lambda = \lambda^{\star}$) fits; the number and percentage of genes in the small-$\omega$ regime, defined by $\hat{\omega}_{\lambda=0} < 0.25$; and the number and percentage of genes within this subset whose selected $\hat{\omega}$ changes after regularization. For genes affected by regularization in the small-$\omega$ subset, the table further reports summary statistics of $\Delta\hat{\omega} = \hat{\omega}_{\lambda=\lambda^{\star}} - \hat{\omega}_{\lambda=0}$, including the mean, standard deviation (SD), median, and interquartile range (IQR).

| Tissue | $N$ | Changed $\hat{\omega}$, $n$ (%) | Small-$\omega$ genes, $n$ (%) | Changed in subset, $n$ (%) | Mean $\Delta\hat{\omega}\pm$ SD | Median $\Delta\hat{\omega}$ (IQR) |
|---|---|---|---|---|---|---|
| SUB | 15002 | 2265 (15.10%) | 2089 (13.92%) | 1652 (79.08%) | 0.10 ± 0.12 | 0.06 (0.11) |
| VIS | 15000 | 1335 (8.90%) | 1129 (7.53%) | 916 (81.13%) | 0.11 ± 0.12 | 0.06 (0.13) |
| AOR | 15002 | 1856 (12.37%) | 1556 (10.37%) | 1278 (82.13%) | 0.11 ± 0.12 | 0.07 (0.12) |
| COR | 15000 | 1749 (11.66%) | 1773 (11.82%) | 1375 (77.55%) | 0.11 ± 0.14 | 0.05 (0.10) |
| TIB | 15001 | 1610 (10.73%) | 1363 (9.09%) | 1102 (80.85%) | 0.11 ± 0.12 | 0.07 (0.12) |
| COL | 15002 | 1866 (12.44%) | 1826 (12.17%) | 1525 (83.52%) | 0.12 ± 0.15 | 0.06 (0.13) |
| ESO | 15001 | 3257 (21.71%) | 3271 (21.81%) | 2764 (84.50%) | 0.14 ± 0.15 | 0.08 (0.16) |
| HEA | 15000 | 2056 (13.71%) | 1776 (11.84%) | 1360 (76.58%) | 0.10 ± 0.11 | 0.06 (0.11) |
| LIV | 15001 | 3942 (26.28%) | 3985 (26.56%) | 3125 (78.42%) | 0.13 ± 0.14 | 0.07 (0.15) |
| LUN | 15000 | 2099 (13.99%) | 3264 (21.76%) | 2003 (61.37%) | 0.10 ± 0.13 | 0.04 (0.09) |
| NER | 15000 | 1709 (11.39%) | 2231 (14.87%) | 1511 (67.73%) | 0.07 ± 0.10 | 0.03 (0.06) |
| PIT | 15001 | 2051 (13.67%) | 1862 (12.41%) | 1476 (79.27%) | 0.10 ± 0.12 | 0.06 (0.11) |
| THY | 15001 | 1459 (9.73%) | 1241 (8.27%) | 1034 (83.32%) | 0.13 ± 0.14 | 0.08 (0.17) |

## S4.1 Null representation of the CircaFMM statistic conditional on the design

The CircaFMM weighted LRT-type statistic $\Lambda_w$ is obtained through a penalized grid-based selection of $(\alpha, \omega)$, followed by a weighted least-squares refit under the alternative. Because this construction involves a nonregular null hypothesis, a design-dependent weighting scheme, and a data-driven selection step, its null distribution is not expected to follow a standard reference law (Self and Liang, 1987). The following result formalizes its structure under $H_0$.

**Proposition 2** (Null representation conditional on the design). *Fix the sampling times $\{t_i\}_{i=1}^n$, the nonnegative weights $\{w_i\}_{i=1}^n$, the finite grid $\mathcal{A} \times \Omega \subset [0, 2\pi) \times (0, 1]$, the penalty level $\lambda \geq 0$, the modulation function $g$, and the roughness functional $R(\alpha, \omega)$. Under the null model*

$$x_i = \mu + \varepsilon_i, \qquad i = 1, \ldots, n,$$

*assume that*

$$\varepsilon = (\varepsilon_1, \ldots, \varepsilon_n)^\top$$

*is a zero-mean random vector. Let $\bar{x}_w$ be the weighted mean of $x_1, \ldots, x_n$, let*

$$\tilde{x} = (x_1 - \bar{x}_w, \ldots, x_n - \bar{x}_w)^\top,$$

*and define the penalized criterion $J(\alpha, \omega)$ over $\mathcal{A} \times \Omega$ as in the fitting procedure. Let $(\hat{\alpha}, \hat{\omega})$ be the selected pair, and let*

$$\Lambda_w = n \log\left( \frac{\mathrm{WRSS}_0}{\mathrm{WRSS}_1(\hat{\alpha}, \hat{\omega})} \right)$$

*be the corresponding CircaFMM weighted LRT-type statistic.*

*Then, conditional on the design quantities*

$$(\{t_i\}, \{w_i\}, \mathcal{A} \times \Omega, \lambda, g, R),$$

*there exists a measurable map $\Psi$ such that*

$$\Lambda_w = \Psi(\varepsilon) \qquad \text{almost surely under } H_0.$$

*Hence, the conditional null distribution of $\Lambda_w$ is well defined, but generally nonstandard and design-dependent.*

*Proof.* Under $H_0$, the observation vector satisfies

$$x = \mu_0 \mathbf{1} + \varepsilon.$$

Under $H_0$, the centered observations $\tilde{x}$ are completely determined by the noise vector $\varepsilon$ through a deterministic centering operation, hence $\tilde{x}$ is a measurable function of $\varepsilon$. For each fixed $(\alpha, \omega)$, the criterion $J(\alpha, \omega)$ is a measurable function of $\tilde{x}$. Because the grid is finite, the selected pair $(\hat{\alpha}, \hat{\omega})$ is also measurable. The weighted least-squares refit and the resulting quantities

$$\mathrm{WRSS}_1(\hat{\alpha}, \hat{\omega}) \qquad \text{and} \qquad \Lambda_w$$

are then obtained through deterministic operations. Therefore, $\Lambda_w$ is a measurable function of $\varepsilon$. □

Proposition 2 shows that, conditional on the sampling design, weighting scheme, search grid, and regularization specification, the null distribution of $\Lambda_w$ is completely determined by the random error component. The sampling times, weights, candidate grid, and penalty structure remain fixed and act only through the transformation that maps the noise realization into the final statistic. As a result, the null distribution is generally design-dependent and nonstandard, motivating calibration through a design-preserving resampling scheme.

## S4.2 Wild bootstrap under $H_0$ with design weights

To approximate the null distribution of the LRT-type statistic $\Lambda_w$, we employ a design-preserving wild bootstrap with Rademacher sign multipliers (Wu, 1986; Liu, 1988; Davidson and Flachaire, 2008). This choice is particularly useful when error variances may differ across observations. The weights $\{w_i\}_{i=1}^n$ are treated as fixed design weights and are used both in model fitting and in the definition of the test statistic.

Let

$$e_i = x_i - \hat{\mu}_0, \qquad \hat{\mu}_0 = \bar{x}_w,$$

denote the residuals under the null model. We draw i.i.d. Rademacher multipliers $\xi_i$ such that

$$\mathbb{P}(\xi_i = +1) = \mathbb{P}(\xi_i = -1) = \frac{1}{2},$$

and define bootstrap residuals

$$e_i^* = \xi_i e_i.$$

The bootstrap residuals may be recentered to enforce the null constraint under the same weighting scheme:

$$e_i^* \leftarrow e_i^* - \bar{e}_w^*, \qquad \bar{e}_w^* = \frac{\sum_{j=1}^n w_j e_j^*}{\sum_{j=1}^n w_j},$$

so that

$$\sum_{i=1}^n w_i e_i^* = 0$$

holds exactly. The bootstrap sample under $H_0$ is then defined by

$$x_i^* = \hat{\mu}_0 + e_i^*, \qquad i = 1, \ldots, n.$$

For each bootstrap sample $\{x_i^*\}_{i=1}^n$, we recompute the null fit

$$\hat{\mu}_0^* = \frac{\sum_{i=1}^n w_i x_i^*}{\sum_{i=1}^n w_i}, \qquad \mathrm{WRSS}_0^* = \sum_{i=1}^n w_i (x_i^* - \hat{\mu}_0^*)^2.$$

We then refit the $\mathrm{FMM}_1$ alternative using the same complete fitting pipeline as for the observed data, including the same weights, the same $(\alpha, \omega)$ grid, the same roughness mode, and the same dataset-level penalty $\lambda^\star$. Let $(\hat{\alpha}^*, \hat{\omega}^*)$ denote the selected pair, and let

$$\hat{\mu}_1^*(t_i; \hat{\alpha}^*, \hat{\omega}^*)$$

be the corresponding fitted values under the alternative. We define

$$\mathrm{WRSS}_1^*(\hat{\alpha}^*, \hat{\omega}^*) = \sum_{i=1}^n w_i (x_i^* - \hat{\mu}_1^*(t_i; \hat{\alpha}^*, \hat{\omega}^*))^2.$$

The bootstrap replicate statistic is

$$\Lambda_w^* = n \log\left( \frac{\mathrm{WRSS}_0^*}{\mathrm{WRSS}_1^*(\hat{\alpha}^*, \hat{\omega}^*)} \right).$$

## S4.3 Sequential Monte Carlo $p$-values with two-stage stopping

Let $\Lambda_{w,\mathrm{obs}}$ denote the observed test statistic. Under $H_0$, we generate bootstrap samples $\{x_i^*\}$ using the wild bootstrap scheme described in Section S4.2. For each bootstrap sample, we refit the alternative model using the same fitting pipeline and compute the corresponding bootstrap replicate statistic $\Lambda_w^*$.

Define the exceedance indicators

$$I_b = \mathbb{I}\{\Lambda_{w,b}^* \geq \Lambda_{w,\mathrm{obs}}\}, \qquad b = 1, 2, \ldots.$$

If $B$ bootstrap replicates have been generated, let

$$E_B = \sum_{b=1}^B I_b$$

denote the cumulative number of exceedances. The finite-sample corrected Monte Carlo $p$-value is

$$\hat{p} = \frac{E_B + 1}{B + 1}.$$

Note that the correction above avoids zero Monte Carlo $p$-values.

### S4.3.1 Sequential stopping

Rather than fixing $B$ a priori, we employ the sequential Monte Carlo testing framework of Besag–Clifford (Besag and Clifford, 1991). Simulation is continued until either the cumulative exceedance count $E_B$ reaches a prespecified stopping limit, or the number of generated bootstrap replicates $B$ reaches a maximum budget. The resulting pair $(E_B, B)$ at the stopping time is then inserted into the Monte Carlo $p$-value formula above.

This strategy saves computation for clearly non-significant cases while retaining the finite-sample Monte Carlo correction at the stopping time.

### S4.3.2 Two-stage design for large-scale testing

To further reduce computational cost across thousands of genes while retaining the ability to resolve very small $p$-values for strong signals, we use a two-stage version of the sequential procedure. Here, $N$ denotes the maximum bootstrap budget and $S$ the exceedance stopping limit at each stage.

- Stage 1 (screening). Run the sequential procedure with limits $(N_1, S_1)$, stopping when either $E_B = S_1$ or $B = N_1$. Let $B_1 \leq N_1$ denote the number of bootstrap replicates actually generated in this stage, and let $E_{B_1}$ denote the corresponding number of exceedances.
- Promotion rule. Only genes with $E_{B_1} \leq s_{\text{prom}}$ at the end of Stage 1 are promoted to Stage 2.
- Stage 2 (refinement). For promoted genes, continue the sequential procedure from the Stage 1 counts using larger limits $(N_2, S_2)$. The final $p$-value is computed using the total number of bootstrap replicates and the total number of exceedances accumulated across both stages.

Our two-stage strategy is a practical implementation of adaptive sequential resampling ideas, which allocate substantially more Monte Carlo effort to tests with potentially very small $p$-values while stopping early for clearly non-significant cases (Sandve et al., 2011; Che et al., 2014). Unless stated otherwise, we used $N_1 = 200$ and $S_1 = 10$ in Stage 1, promoted genes with $E_{B_1} \leq s_{\text{prom}} = 5$, and used $N_2 = 50{,}000$ and $S_2 = 20$ in Stage 2. The values of $(N_1, S_1, s_{\text{prom}}, N_2, S_2)$ were chosen as a practical compromise between computational efficiency at the screening stage and sufficient Monte Carlo resolution for strong signals in the refinement stage.

## S4.4 Runtime profiling of the sequential bootstrap procedure in GTEx-13T

Because runtime is dominated by bootstrap calibration, computational performance is largely determined by the number of bootstrap replicates generated by the sequential Monte Carlo procedure. To provide a practical estimate of computational cost, we profiled a stratified subset of genes from the GTEx-13T dataset representing the main stopping regimes of the sequential procedure.

Genes were stratified according to whether inference stopped in Stage 1 or was promoted to Stage 2, and according to whether stopping occurred by reaching the exceedance stopping limit or the maximum bootstrap budget. This sampling scheme captures the main drivers of runtime, since each bootstrap replicate requires repeating the grid-based selection of $(\alpha, \omega)$ and the WLS refit under the alternative. For each sampled gene, we recorded the tissue, final rhythmicity status, number of generated bootstrap replicates, number of exceedances, stopping regime, and elapsed runtime.

Runtime profiling was performed on a Windows 11 x64 workstation equipped with a 13th Gen Intel(R) Core(TM) i9-13900K processor, 32 logical cores, and 31.7 GB of RAM, using R version 4.5.3. Gene-level profiling was run with the `serial` backend, meaning that each profiled gene was processed on a single worker without within-gene parallelization. Therefore, the reported runtimes in Table S4 correspond to single-gene sequential execution. Since CircaFMM inference is independent across genes, the full GTEx-13T analysis can be parallelized at the gene level across available workers.

Because runtime is mainly determined by the number of bootstrap replicates actually generated, we summarize computational cost by the stopping regime observed during the profiling run rather than by tissue. This provides a hardware-dependent but practically interpretable summary of the cost of the sequential bootstrap procedure.

Table S4: Runtime profiling of the CircaFMM sequential bootstrap procedure in the GTEx-13T analysis. Genes were sampled across tissues and stopping regimes, and runtimes were summarized according to the stopping regime observed during the profiling run. Reported runtimes correspond to single-gene execution using the `serial` backend. $B$ denotes the number of bootstrap replicates generated.

| Stopping regime | Genes in GTEx-13T | Genes profiled | Median time (s) | IQR time (s) | $p_{10}$–$p_{90}$ time (s) | Median $B$ | IQR $B$ |
|---|---|---|---|---|---|---|---|
| Stage 1: stopped by $S_1$ | 88652 (51.7%) | 213 | 0.100 | 0.150 | 0.020–0.220 | 93 | 132 |
| Stage 1: reached $N_1$ | 7787 (4.5%) | 110 | 0.230 | 0.070 | 0.170–0.260 | 200 | 0 |
| Stage 2: stopped by $S_2$ | 37345 (21.8%) | 180 | 1.64 | 4.49 | 0.670–13.5 | 1534 | 4425 |
| Stage 2: reached $N_2$ | 37619 (22.0%) | 129 | 60.2 | 6.59 | 39.2–61.0 | 50000 | 0 |

As expected, runtime increased with the number of bootstrap replicates generated by the sequential procedure. Genes stopped during Stage 1 required negligible computation, with median runtimes below 0.25 seconds. Genes

promoted to Stage 2 but stopped by the exceedance limit $S_2$ showed intermediate runtimes, with a median of 1.64 seconds and a wider range reflecting variability in the number of generated bootstrap replicates. The most computationally demanding cases were those reaching the maximum second-stage budget $N_2 = 50{,}000$, with a median runtime of approximately 60 seconds per gene under serial execution. These results support the use of the two-stage sequential design, which avoids applying the maximum bootstrap budget to all genes.

# S5 Simulation settings and additional results

This section provides additional simulation details and supplementary results supporting the main findings of the paper. We first describe the simulation framework used to evaluate the proposed methodology across a range of sampling designs, sample sizes, and signal characteristics. We then examine the contribution of the weighting and regularization components through ablation analyses, assess robustness to departures from Gaussian errors, and investigate their impact on both type-I error control and power. Finally, we present an additional study of the classical Cosinor test to illustrate the relationship between signal strength and rhythmicity detection under a correctly specified sinusoidal model.

## S5.1 Main simulation settings

All simulations were performed on circadian times generated on the 24-hour scale and mapped to radians by $t_i = 2\pi(h_i/24)$, where $h_i$ denotes the sampling time on the 24-hour scale and is generated according to the corresponding sampling regime. For sample sizes $n \in \{24, 48, 96, 192\}$, we considered four sampling regimes designed to represent increasing departures from regular circadian sampling: equispaced schedules, jittered schedules with small timing deviations, clustered designs with locally dense temporal regions, and gappy designs combining clustering with a poorly sampled circadian window. Specifically,

1. Equispaced: $h_i = 24(i-1)/n$, $i = 1, \ldots, n$.
2. Jittered equispaced: to mimic scheduled sampling with timing variability, we generated

$$h_i = 24(i-1)/n + \eta_i, \qquad \eta_i \sim \text{Unif}(-0.5, 0.5)$$

   hours. Times were wrapped to $[0, 24)$ and sorted.
3. Clustered without gaps: starting from the near-uniform schedule above, we reassigned 35% of time points by sampling from a Gaussian cluster $\mathcal{N}(13, 1.2^2)$ hours. Times were wrapped to $[0, 24)$ and sorted, yielding a locally dense region while preserving nonzero coverage across the full 24h cycle.
4. Clustered with one gap: as in the clustered-without-gaps regime, but introducing a single low-coverage window by thinning samples inside the 2-hour interval $[16, 18]$ hours. Proposed times falling in this interval were retained with probability $p_{\text{gap}} = 0.08$, yielding a very sparse gap.

For each sampling design and sample size, time-density weights and the dataset-level roughness parameter $\lambda^\star$ were computed once and kept fixed across simulated genes.

Under $H_0$, expression values were generated from flat profiles

$$x_i = \mu + \varepsilon_i, \qquad \mu \sim \text{Unif}(0, 3), \qquad \varepsilon_i \sim N(0, 1).$$

Under $H_1$, expression values were generated from the $\text{FMM}_1$ model with $M = 0$, gene-specific $\alpha \sim \text{Unif}(0, 2\pi)$ and $\beta \sim \text{Unif}(0, 2\pi)$, Gaussian noise $\varepsilon_i \sim N(0, 1)$, and amplitudes $A \in \{0.5, 0.75, 1\}$. Since $\sigma = 1$, $A$ corresponds to the amplitude-based signal-to-noise (SNR) ratio $A/\sigma$ (Ding et al., 2021). For the sharpness parameter, we considered $\omega \in \{0.05, 0.15, 0.25, 0.4, 0.7, 1\}$. The case $\omega = 1$ corresponds to a sinusoidal waveform, whereas smaller values of $\omega$ produce increasingly localized and asymmetric profiles.

For each scenario, $G = 2000$ genes were simulated and the experiment was repeated over $B = 10$ Monte Carlo repetitions. CircaFMM $p$-values were computed using the sequential wild-bootstrap calibration described in Supplementary Section S4. For power comparisons, Cosinor was fitted by least-squares regression on $\{1, \cos(t_i), \sin(t_i)\}$, and rhythmicity was assessed using the classical F-test for the joint null of zero cosine and sine coefficients (Cornelissen, 2014).

## S5.2 Additional type-I error results under $H_0$

### S5.2.1 Ablation: time-density weights and roughness penalization

To assess the impact of time-density weighting and roughness penalization on type-I error control, we conducted an ablation study based on the simulation framework described in Section S3.1. Specifically, we considered simplified variants of the procedure designed to isolate the contributions of time-density weighting and the curvature-type penalty used to stabilize grid-based selection in the small-$\omega$ regime. We compared three variants: (A) an unweighted and unpenalized procedure ($w_i = 1$, $\lambda = 0$); (B) a weighting-only procedure using time-density weights

Table S5: Ablation study under $H_0 : \omega = 0$. Empirical type-I error (mean $\pm$ standard deviation over $B = 10$ repetitions; $G = 2000$ genes per repetition) for three variants: (A) no weights and no penalization ($w_i = 1$, $\lambda = 0$), (B) time-density weights only ($w_i$ from time density, $\lambda = 0$), and (C) full method (weights $w_i$ + calibrated penalization, $\lambda = \lambda^\star$). Results for (C) coincide with Table 1 and are reproduced here to facilitate side-by-side comparison.

| $n$ | design | A: $w_i = 1$, $\lambda = 0$ | B: $w_i$, $\lambda = 0$ | C: $w_i$, $\lambda = \lambda^\star$ |
|---|---|---|---|---|
| 24 | equispaced | $0.0482 \pm 0.0034$ | $0.0454 \pm 0.0059$ | $0.0506 \pm 0.0034$ |
| 24 | jittered | $0.0484 \pm 0.0030$ | $0.0452 \pm 0.0047$ | $0.0508 \pm 0.0033$ |
| 24 | clustered | $0.0471 \pm 0.0050$ | $0.0545 \pm 0.0048$ | $0.0540 \pm 0.0045$ |
| 24 | gappy | $0.0460 \pm 0.0050$ | $0.0532 \pm 0.0046$ | $0.0543 \pm 0.0040$ |
| 48 | equispaced | $0.0498 \pm 0.0052$ | $0.0498 \pm 0.0043$ | $0.0493 \pm 0.0047$ |
| 48 | jittered | $0.0498 \pm 0.0064$ | $0.0495 \pm 0.0047$ | $0.0499 \pm 0.0063$ |
| 48 | clustered | $0.0493 \pm 0.0029$ | $0.0547 \pm 0.0047$ | $0.0554 \pm 0.0044$ |
| 48 | gappy | $0.0490 \pm 0.0040$ | $0.0549 \pm 0.0025$ | $0.0531 \pm 0.0050$ |
| 96 | equispaced | $0.0497 \pm 0.0043$ | $0.0491 \pm 0.0066$ | $0.0497 \pm 0.0057$ |
| 96 | jittered | $0.0482 \pm 0.0050$ | $0.0464 \pm 0.0041$ | $0.0494 \pm 0.0042$ |
| 96 | clustered | $0.0484 \pm 0.0033$ | $0.0541 \pm 0.0066$ | $0.0526 \pm 0.0066$ |
| 96 | gappy | $0.0486 \pm 0.0045$ | $0.0521 \pm 0.0048$ | $0.0516 \pm 0.0097$ |
| 192 | equispaced | $0.0523 \pm 0.0048$ | $0.0482 \pm 0.0050$ | $0.0513 \pm 0.0033$ |
| 192 | jittered | $0.0485 \pm 0.0036$ | $0.0501 \pm 0.0041$ | $0.0513 \pm 0.0053$ |
| 192 | clustered | $0.0484 \pm 0.0052$ | $0.0540 \pm 0.0061$ | $0.0518 \pm 0.0053$ |
| 192 | gappy | $0.0524 \pm 0.0049$ | $0.0515 \pm 0.0048$ | $0.0509 \pm 0.0048$ |

($w_i$, $\lambda = 0$); and (C) the full method, combining time-density weighting with calibrated penalization ($w_i$, $\lambda^\star$). Apart from the weighting and penalization components, all simulation and inference settings were kept unchanged across variants.

All three variants control type-I error close to the nominal level across sample sizes and sampling-time regimes (Table S5), indicating that neither time-density weighting nor roughness penalization compromises calibration under $H_0$. Differences between variants are negligible for regular schedules (equispaced and jittered), and become more apparent only under irregular designs (clustered and gappy), where design-aware weighting changes the effective contribution of densely and sparsely sampled regions of the cycle. Comparing the settings (B) and (C), the calibrated roughness penalty does not induce systematic inflation and is often neutral or slightly conservative in the irregular settings, consistent with its role in suppressing spuriously localized small-$\omega$ fits while preserving type-I error control.

### S5.2.2 Robustness to heavy-tailed noise

To evaluate robustness to violations of Gaussian errors, we repeated the type-I error experiment in main Section 3.1, replacing $\varepsilon_i \sim \mathcal{N}(0, 1)$ with Student $t$ errors. Specifically, we generated $\varepsilon_i \sim t(\nu)$ and rescaled to unit variance, considering $\nu \in \{3, 5, 10\}$. All other settings were kept identical.

We evaluated the same three test variants as in the ablation study: (A) unweighted and unpenalized ($w_i = 1$, $\lambda = 0$), (B) time-density weights only ($w_i$, $\lambda = 0$), and (C) the full method ($w_i$, $\lambda^\star$). Empirical type-I error at nominal level $\alpha = 0.05$ was estimated from $G = 2000$ null genes and $B = 10$ Monte Carlo repetitions per scenario.

Figure S4 shows that calibration remains close to the nominal level $\alpha = 0.05$ across sampling designs and sample sizes, with the largest deviations occurring in the most challenging setting ($n = 24$) under the heaviest tails ($t(3)$). As expected, type-I error approaches 0.05 as $\nu$ increases from 3 to 10, reflecting progressively milder departures from Gaussianity. Across all panels, the full method (C: $w_i$, $\lambda^\star$) behaves similarly to the simplified variants, indicating that time-density weighting and roughness penalization do not introduce systematic inflation under heavy-tailed errors.

## S5.3 Additional power results under $H_1$: ablation analysis

To complement the type-I error analysis, we examined whether time-density weights and calibrated roughness penalization affected power under oscillatory alternatives. Because the full $H_1$ simulation grid is large, we focused on the intermediate SNR setting $A = 0.75$, where power is neither uniformly low nor saturated and differences between variants are most likely to be visible. We compared the same three variants considered under $H_0$: (A) no weights and no penalization ($w_i = 1$, $\lambda = 0$), (B) time-density weights only ($w_i$, $\lambda = 0$), and (C) the full method ($w_i$, $\lambda^\star$).

Under $H_1$, the ablation results (Figure S5) showed that the three variants followed the same overall power pattern, with power increasing as $\omega$ approached the sinusoidal regime and as the sample size increased. However,

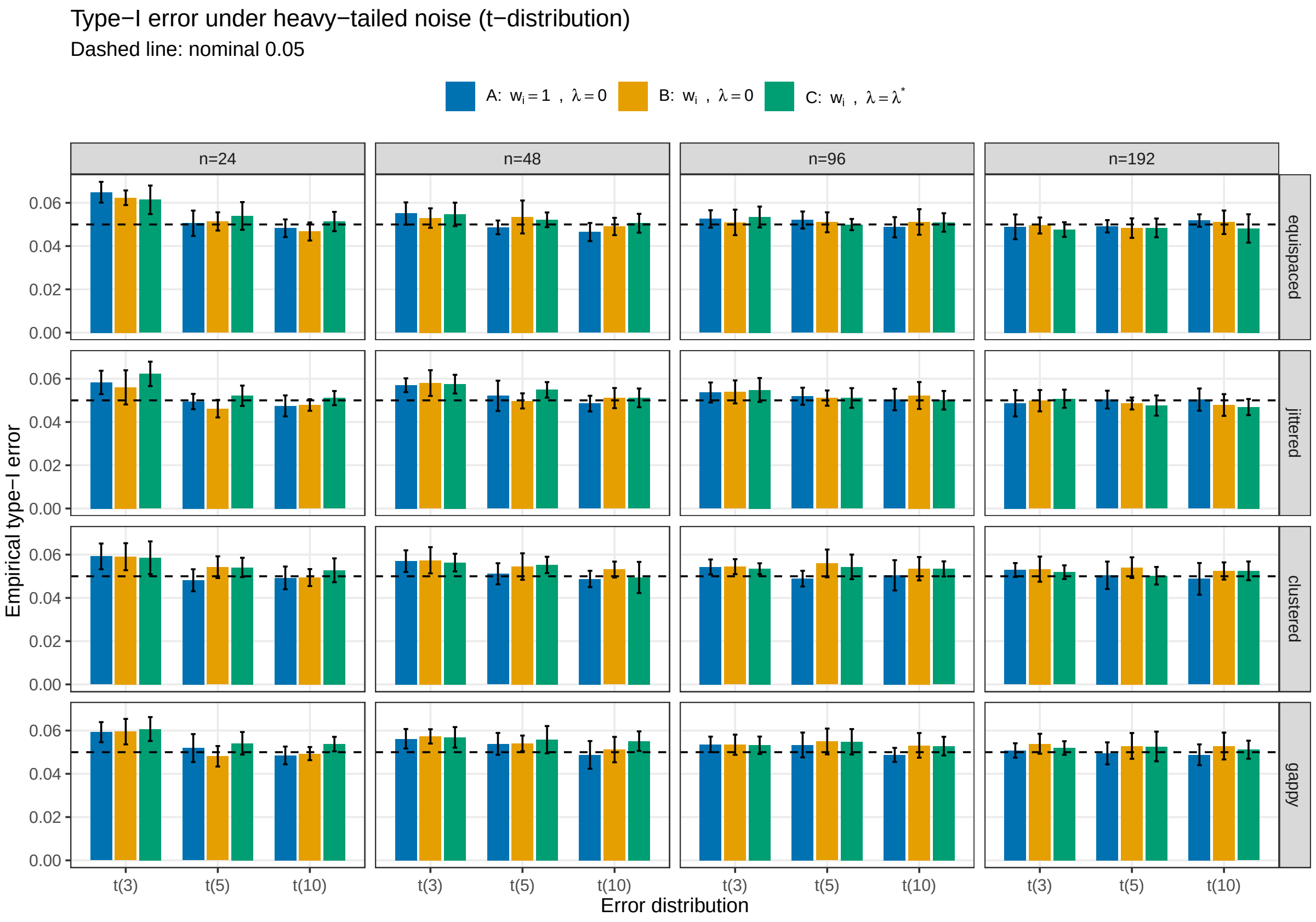


Figure S4: Type-I error under heavy-tailed noise ($t$-distribution) for the three ablation variants: (A) no time-density weights and no penalization ($w_i \equiv 1$, $\lambda = 0$), (B) weights only ($w_i$, $\lambda = 0$), (C) full method (weights + calibrated penalization, $w_i$, $\lambda = \lambda^\star$). Rows correspond to sampling-time designs (equispaced, jittered, clustered, gappy) and columns to sample sizes ($n \in \{24, 48, 96, 192\}$). Bars show the empirical type-I error at nominal level $\alpha = 0.05$ (dashed line), reported as mean ± standard deviation over $B = 10$ Monte Carlo repetitions ($G = 2000$ null genes per repetition).

the comparison also revealed scenario-specific differences among variants. These differences were most visible for $n = 48$ and, to a lesser extent, for $n = 96$, whereas they largely disappeared for $n = 192$ because power was close to saturation in most settings.

The weighted variants (B–C) tended to remain close to, and in some scenarios exceeded, the unweighted variant (A), especially under irregular sampling schedules. This supports the role of time-density weights in reducing the influence of unevenly represented regions of the cycle. More importantly, the full method (C) did not show a systematic loss of power relative to the weighted unpenalized variant (B). On the contrary, in several non-saturated scenarios, (C) achieved slightly higher empirical power, suggesting that the calibrated roughness penalty may improve the stability of grid-based selection and thereby strengthen the resulting test statistic in some non-saturated scenarios.

Overall, these results indicate that the full method preserves the power profile of the simplified procedures and may provide moderate gains in difficult, non-saturated settings. Together with the type-I error results under $H_0$, this supports the use of variant C as the final CircaFMM testing pipeline: it maintains calibration under the null while retaining, and in some cases improving, power under alternatives.

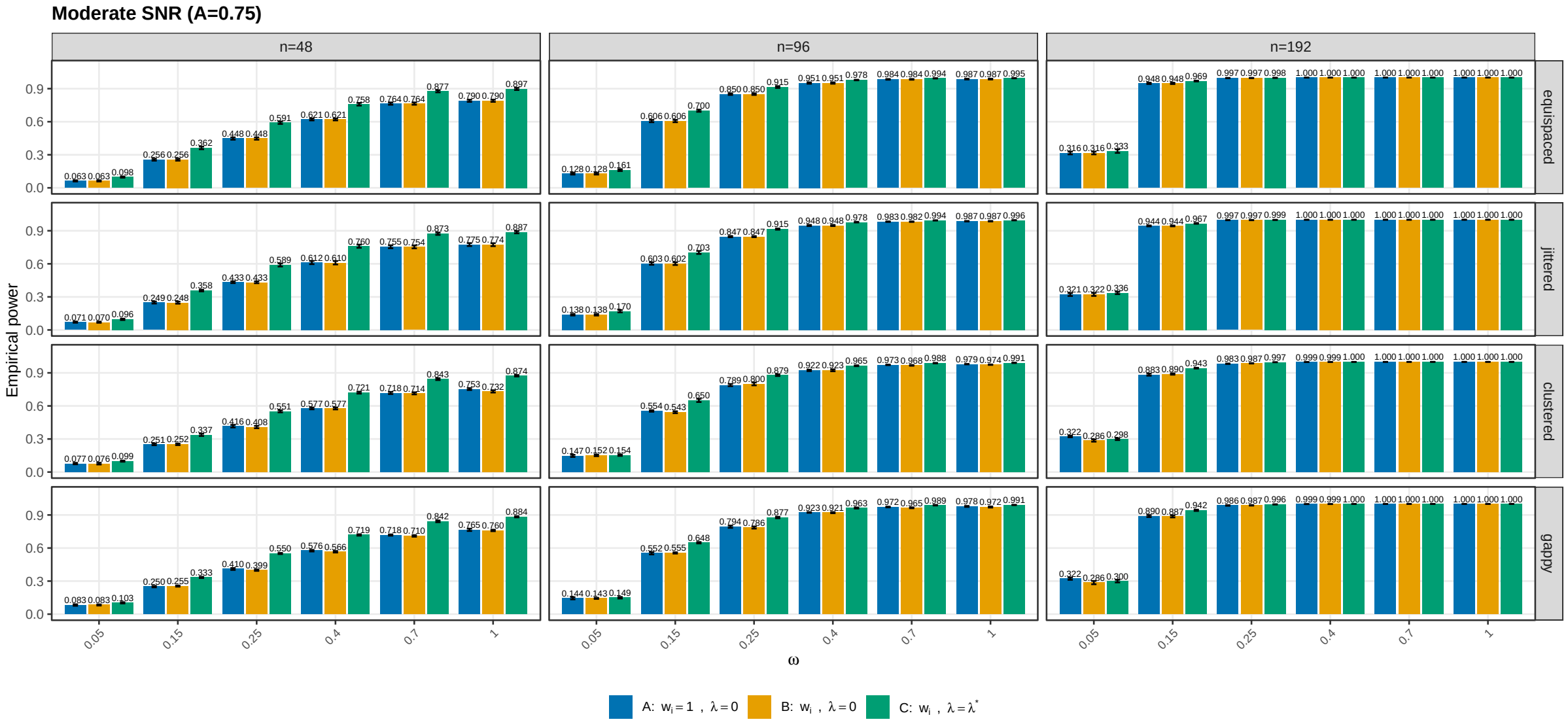


Figure S5: Ablation study of empirical power under $H_1$ for the intermediate SNR setting $A = 0.75$. Bars compare three variants of the proposed CircaFMM testing pipeline: (A) no weights and no penalization ($w_i = 1$, $\lambda = 0$), (B) time-density weights only ($w_i$, $\lambda = 0$), and (C) the full method with time-density weights and calibrated roughness penalization ($w_i$, $\lambda = \lambda^\star$). Empirical power is shown as a function of the sharpness parameter $\omega$ across sampling-time designs and sample sizes. Bars report mean power over $B = 10$ Monte Carlo repetitions with $G = 2000$ genes per repetition; error bars show $\pm 1$ standard deviation.

## S5.4 Cosinor sensitivity under variance-based SNR

This section illustrates, through simulation, the behavior of the classical Cosinor F-test for rhythmicity under a correctly specified sinusoidal model as a function of signal strength. In the simulation study presented in the main text, signal strength was indexed by the amplitude-to-noise ratio $A/\sigma$, corresponding to the amplitude-based SNR measure proposed by Ding et al. (2021). Here, however, to facilitate interpretation of the Cosinor F-statistic in terms of explained variability, we quantify signal strength using the variance-based signal-to-noise ratio

$$\mathrm{SNR}_{\mathrm{var}} = \frac{\mathrm{Var}(\mu(t))}{\mathrm{Var}(\varepsilon)}.$$

This definition is natural for the Cosinor F-test because the statistic is based on the reduction in residual sum of squares relative to residual variance. For a sinusoidal signal observed with approximately uniform temporal coverage,

$$\mathrm{SNR}_{\mathrm{var}} = \frac{A^2}{2\sigma^2} = \frac{1}{2}\left(\frac{A}{\sigma}\right)^2.$$

Thus, both quantities measure signal strength relative to noise, but on different scales.

### S5.4.1 Cosinor rhythmicity test and variance-based SNR

In the Cosinor model (Cornelissen, 2014), a periodic signal of known period $T$ observed at times $\{t_i\}_{i=1}^n$ is modeled as

$$X(t_i) = M + A\cos(\omega t_i + \phi) + \varepsilon_i, \qquad \omega = \frac{2\pi}{T}, \quad \varepsilon_i \sim \mathcal{N}(0, \sigma^2).$$

For estimation, it is convenient to use the linear reparametrization

$$X(t_i) = M + \beta\cos(\omega t_i) + \gamma\sin(\omega t_i) + \varepsilon_i,$$

from which amplitude and phase follow as

$$A = \sqrt{\beta^2 + \gamma^2}, \qquad \phi = \operatorname{atan2}(-\gamma, \beta).$$

This is a standard Gaussian linear model, and the maximum likelihood estimators (MLEs) coincide with ordinary least-squares estimators.

Rhythmicity in the Cosinor model is assessed through the nested hypotheses

$$H_0 : \beta = \gamma = 0 \qquad \text{vs.} \qquad H_1 : (\beta, \gamma) \neq (0, 0).$$

Let $\mathrm{RSS}_0$ and $\mathrm{RSS}_1$ denote the residual sums of squares of the reduced and full models, respectively, and let

$$\mathrm{SSR} = \mathrm{RSS}_0 - \mathrm{RSS}_1$$

denote the regression sum of squares. The classical test statistic is

$$F = \frac{(\mathrm{RSS}_0 - \mathrm{RSS}_1)/2}{\mathrm{RSS}_1/(n-3)} = \frac{\mathrm{SSR}/2}{\mathrm{RSS}_1/(n-3)},$$

which satisfies $F \sim F_{2,n-3}$ under $H_0$.

Assume approximately uniform coverage of time points over the period and moderate-to-large $n$. Let

$$\mu_i = A\cos(\omega t_i + \phi)$$

denote the oscillatory component. Under uniform coverage,

$$\mathbb{E}(\mu_i) = 0, \qquad \mathbb{E}(\mu_i^2) = A^2/2.$$

Hence, the signal variance is

$$\mathrm{Var}(\mu_i) = A^2/2,$$

whereas the noise variance is

$$\mathrm{Var}(\varepsilon_i) = \sigma^2.$$

Therefore,

$$\mathrm{SNR}_{\mathrm{var}} = \frac{\mathrm{Var}(\mu_i)}{\mathrm{Var}(\varepsilon_i)} = \frac{A^2}{2\sigma^2}.$$

In linear regression, the sums of squares scale with $n$ times the corresponding variance, up to centering effects. Hence, under the above idealized conditions,

$$\mathbb{E}(\mathrm{SSR}) \approx n\,\frac{A^2}{2}, \qquad \mathbb{E}(\mathrm{RSS}_1) \approx n\,\sigma^2.$$

Substituting in the definition of $F$ yields the heuristic relation

$$\mathbb{E}(F) \approx \frac{nA^2}{4\sigma^2} = \frac{n}{2}\,\mathrm{SNR}.$$

Thus, for a fixed sample size, the expected magnitude of the Cosinor $F$-statistic increases linearly with the variance-based signal-to-noise ratio. Conversely, when the oscillatory component is weak relative to the residual variability, $F$ remains close to its null behavior and rhythmicity becomes difficult to detect.

This interpretation is consistent with the classical noncentral-$F$ formulation of the test under $H_1$, where the noncentrality parameter is proportional to $n \times \mathrm{SNR}_{\mathrm{var}}$. Consequently, statistical power increases with both sample size and signal strength, whereas low-SNR settings yield limited separation between the null and alternative distributions of the test statistic.

Therefore, when the oscillatory component is small relative to residual variability, the explained variability becomes dominated by noise and the distribution of $F$ under $H_1$ approaches its null distribution, leading to substantial loss of power for classical Cosinor inference.

### S5.4.2 Simulations: Cosinor detection versus variance-based SNR

Let $\{t_i\}_{i=1}^n$ be equally spaced time points over a full 24-hour period with $n \in \{48, 72, 96\}$. Let further assume $\text{SNR}_{var} \in \{0.01, 0.025, 0.05, 0.1, 0.15, 0.2, 0.25, 0.5, 1, 2, 2.5, 5\}$. Note that time points count were chosen to represent moderate sampling depths commonly encountered in transcriptomic circadian studies. While the SNR grid spans very low variance-based SNR values, representative of challenging transcriptomic settings, as well as moderate-to-high SNR levels.

For each $(n, \text{SNR}_{\text{var}})$ configuration, we generated $G = 1000$ independent expression profiles per replicate, with 5% truly rhythmic and 95% non-rhythmic signals. The proportion of truly rhythmic profiles was fixed at 5% to mimic a sparse-detection setting and to evaluate the effect of FDR correction on Cosinor sensitivity.

Under $H_1$, expression date were simulated under the Cosinor model with $M = 0$, $A = 1$, $\phi = \pi$ as follows:

$$X(t_i) = A\cos(\omega t_i + \phi) + \varepsilon_i, \qquad \varepsilon_i \sim \mathcal{N}(0, \sigma^2), \quad i = 1, \dots, n.$$

Under $H_0$, data were generated as flat noisy signals as follows:

$$X(t_i) \sim \mathcal{N}(0, \sigma^2).$$

To achieve the target variance-based SNR, we set in both cases:

$$\sigma^2 = \frac{A^2}{2\,\text{SNR}}.$$

Each configuration was replicated $B = 30$ times. In each replicate, Cosinor models were fitted by ordinary least squares, and $H_0 : \beta = \gamma = 0$ was tested for each profile. The resulting $p$-values were adjusted across the $G$ profiles using the Benjamini–Hochberg procedure to control FDR at 5%. We report the true positive rate (TPR), defined as the proportion of truly rhythmic profiles declared significant at 5% FDR (Figure S6).

These simulations illustrate that Cosinor sensitivity deteriorates rapidly in low-$\text{SNR}_{\text{var}}$ regimes: detection remains near zero for very low SNR and increases substantially only once SNR becomes moderate-to-high. This behavior is expected from the scaling argument above and supports the need for alternative procedures (e.g., resampling-based or robust methods) in low-SNR transcriptomic contexts (Hollis et al., 2025; Burns et al., 2025; Zulfiqar et al., 2025).

The above simulation study is intended to illustrate the intrinsic dependence of Cosinor detection on signal strength under a correctly specified sinusoidal model. In contrast, the reduced sensitivity of the Cosinor test observed in the power simulations of the main text (Section S3.2) reflects not only finite signal strength but also the additional loss of efficiency arising from fitting a sinusoidal model to non-sinusoidal FMM waveforms.

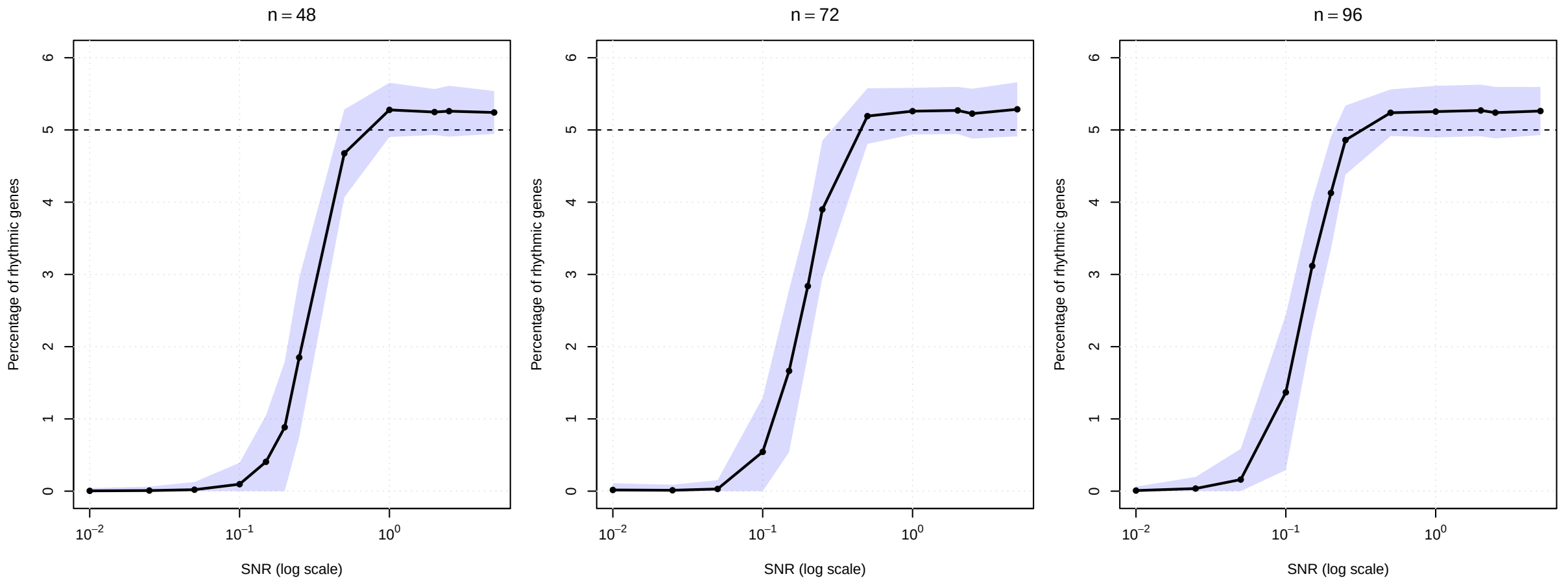


Figure S6: Cosinor detection vs. variance-based SNR on simulated data. The figure reports the true positive rate (TPR) at 5% FDR (BH-adjusted), averaged over $B = 30$ replicates, for $n \in \{48, 72, 96\}$. Shaded bands represent mean ±2 standard deviation across replicates. The x-axis is shown on a base-10 logarithmic scale; the dashed horizontal line marks the nominal 5% level.

# S6 GTEx-13T application: additional results and methodological details

This section provides supplementary methodological details, figures, and tables supporting the GTEx-13T application presented in the main text. In particular, it expands upon the GTEx-13T-specific results for sampling

weights, calibration of the roughness parameter, and sequential bootstrap runtime analyses across tissues presented in Sections S2.3, S3.5, and S4.4, respectively. Additional materials document data preparation, temporal coverage and expression heterogeneity metrics, detailed rhythmicity comparisons, waveform characterization, functional clustering results, and circadian phase analyses.

Unless otherwise stated, multiple testing correction in the supplementary analyses was performed using the Benjamini–Hochberg false discovery rate (FDR) procedure (Benjamini and Hochberg, 1995).

## S6.1 Data preparation, temporal coverage and expression variability

This section describes the preparation of the GTEx-13T expression matrices used in the real-data application and the tissue-level metrics used to summarize sampling irregularity and expression heterogeneity. We first describe the robust preprocessing step applied to attenuate isolated atypical observations, and then define the descriptive metrics reported in Table S6.

### S6.1.1 Data preprocessing and outlier attenuation

The GTEx-13T expression matrices used in the real-data application were prepared from the tissue-specific datasets described by Ruben et al. (2018). In the original preprocessing, genes with zero expression in at least one sample were removed and, for each tissue, the 15000 most highly expressed genes with Entrez identifiers were retained. After matching expression profiles with the available circadian ordering and gene annotation information, this resulted in approximately 13000–13400 analyzed genes per tissue.

Because isolated atypical observations can strongly influence rhythmicity detection, especially when the detection procedure is designed to accommodate sparse or irregular temporal sampling, we applied an additional gene-wise robust preprocessing step aimed at attenuating extreme observations while preserving the overall temporal structure of each gene profile. This step was not intended to remove rhythmic patterns or modify global expression trends, but to reduce the leverage of isolated expression values that were locally inconsistent with neighboring observations in circadian time.

The procedure consisted of two complementary stages. First, extreme local outliers were identified within circular time bins using a median absolute deviation (MAD)-based rule. Let $x = (x_1, \ldots, x_n)$ denote the vector of expression values for a given gene within a tissue, where $x_i = X(t_i)$. The $i-$th observation was considered an extreme outlier if:

$$\frac{|x_i - \mathrm{Me}(x)|}{\mathrm{MAD}(x)} > c,$$

where $\mathrm{Me}(x)$ denotes the sample median and

$$\mathrm{MAD}(x) = \mathrm{median}_i\{|x_i - \mathrm{Me}(x)|\}.$$

We used $c = 10$ by default.

Rather than removing samples, identified extreme values were locally adjusted. Each observation was assigned to a circular time bin according to its circadian phase. By default, six bins of 4 hours each were used to span the full circadian cycle $[0, 2\pi)$. Extreme observations were replaced by the median expression within their corresponding circular time bin. To ensure robustness, bin-specific medians were computed after excluding previously identified extreme observations.

In a second step, moderate isolated outliers were identified using the same gene-wise robust standardization, but with a less stringent threshold ($c = 5$ by default). After the extreme-outlier correction, a LOESS curve was fitted for each gene over circadian time, and only the flagged moderate outliers were replaced by their corresponding fitted values. This procedure reduces the influence of isolated deviations while preserving the sample size and the overall temporal structure of the data.

### S6.1.2 Temporal coverage and expression variability metrics

To characterize the tissue-specific structure of the GTEx-13T dataset, we computed descriptive metrics summarizing two complementary aspects of the data: temporal coverage of the circadian cycle and gene-expression variability. These metrics were used to contextualize tissue-specific results in the main text, particularly when interpreting analyses involving irregular sampling or low signal strength.

Temporal coverage was summarized using three metrics computed from the circularly ordered sampling times. First, $\%Max(Gaps)$ denotes the largest circular gap between consecutive observations, expressed as a percentage of the full circadian cycle. Larger values indicate the presence of poorly covered temporal regions. Second, $Cv(Gaps)$ denotes the coefficient of variation of circular inter-sample distances, with larger values indicating more uneven temporal spacing. Third, $\%Max(Bins)$ is the maximum percentage of observations falling within any 4-hour circular time bin, and therefore captures local concentration of sampling times.

Expression variability was summarized using two tissue-level metrics computed from gene-wise quantities. $Me(Cv)$ denotes the median coefficient of variation of gene expression values across genes. $RobSNR$ is a robust SNR proxy, defined as the median across genes of the ratio between the interquartile range (IQR) and the MAD.

The resulting tissue-level summaries are reported in Table S6. They show that temporal coverage and expression variability differ substantially across tissues. For example, COR shows marked temporal irregularity, whereas TIB, COL and NER correspond to tissues with comparatively low signal-strength regimes. These descriptive differences provide context for the tissue-specific analyses reported in the main text.

Table S6: Tissue-level descriptive metrics for temporal coverage and expression variability in the GTEx-13T dataset. Temporal coverage is summarized by the largest circular gap between consecutive observations $\%Max(Gaps)$, the coefficient of variation of circular inter-sample distances $Cv(Gaps)$, and the maximum percentage of observations within any 4-hour circular time bin $\%Max(Bins)$. Expression variability is summarized by the median coefficient of variation across genes $Me(Cv)$ and the robust SNR proxy $RobSNR$.

| Source | Metric | SUB | VIS | AOR | COR | TIB | COL | ESO | HEA | LIV | LUN | NER | PIT | THY |
|---|---|---|---|---|---|---|---|---|---|---|---|---|---|---|
| **Sampling Irregularity** | $\%Max(Gaps)$ | 7.189 | 6.204 | 6.429 | 28.748 | 5.690 | 11.618 | 7.655 | 9.342 | 4.547 | 7.939 | 4.341 | 4.765 | 8.760 |
| | $Cv(Gaps)$ | 1.155 | 1.496 | 1.511 | 3.720 | 1.320 | 2.010 | 1.877 | 1.939 | 1.545 | 1.712 | 1.221 | 1.352 | 1.915 |
| | $\%Max(Bins)$ | 35.625 | 43.750 | 44.375 | 50.000 | 34.375 | 42.500 | 50.325 | 40.000 | 49.143 | 46.250 | 34.375 | 41.250 | 51.250 |
| **Expression Heterogeneity** | $Me(Cv)$ | 0.293 | 0.380 | 0.284 | 0.316 | 0.298 | 0.295 | 0.332 | 0.379 | 0.425 | 0.352 | 0.261 | 0.279 | 0.276 |
| | $RobSNR$ | 1.354 | 1.367 | 1.355 | 1.360 | 1.355 | 1.356 | 1.367 | 1.356 | 1.376 | 1.363 | 1.350 | 1.353 | 1.356 |

## S6.2 Cross-tissue overlap and comparison with Cosinor-based detections

This section provides additional details on the comparison between CircaFMM detections and the Cosinor-based rhythmicity calls reported by Ruben et al. (2018) for the GTEx-13T dataset. Genes were classified as shared rhythmic (SR), Ruben-only rhythmic (RR), or CircaFMM-only rhythmic (FR), as described in the main text. Peak phases were aligned by setting the peak of *ARNTL* to phase 0 in both approaches.

### S6.2.1 Cross-tissue overlap of SR and FR detections

Consistent with previous large-scale circadian transcriptomic studies (Zhang et al., 2014; Ruben et al., 2018; Seney et al., 2019), the degree of overlap across tissues was moderate to low, highlighting the strong tissue specificity of circadian regulation. The cross-tissue overlap structure of SR and FR CircaFMM detections across tissues is summarized in Figure S7. Panel A displays genes detected as rhythmic in at least 10 tissues, whereas panels B and C show the distribution of the number of tissues in which SR and FR genes were classified as rhythmic, respectively.

At the distribution level, SR detections were generally more tissue-restricted, with only 7.66% detected as rhythmic in four or more tissues, compared with 40.67% among FR detections. Overall, we identified a set of 8 SR genes and 22 FR genes rhythmic in at least 10 tissues. The SR group mainly comprised canonical clock-associated genes, members of the *PER*, *CRY* and REV-ERB families, which are typically characterized by well-established near-sinusoidal oscillatory profiles. In contrast, recurrent FR detections included genes such as *TMED4* or *SERPINA3*, involved in intracellular transport, inflammatory signaling and stress-response pathways (de Mezer et al., 2023; Casanova et al., 2024). Across tissues, these genes displayed heterogeneous non-sinusoidal temporal patterns, as shown for *TMED4* and *SERPINA3* in Figures S8 and S9, respectively.

### S6.2.2 Effect of Cosinor rhythmicity criteria

The Cosinor-based framework used by Ruben et al. (2018) defined rhythmicity through multiple criteria: FDR $< 0.01$, relative amplitude $rAmp \geq 0.1$, and coefficient of determination $R^2 \geq 0.1$. In contrast, CircaFMM defines rhythmicity using FDR-adjusted $p$-values ($p.adj$) from the proposed weighted LRT-type statistic. Differences between the two sets of detections should therefore be interpreted in light of both model class and calling criteria.

The additional $rAmp$ and $R^2$ filters used in the Cosinor-based framework can play a stabilizing role in settings with pronounced temporal irregularity. To illustrate this, we compared Cosinor detections obtained using FDR alone with those retained after applying the full set of Cosinor criteria. Representative examples from COL, LIV and LUN are shown in Figure S10. These tissues show distinct temporal coverage patterns according to the descriptive metrics reported in Table S6: COL displays large temporal gaps, LIV shows stronger local concentration of observations, and LUN shows heterogeneous spacing without extreme gaps. In such settings, FDR-only Cosinor calls may include potentially unstable or spurious detections. The examples shown were neither retained by CircaFMM nor by the full multi-criteria Cosinor procedure.

Conversely, the same additional filters may reduce sensitivity for rhythmic profiles with low amplitude, low explained variance, or non-sinusoidal shape. The $rAmp$ threshold can penalize low-amplitude rhythms, whereas the $R^2$ criterion can be sensitive to irregular sampling and isolated atypical observations. Figure S11 illustrates representative CircaFMM detections in SUB, HEA and NER that were not retained by the multi-criteria Cosinor framework after applying the $rAmp$ and $R^2$ thresholds. These tissues were selected to represent complementary challenging settings: SUB and NER showed low $Me(Cv)$ and relatively low $RobSNR$, consistent with weaker signal-strength regimes, whereas HEA combined higher expression variability with irregular temporal coverage

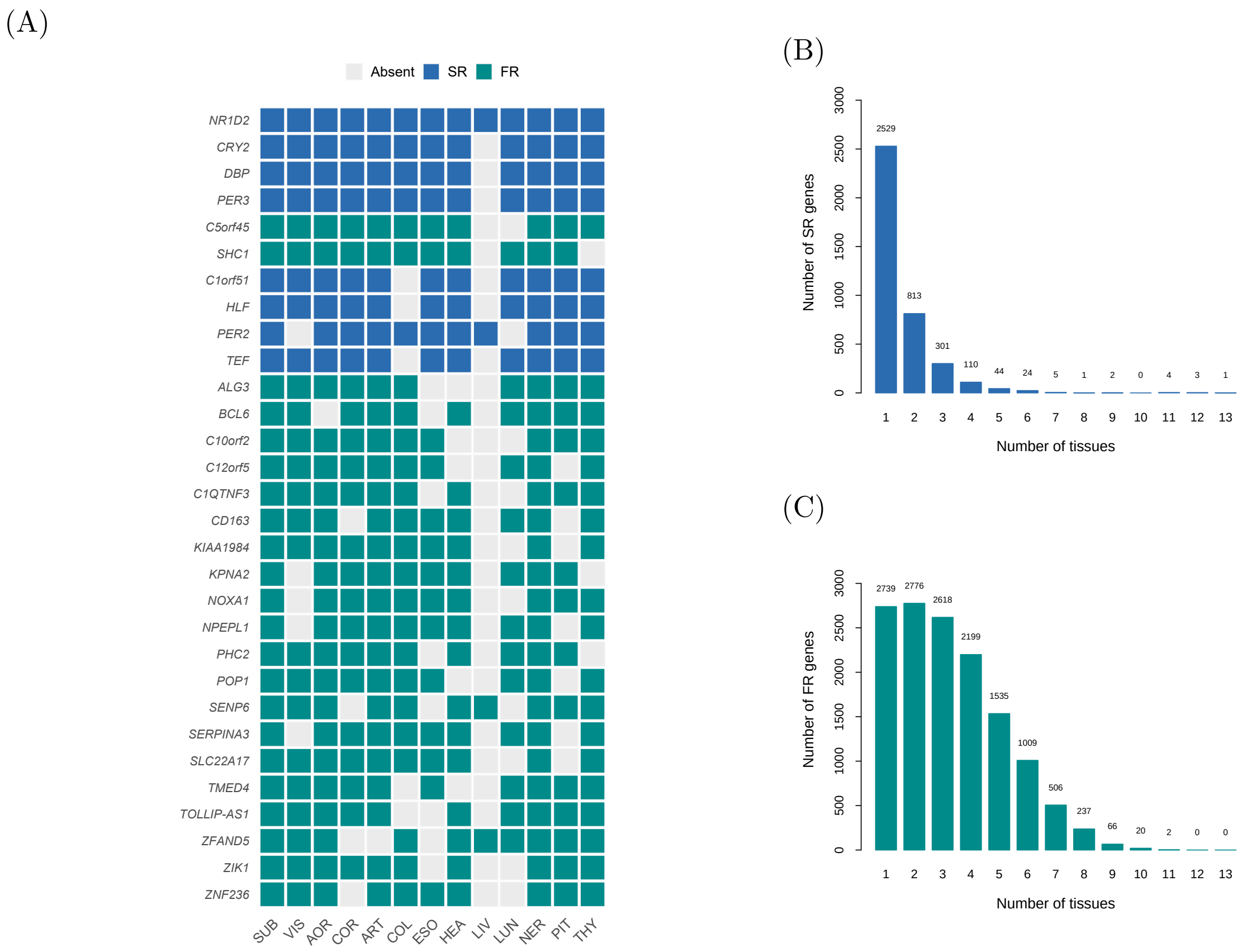


Figure S7: Cross-tissue overlap structure of SR and FR CircaFMM detections. (A) Heatmap for genes classified as rhythmic in at least 10 tissues, with rows representing genes and columns representing tissues. (B) Distribution of the number of tissues in which SR genes were classified as rhythmic by CircaFMM. (C) Distribution of the number of tissues in which FR genes were classified as rhythmic by CircaFMM.

(Table S6). These features can reduce the ability of a sinusoidal fit to satisfy amplitude- and goodness-of-fit-based criteria, even when structured temporal variation is present.

Table S7 summarizes the number of Cosinor detections obtained under the FDR-only and multi-criteria frameworks at nominal FDR level 0.01.

Table S7: Number of circadian detections across tissues for the FDR-only and multi-criteria Cosinor frameworks at nominal level $\alpha = 0.01$. The multi-criteria framework corresponds to the criteria used by Ruben et al. (2018): FDR control, $R^2 \geq 0.1$, and relative amplitude $rAmp \geq 0.1$.

| Criteria | SUB | VIS | AOR | COR | TIB | COL | ESO | HEA | LIV | LUN | NER | PIT | THY |
|---|---|---|---|---|---|---|---|---|---|---|---|---|---|
| $FDR$ | 150 | 1777 | 115 | 1430 | 128 | 282 | 713 | 865 | 244 | 180 | 598 | 156 | 168 |
| $FDR + R^2 + rAmp$ | 149 | 1770 | 115 | 1417 | 128 | 263 | 666 | 861 | 179 | 163 | 591 | 154 | 167 |

## S6.3 Characterization of CircaFMM detections

This section provides additional details on the characterization of CircaFMM detections in the GTEx-13T dataset, focusing on two properties used in the main text to interpret CircaFMM-specific rhythmic genes: waveform asymmetry and signal strength. We first define the duration asymmetry index $AD_{\text{Dur}}$ and examine its relationship with the selected sharpness parameter $\hat{\omega}$. We then describe the spline-based variance SNR used to compare shared rhythmic genes (SR) and CircaFMM-only rhythmic genes (FR).

### S6.3.1 Waveform asymmetry and sharpness

To quantify waveform asymmetry, we used the duration asymmetry index $AD_{\text{Dur}}$, defined from the relative duration of the ascending and descending phases within the circadian period $T = 2\pi$. These phases are determined by the locations of the maximum and minimum of the fitted FMM curve, denoted by $t^U$ and $t^L$, respectively.

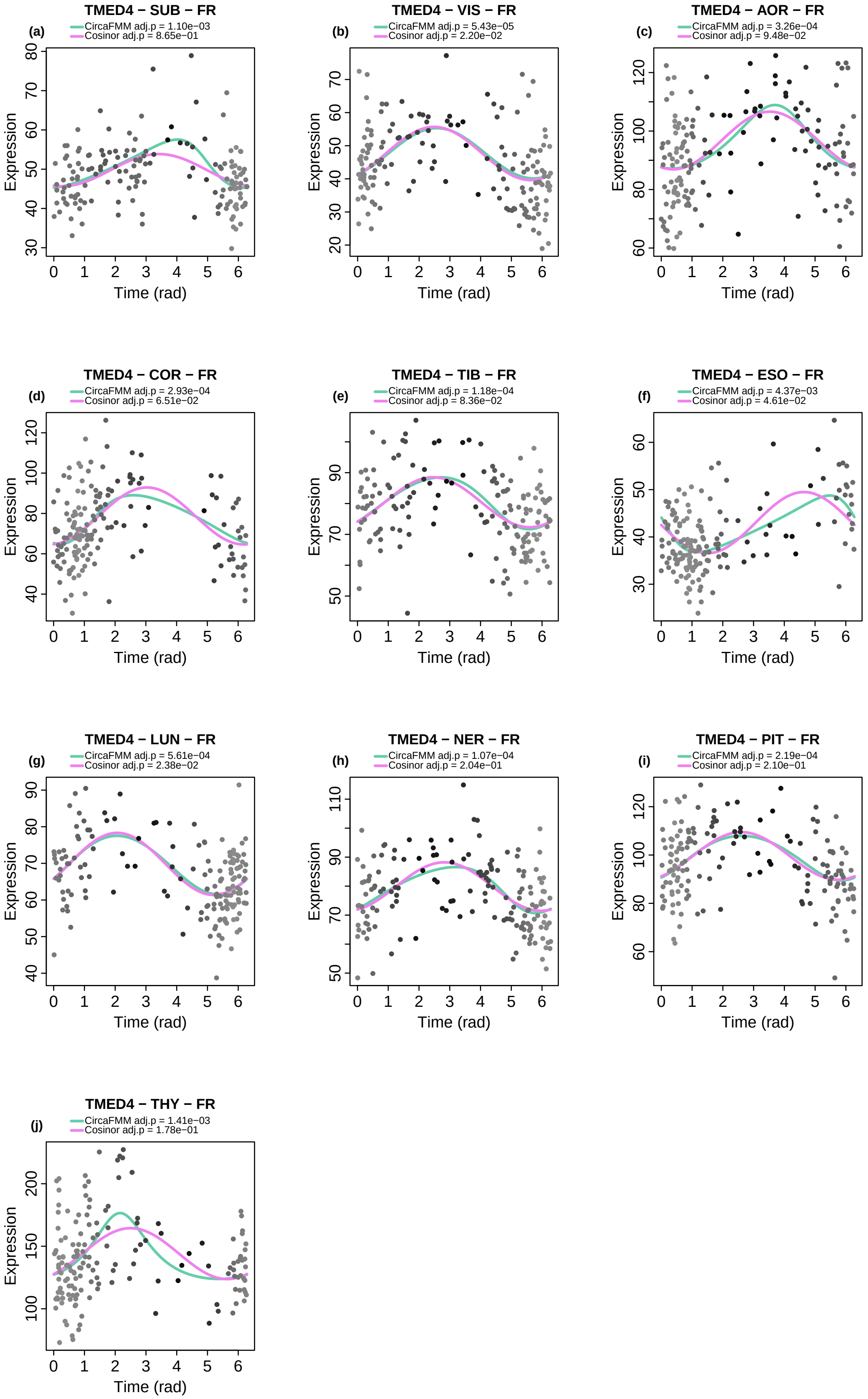


Figure S8: *TMED4* CircaFMM detections across the 10 tissues in which the gene was classified as rhythmic. Grayscale intensity represents observation weights $w_i$, with darker points indicating higher weights.

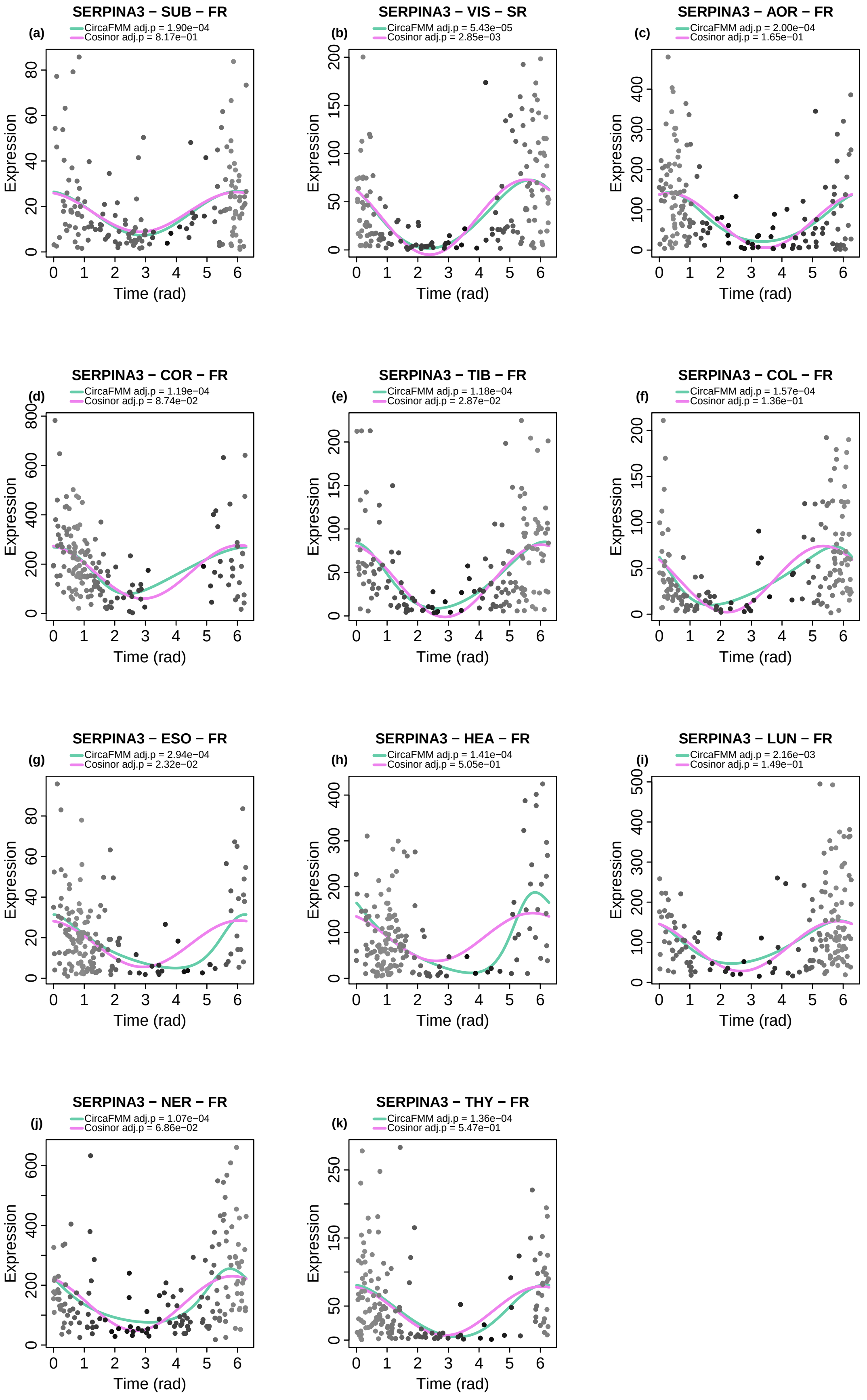


Figure S9: *SERPINA3* CircaFMM detections across the 10 tissues in which the gene was classified as rhythmic. Grayscale intensity represents observation weights $w_i$, with darker points indicating higher weights.

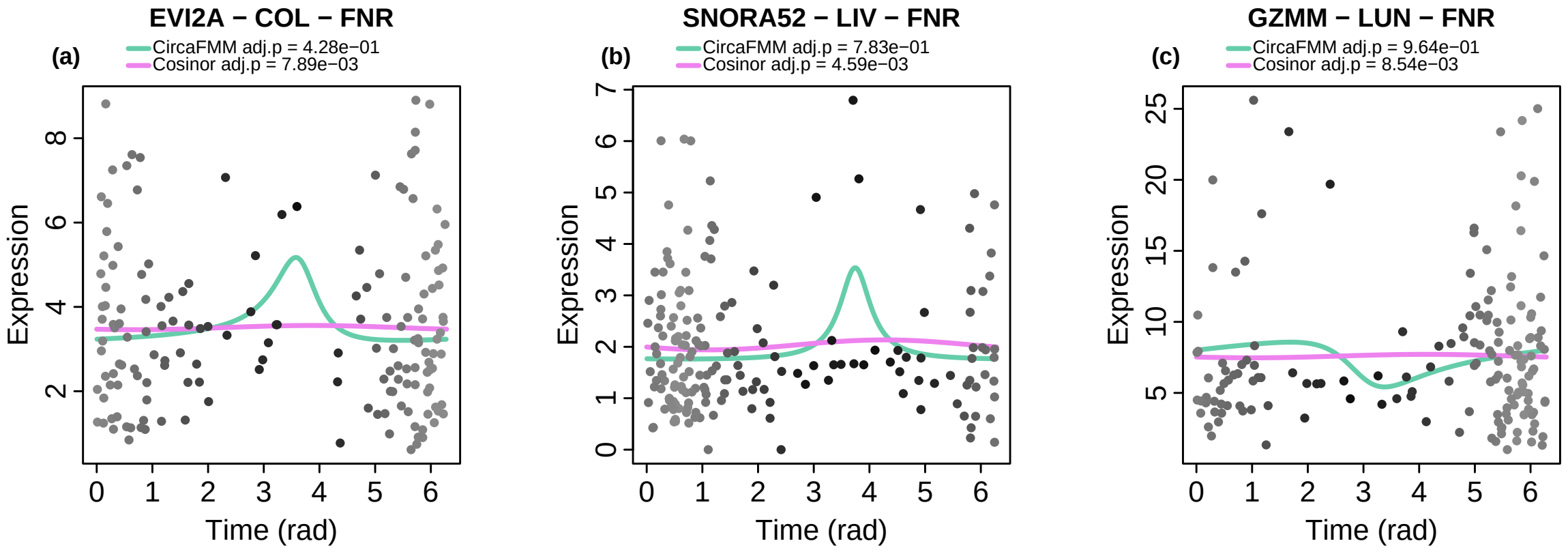


Figure S10: Representative circadian detections identified under the FDR-only Cosinor criterion but not retained by CircaFMM in COL, LIV and LUN tissues. Grayscale intensity represents observation weights $w_i$, with darker points indicating higher weights. FNR denotes genes not detected as rhythmic by CircaFMM.

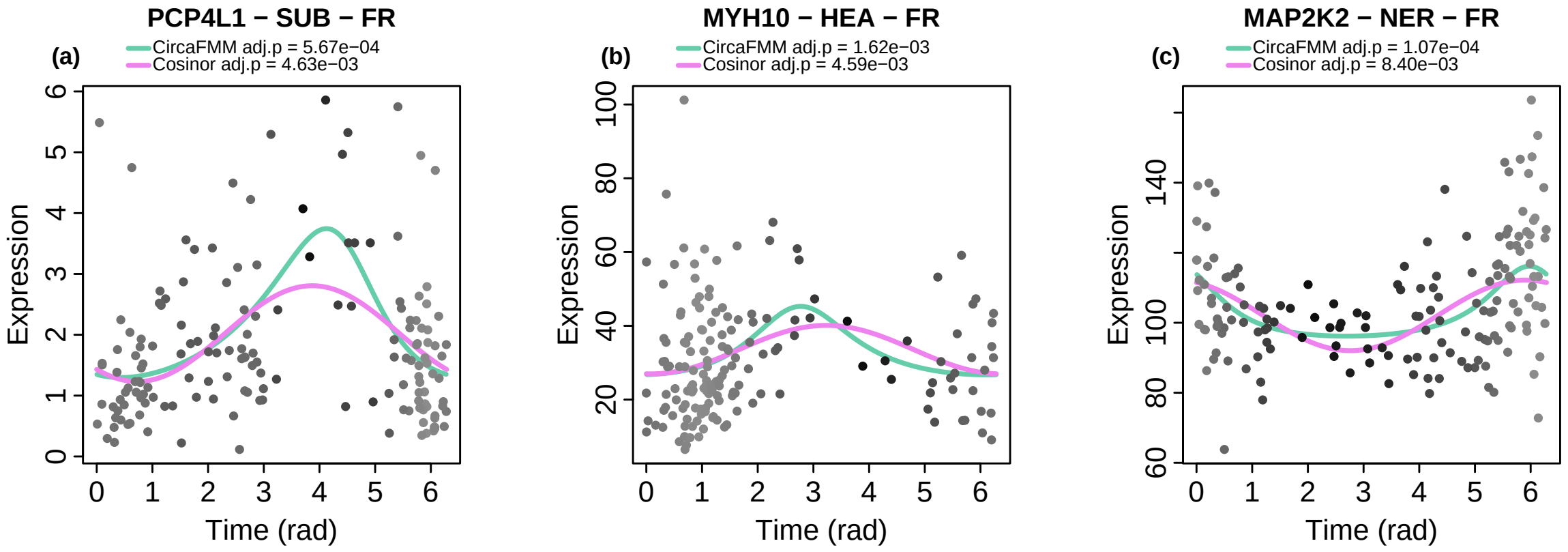


Figure S11: Representative profiles detected by CircaFMM and by the FDR-only Cosinor analysis, but not retained under the multi-criteria Cosinor framework, in SUB, HEA and NER tissues. Grayscale intensity represents observation weights $w_i$, with darker points indicating higher weights.

Following Rueda et al. (2019), these circadian landmarks can be obtained from the FMM parameters as

$$t^U = \alpha + 2\arctan\left(\frac{1}{\omega}\tan\left(\frac{-\beta}{2}\right)\right) \mod 2\pi,$$
$$t^L = \alpha + 2\arctan\left(\frac{1}{\omega}\tan\left(\frac{\pi-\beta}{2}\right)\right) \mod 2\pi.$$

The ascending duration was defined as the circular distance from trough to peak,

$$\text{asc} = (t^U - t^L) \mod 2\pi,$$

whereas the descending duration corresponds to the remaining part of the cycle,

$$\text{desc} = T - \text{asc}.$$

The duration asymmetry index is then

$$AD_{\text{Dur}} = \frac{\text{desc} - \text{asc}}{T} = 1 - 2\frac{\text{asc}}{T}.$$

Since asc $\in [0, T]$, it follows that $AD_{\text{Dur}} \in [-1, 1]$. Values close to zero indicate approximately symmetric profiles, whereas larger absolute values indicate stronger asymmetry. The sign indicates which phase is relatively longer.

The joint analysis $AD_{\text{Dur}}$ and $\hat{\omega}$ is informative because $\hat{\omega}$ constrains the range of feasible duration asymmetries. As $\omega \to 1$, the waveform approaches a symmetric cosine-like profile, implying asc $\approx T/2$ and therefore $AD_{\text{Dur}} \approx 0$. Smaller values of $\omega$ allow a wider range of asymmetry shapes, which gives rise to the characteristic triangular admissible region in the $(\omega, AD_{\text{Dur}})$ plane.

Figures S12–S13 illustrate the tissue-specific distribution of $(\hat{\omega}, AD_{\text{Dur}})$ values for SR and FR genes. SR genes were concentrated closer to the symmetric regime, whereas FR genes spanned a broader portion of the admissible region, including lower $\hat{\omega}$ values and more extreme $|AD_{\text{Dur}}|$ values. Consistently, $|AD_{\text{Dur}}|$ was significantly higher for FR genes than for SR genes in most tissues according to Wilcoxon rank-sum tests with FDR adjustment, $p.adj < 0.01$. The only exceptions were LIV and COR, where the reduced number of rhythmic detections (see Table 2 in the main text) and highly irregular sampling distribution (Table S6), respectively, may limit the ability to detect global differences in asymmetry.

The broader asymmetry of FR detections was also evident across tissues. The IQR of $AD_{\text{Dur}}$ was consistently larger for FR genes than for SR genes across most $\hat{\omega}$ bins (Table S8). These results indicate that CircaFMM-specific detections are enriched for asymmetric temporal profiles that are not well represented by a classical sinusoidal basis.

To further illustrate the ability of CircaFMM to capture asymmetric temporal expression patterns, Figures S14–S26 display representative FR genes selected across tissues among those with $\hat{\omega} < 0.5$, covering a broad range of $AD_{\text{Dur}}$ values from $-0.577$ to $0.807$ (Table S9).

### S6.3.2 Low-SNR CircaFMM detections

To assess the signal-strength regime of SR and FR detections, we computed a gene-wise variance-based SNR using a nonparametric periodic-spline fit. This metric differs from the tissue-level *RobSNR* proxy reported in Table S6. Whereas *RobSNR* provides a descriptive tissue-level summary of expression variability, the SNR used here is computed separately for each gene as the ratio between the variance explained by a smooth periodic trend and the residual variance.

ºtextcolorredFor each gene, a generalized additive model (GAM) was fitted to the observed data $x_i$ as follows:

$$x_i = f(t_i) + \varepsilon_i,$$

where $f(\cdot)$ is a smooth periodic function estimated using a cyclic cubic spline basis, and $\varepsilon_i$ represents residual noise. The spline basis dimension was set to $k = 10$, providing sufficient flexibility to capture smooth oscillatory trends while avoiding overly complex fits. Models were fitted using restricted maximum likelihood (REML).

The gene-wise variance-based SNR was defined as

$$\text{SNR} = \frac{\text{Var}(\hat{f}(t))}{\text{Var}(\hat{\varepsilon})},$$

where $\text{Var}(\hat{f}(t))$ denotes the variance of the fitted periodic trend and $\text{Var}(\hat{\varepsilon})$ denotes the variance of the residual component. To improve numerical stability and reduce skewness, we used the log-transformed measure

$$\log(1 + \text{SNR}).$$

This procedure was applied gene-wise within each tissue. According to the one-sided Wilcoxon rank-sum tests, FR genes exhibited consistently lower $\log(1 + \text{SNR})$ values than SR genes across tissues, with statistically

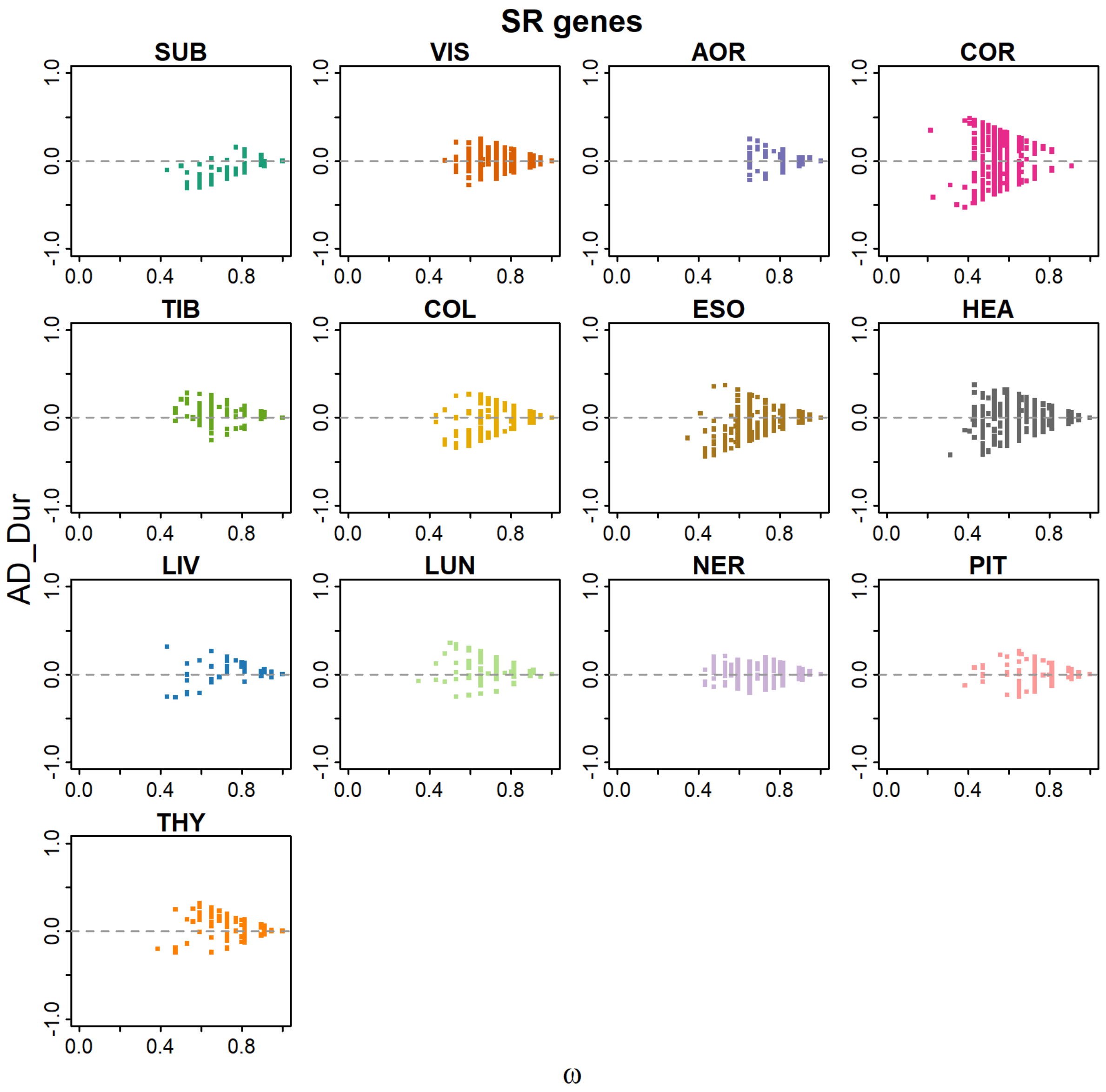


Figure S12: Relationship between the duration asymmetry index $AD_{\mathrm{Dur}}$ and the selected sharpness parameter $\hat{\omega}$ for shared rhythmic genes (SR) across GTEx-13T tissues. Values of $AD_{\mathrm{Dur}}$ close to zero indicate approximately symmetric profiles, whereas larger absolute values indicate stronger waveform asymmetry.

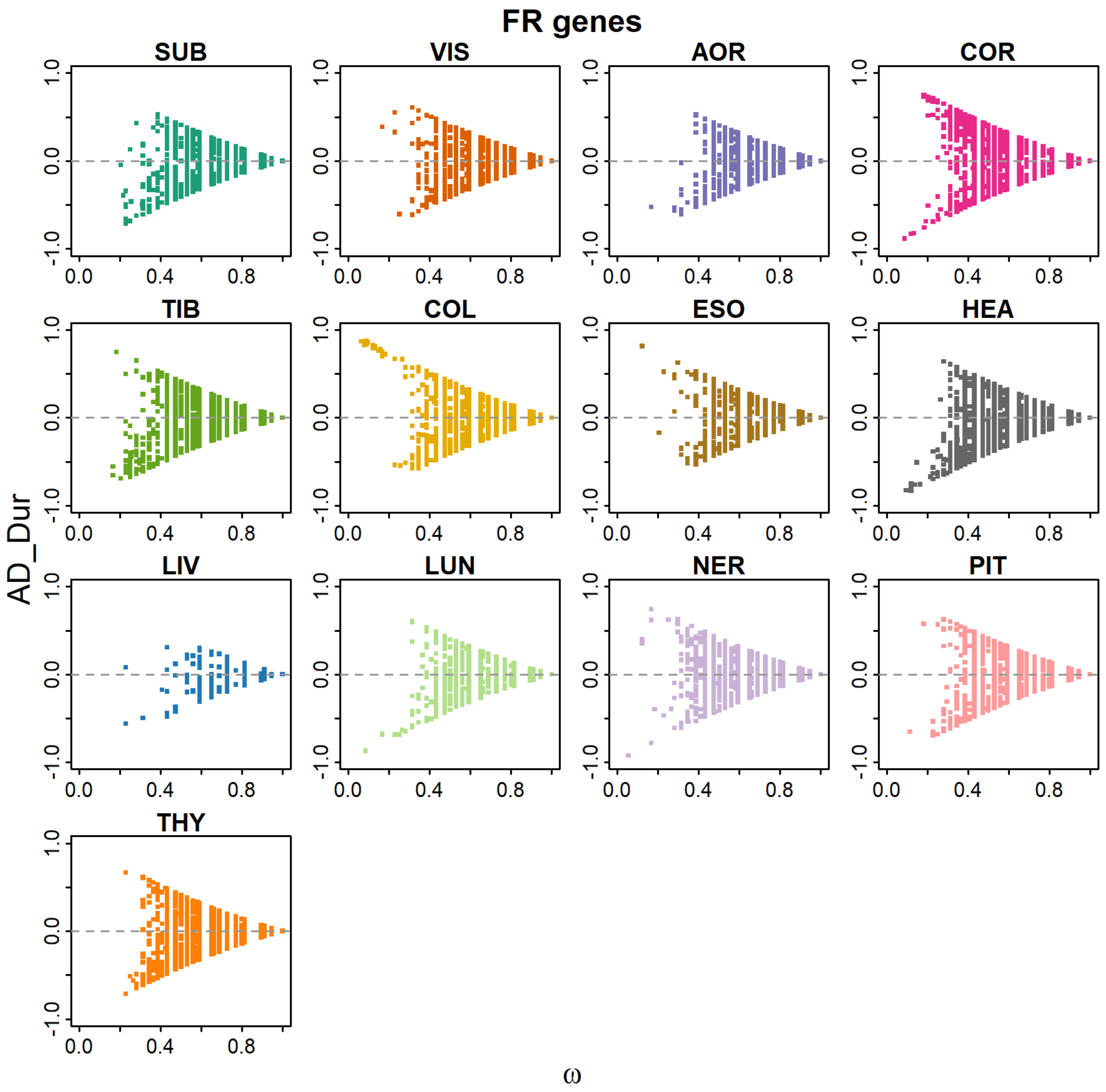


Figure S13: Relationship between the duration asymmetry index $AD_{\text{Dur}}$ and the selected sharpness parameter $\hat{\omega}$ for CircaFMM-only rhythmic genes (FR) across GTEx-13T tissues. Compared with SR genes, FR genes occupy a broader region of the $(\hat{\omega}, AD_{\text{Dur}})$ space, including lower $\hat{\omega}$ values and more asymmetric waveform profiles.

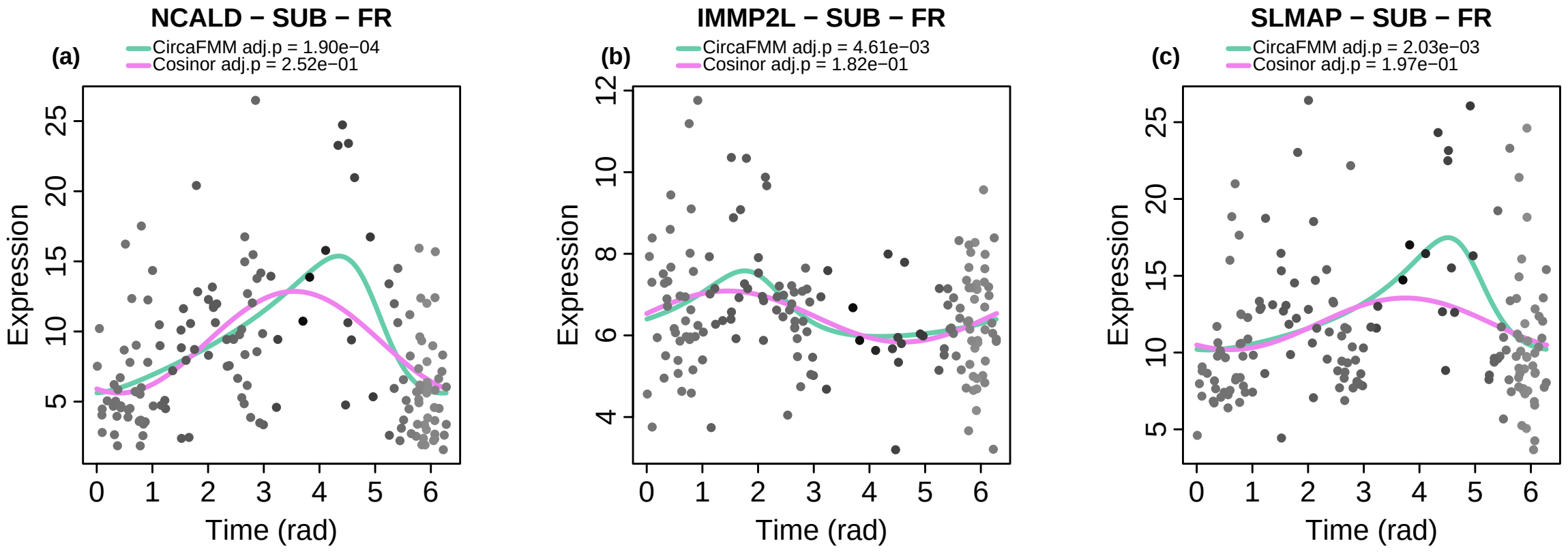


Figure S14: Representative FR genes selected to illustrate asymmetric CircaFMM profiles in SUB. Genes were selected among FR detections with $\hat{\omega} < 0.5$. Points represent observed expression values, and grayscale intensity indicates observation weights $w_i$, with darker points corresponding to higher weights.

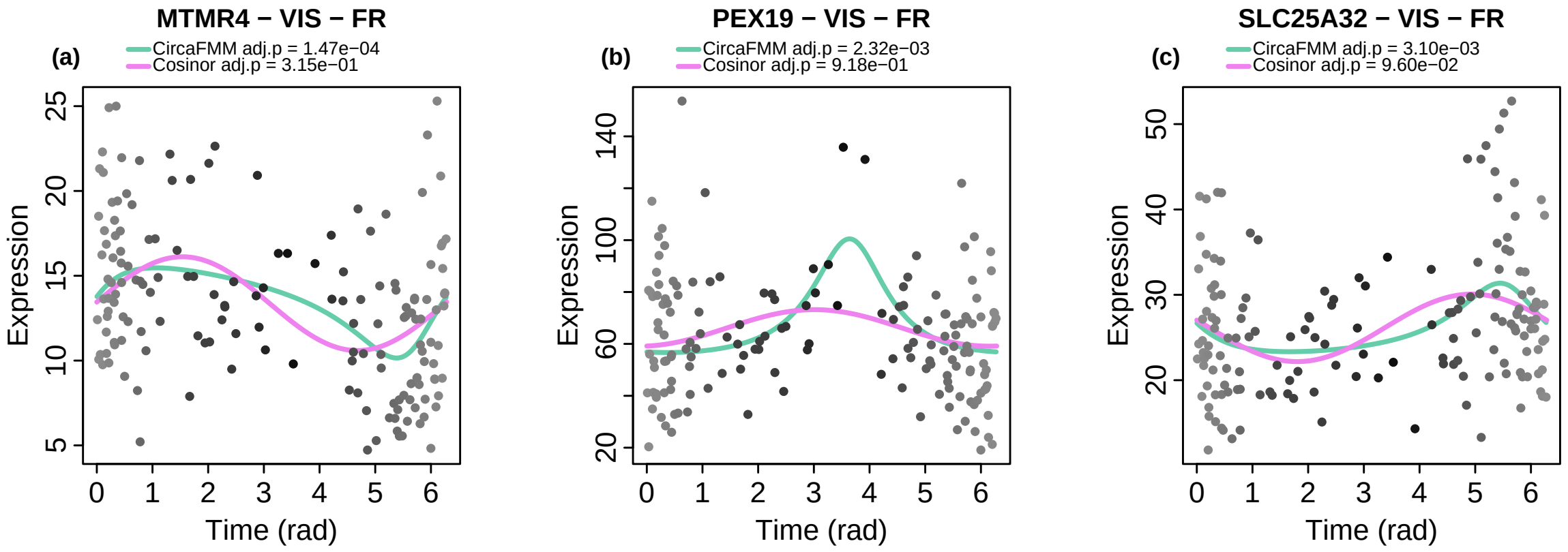


Figure S15: Representative FR genes selected to illustrate asymmetric CircaFMM profiles in VIS. Genes were selected among FR detections with $\hat{\omega} < 0.5$. Points represent observed expression values, and grayscale intensity indicates observation weights $w_i$, with darker points corresponding to higher weights.

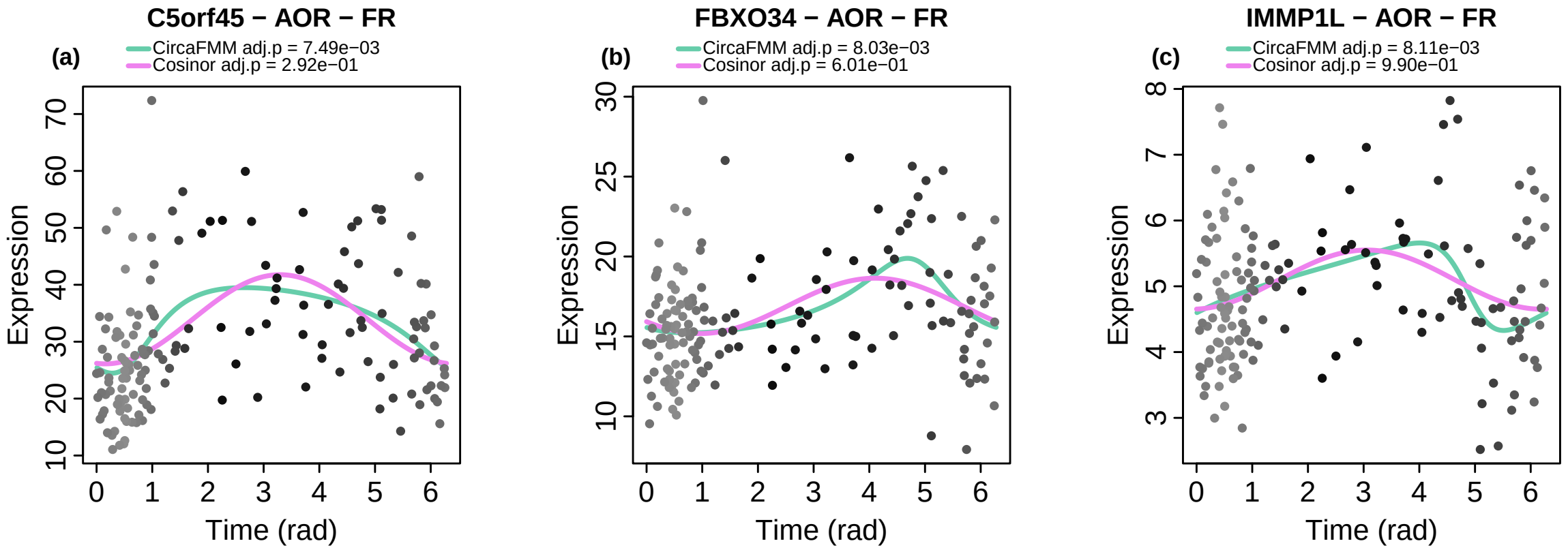


Figure S16: Representative FR genes selected to illustrate asymmetric CircaFMM profiles in AOR. Genes were selected among FR detections with $\hat{\omega} < 0.5$. Points represent observed expression values, and grayscale intensity indicates observation weights $w_i$, with darker points corresponding to higher weights.

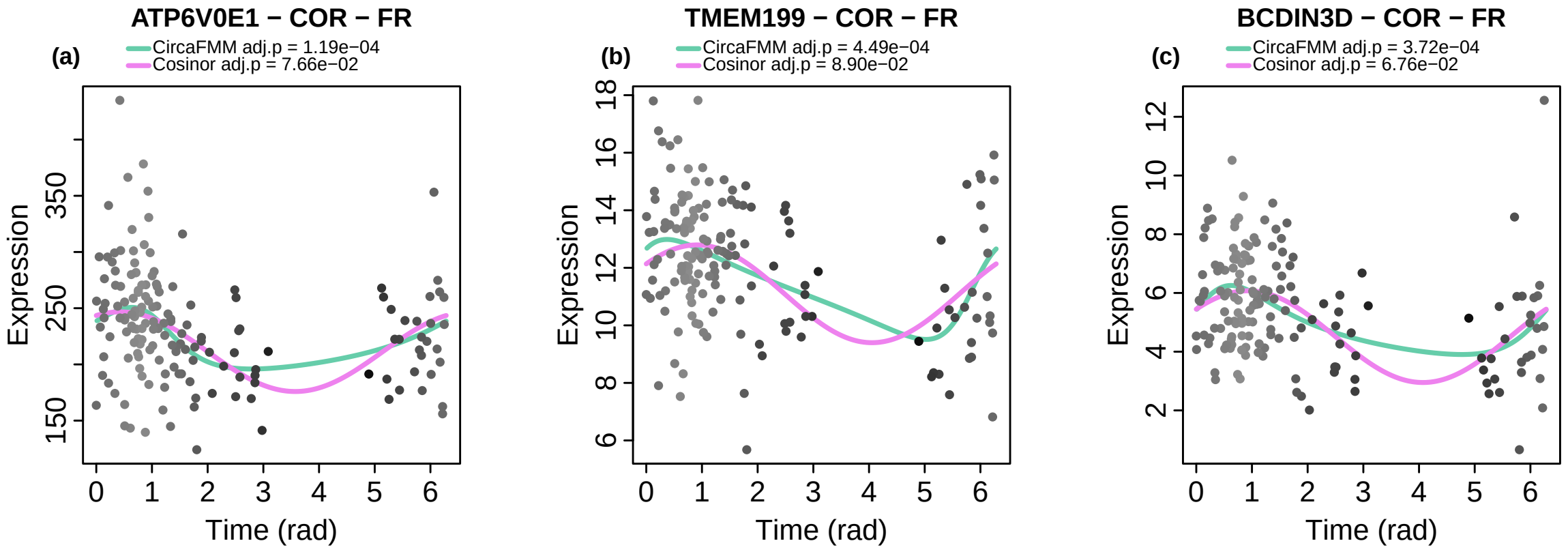


Figure S17: Representative FR genes selected to illustrate asymmetric CircaFMM profiles in COR. Genes were selected among FR detections with $\hat{\omega} < 0.5$. Points represent observed expression values, and grayscale intensity indicates observation weights $w_i$, with darker points corresponding to higher weights.

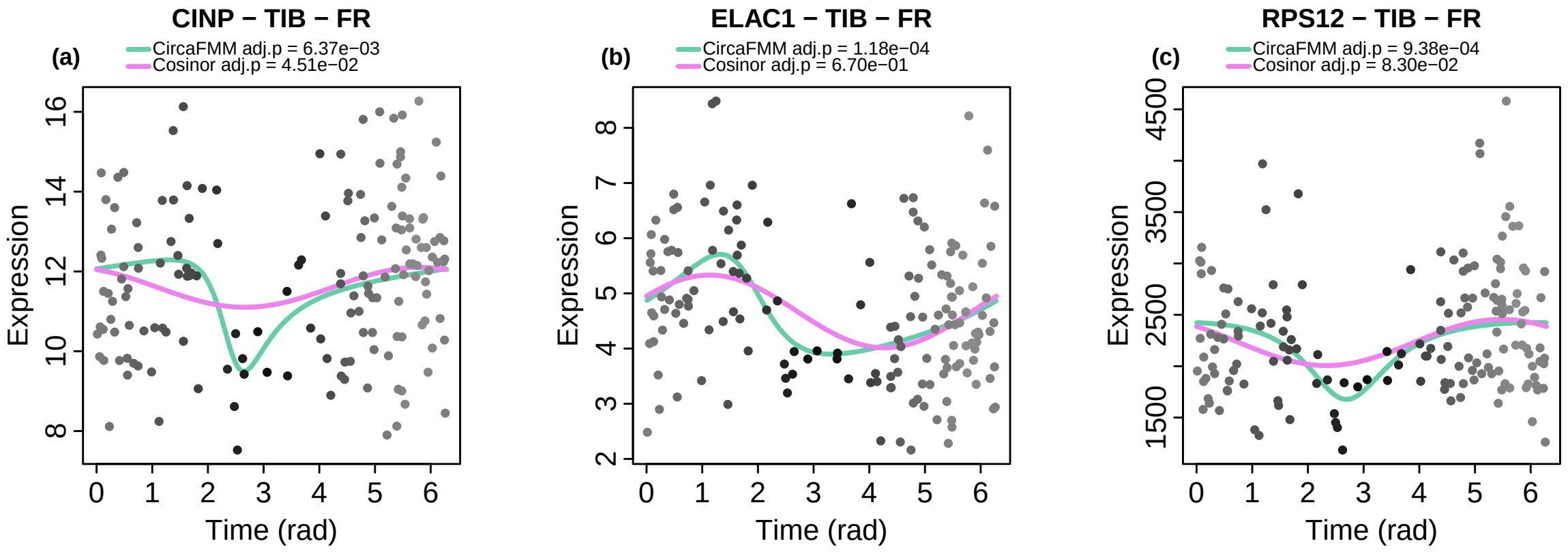


Figure S18: Representative FR genes selected to illustrate asymmetric CircaFMM profiles in TIB. Genes were selected among FR detections with $\hat{\omega} < 0.5$. Points represent observed expression values, and grayscale intensity indicates observation weights $w_i$, with darker points corresponding to higher weights.

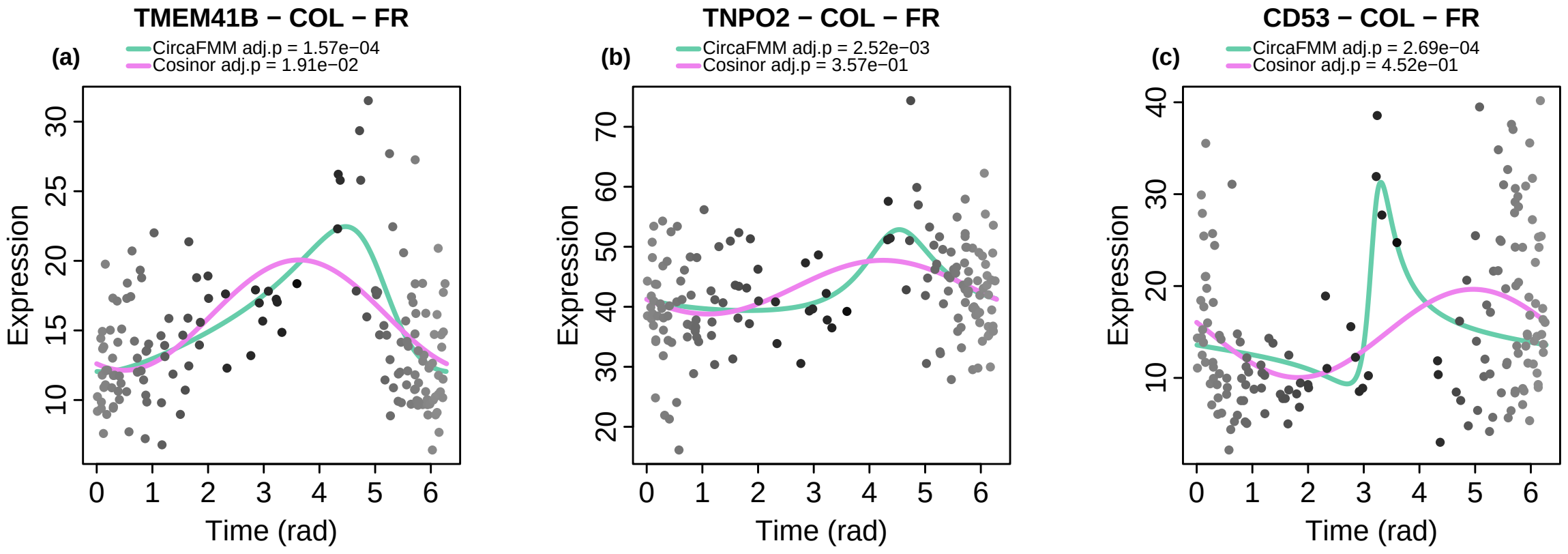


Figure S19: Representative FR genes selected to illustrate asymmetric CircaFMM profiles in COL. Genes were selected among FR detections with $\hat{\omega} < 0.5$. Points represent observed expression values, and grayscale intensity indicates observation weights $w_i$, with darker points corresponding to higher weights.

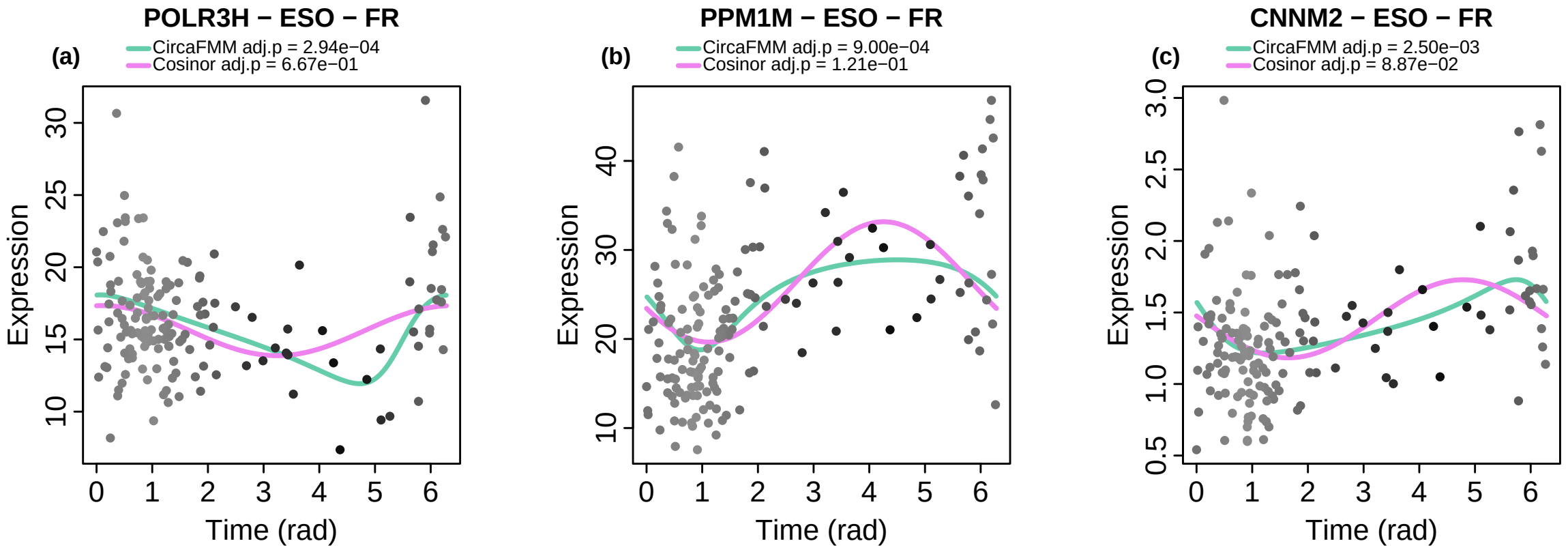


Figure S20: Representative FR genes selected to illustrate asymmetric CircaFMM profiles in ESO. Genes were selected among FR detections with $\hat{\omega} < 0.5$. Points represent observed expression values, and grayscale intensity indicates observation weights $w_i$, with darker points corresponding to higher weights.

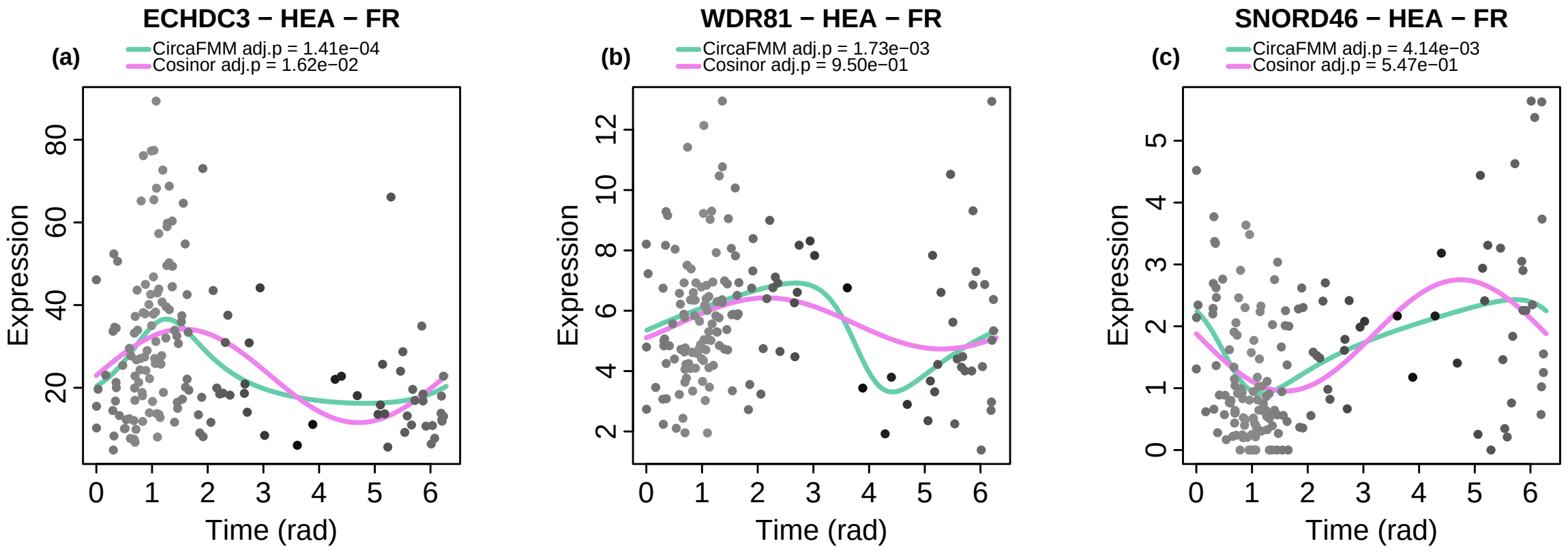


Figure S21: Representative FR genes selected to illustrate asymmetric CircaFMM profiles in HEA. Genes were selected among FR detections with $\hat{\omega} < 0.5$. Points represent observed expression values, and grayscale intensity indicates observation weights $w_i$, with darker points corresponding to higher weights.

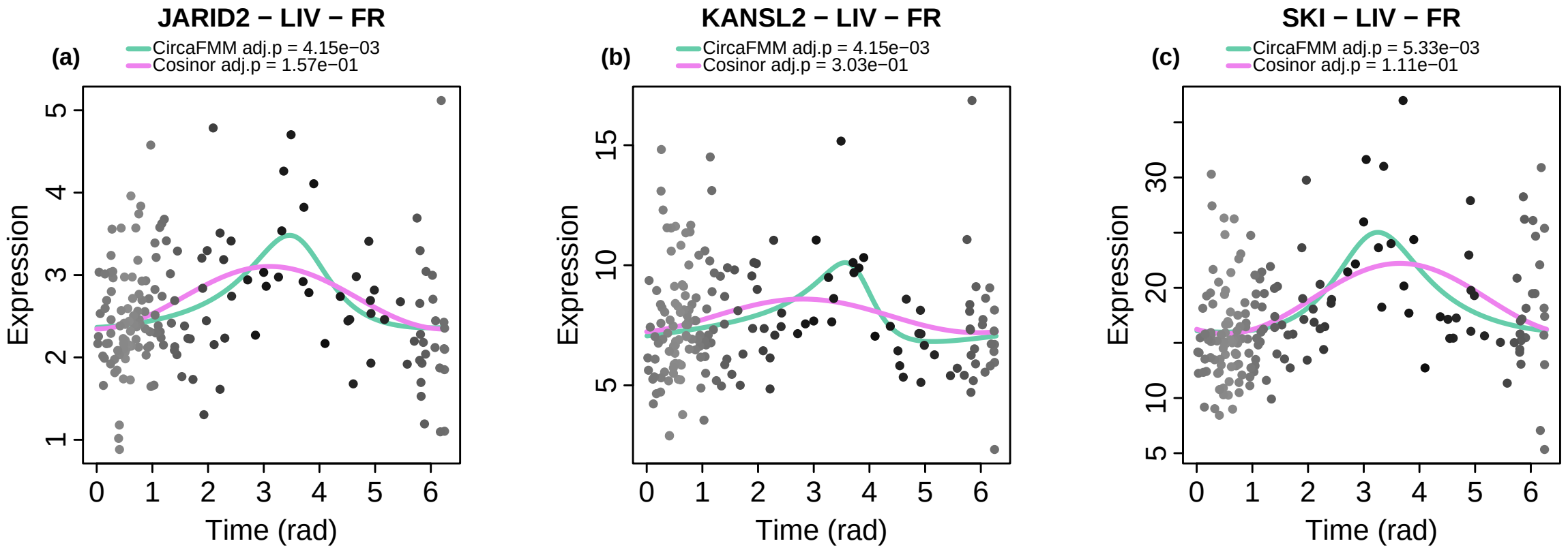


Figure S22: Representative FR genes selected to illustrate asymmetric CircaFMM profiles in LIV. Genes were selected among FR detections with $\hat{\omega} < 0.5$. Points represent observed expression values, and grayscale intensity indicates observation weights $w_i$, with darker points corresponding to higher weights.

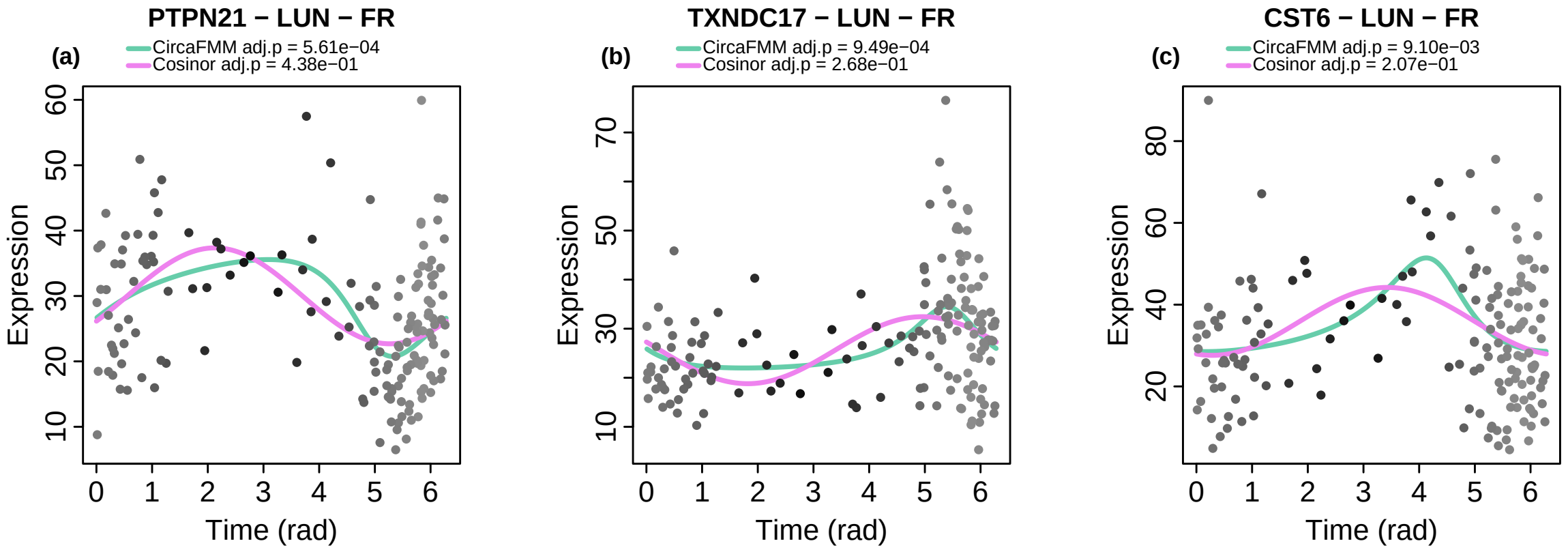


Figure S23: Representative FR genes selected to illustrate asymmetric CircaFMM profiles in LUN. Genes were selected among FR detections with $\hat{\omega} < 0.5$. Points represent observed expression values, and grayscale intensity indicates observation weights $w_i$, with darker points corresponding to higher weights.

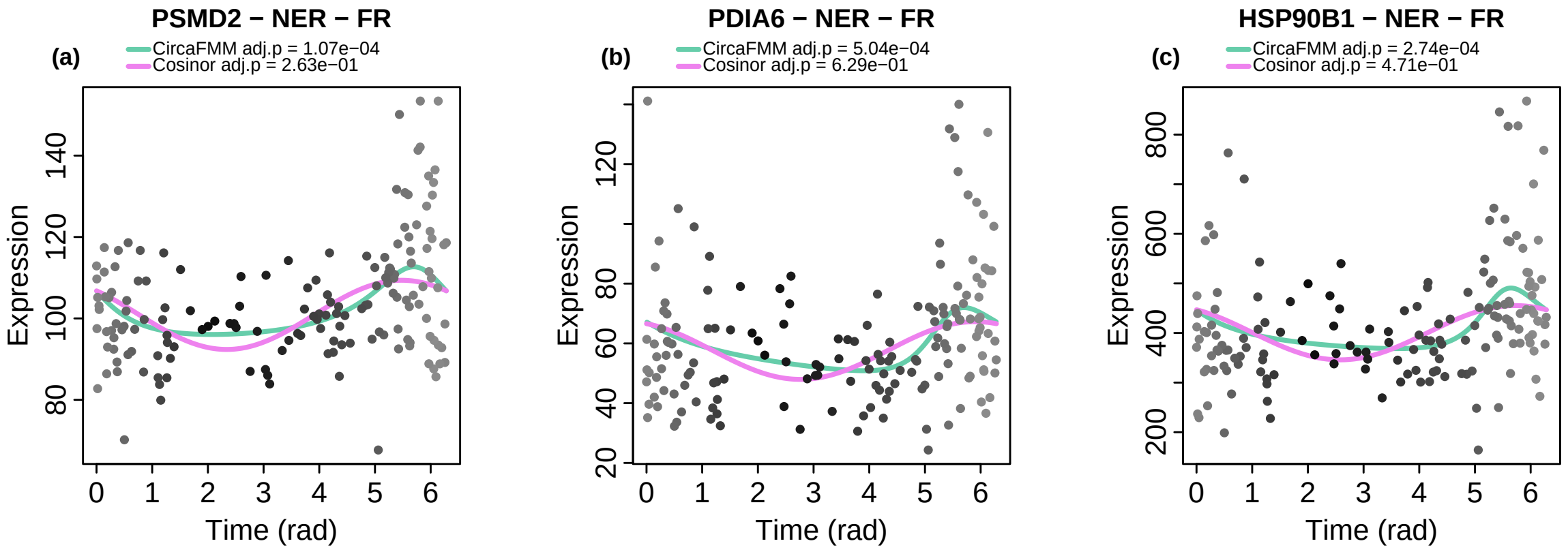


Figure S24: Representative FR genes selected to illustrate asymmetric CircaFMM profiles in NER. Genes were selected among FR detections with $\hat{\omega} < 0.5$. Points represent observed expression values, and grayscale intensity indicates observation weights $w_i$, with darker points corresponding to higher weights.

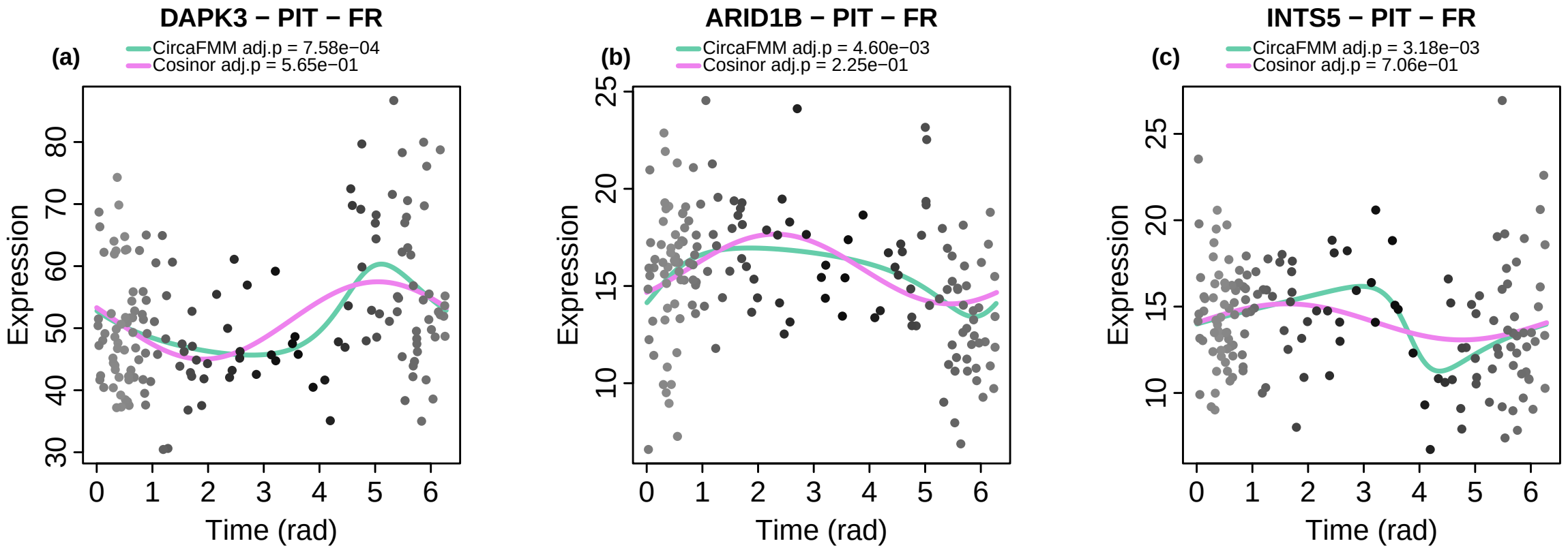


Figure S25: Representative FR genes selected to illustrate asymmetric CircaFMM profiles in PIT. Genes were selected among FR detections with $\hat{\omega} < 0.5$. Points represent observed expression values, and grayscale intensity indicates observation weights $w_i$, with darker points corresponding to higher weights.

Table S8: Interquartile range (IQR) of the duration asymmetry index $AD_{\text{Dur}}$ by tissue, gene group and $\hat{\omega}$ bin. Empty cells indicate that no genes were present in the corresponding bin. Asterisks denote tissues for which no statistically significant differences in $|AD_{\text{Dur}}|$ were observed between SR and FR genes after FDR adjustment.

| Tissue | Group | (0,0.2] | (0.2,0.3] | (0.3,0.4] | (0.4,0.5] | (0.5,0.6] | (0.6,0.7] | (0.7,0.8] | (0.8,0.9] | (0.9,1] |
|---|---|---|---|---|---|---|---|---|---|---|
| SUB | SR | | | | 0.000 | 0.114 | 0.069 | 0.059 | 0.083 | 0.000 |
| | FR | | 0.227 | 0.361 | 0.552 | 0.462 | 0.358 | 0.243 | 0.151 | 0.000 |
| VIS | SR | | | | 0.000 | 0.106 | 0.101 | 0.096 | 0.094 | 0.000 |
| | FR | 0.000 | 0.579 | 0.633 | 0.453 | 0.340 | 0.287 | 0.231 | 0.136 | 0.000 |
| AOR | SR | | | | | | 0.199 | 0.210 | 0.140 | 0.000 |
| | FR | 0.000 | 0.016 | 0.746 | 0.622 | 0.459 | 0.383 | 0.266 | 0.159 | 0.000 |
| COR* | SR | | 0.381 | 0.236 | 0.558 | 0.282 | 0.110 | 0.270 | 0.214 | 0.002 |
| | FR | 1.559 | 0.614 | 0.701 | 0.640 | 0.366 | 0.295 | 0.212 | 0.163 | 0.022 |
| TIB | SR | | | | 0.070 | 0.158 | 0.186 | 0.084 | 0.067 | 0.000 |
| | FR | 0.702 | 0.268 | 0.747 | 0.626 | 0.408 | 0.275 | 0.223 | 0.156 | 0.000 |
| COL | SR | | | | 0.291 | 0.281 | 0.356 | 0.275 | 0.204 | 0.022 |
| | FR | 0.056 | 1.143 | 0.594 | 0.637 | 0.451 | 0.348 | 0.259 | 0.159 | 0.000 |
| ESO | SR | | | 0.000 | 0.263 | 0.239 | 0.245 | 0.179 | 0.166 | 0.000 |
| | FR | 0.000 | 0.349 | 0.892 | 0.535 | 0.473 | 0.364 | 0.261 | 0.150 | 0.000 |
| HEA | SR | | | 0.140 | 0.232 | 0.377 | 0.344 | 0.198 | 0.129 | 0.012 |
| | FR | 0.069 | 0.250 | 0.406 | 0.501 | 0.423 | 0.310 | 0.236 | 0.128 | 0.008 |
| LIV* | SR | | | | 0.287 | 0.270 | 0.148 | 0.115 | 0.068 | 0.027 |
| | FR | | 0.321 | 0.000 | 0.453 | 0.356 | 0.290 | 0.193 | 0.078 | 0.003 |
| LUN | SR | | | 0.000 | 0.220 | 0.308 | 0.225 | 0.176 | 0.087 | 0.000 |
| | FR | 0.097 | 0.040 | 0.685 | 0.361 | 0.361 | 0.242 | 0.208 | 0.123 | 0.000 |
| NER | SR | | | | 0.199 | 0.068 | 0.115 | 0.135 | 0.074 | 0.019 |
| | FR | 1.099 | 0.967 | 0.429 | 0.317 | 0.296 | 0.258 | 0.205 | 0.125 | 0.000 |
| PIT | SR | | | 0.000 | 0.088 | 0.146 | 0.309 | 0.266 | 0.154 | 0.018 |
| | FR | 0.613 | 0.841 | 0.778 | 0.761 | 0.509 | 0.369 | 0.268 | 0.165 | 0.000 |
| THY | SR | | | 0.000 | 0.245 | 0.113 | 0.111 | 0.186 | 0.130 | 0.000 |
| | FR | | 0.101 | 0.867 | 0.571 | 0.508 | 0.378 | 0.280 | 0.151 | 0.003 |

significant differences in all cases after FDR adjustment ($adj.p < 0.01$). The corresponding distributions are shown in Figure S27.

To further illustrate the ability of CircaFMM to detect rhythmic expression under low-SNR conditions, we randomly selected FR genes from the lower quartile of log(1 + SNR) in TIB, COL and NER. These tissues showed the lowest average log(1+SNR) values among FR genes (all below 0.15; Table S10) while still exhibiting CircaFMM rhythmicity rates above 30% in the main analysis. The selected examples are shown in Figures S28–S30, and their corresponding SNR values are reported in Table S11.

## S6.4 Functional clustering analysis

### S6.4.1 Clustering procedure

Functional clustering analysis (FCA) was performed independently within each tissue using the fitted temporal expression profiles obtained from Cosinor and CircaFMM. For the Cosinor analysis, SR and RR detections were included, whereas for the CircaFMM analysis, SR and FR detections were included.

For each gene, fitted curves were evaluated on a dense circular time grid and rescaled to the interval $[-1, 1]$ through gene-wise min–max normalization, reducing the influence of amplitude differences. Functional principal component analysis (FPCA) was then applied to the normalized curves, retaining the first four components. Clustering was performed on the FPCA scores using the $k$-medoids algorithm, implemented through PAM (Partitioning Around Medoids). The number of clusters was selected using an elbow criterion applied to the silhouette curve over candidate values $k = 2, \ldots, 10$.

To improve interpretability and reduce redundant patterns, a post-clustering aggregation step was applied. Cluster-specific mean curves were compared using squared $\ell_2$ distances, and clusters with pairwise distances below 0.05 were iteratively merged using a transitive union–find strategy. This step reduced over-segmentation caused by minor phase shifts or negligible waveform differences.

Table S9: Representative FR genes selected to illustrate asymmetric CircaFMM profiles across tissues. Genes were selected among FR detections with $\hat{\omega} < 0.5$, covering a broad range of duration asymmetry values. The table reports the selected sharpness parameter $\hat{\omega}$ and the duration asymmetry index $AD_{\text{Dur}}$.

| Tissue | Gene | $\hat{\omega}$ | $AD_{\text{Dur}}$ |
|---|---|---|---|
| SUB | *NCALD* | 0.474 | -0.407 |
| SUB | *IMMP2L* | 0.474 | -0.231 |
| SUB | *SLMAP* | 0.431 | -0.375 |
| VIS | *MTMR4* | 0.431 | 0.374 |
| VIS | *PEX19* | 0.386 | -0.040 |
| VIS | *SLC25A32* | 0.431 | -0.203 |
| AOR | *C5orf45* | 0.474 | 0.244 |
| AOR | *FBXO34* | 0.474 | -0.248 |
| AOR | *IMMP1L* | 0.386 | -0.520 |
| COR | *ATP6V0E1* | 0.474 | -0.316 |
| COR | *TMEM199* | 0.431 | 0.481 |
| COR | *BCDIN3D* | 0.474 | 0.320 |
| TIB | *CINP* | 0.251 | -0.577 |
| TIB | *ELAC1* | 0.474 | -0.365 |
| TIB | *RPS12* | 0.431 | -0.067 |
| COL | *TMEM41B* | 0.474 | -0.390 |
| COL | *TNPO2* | 0.386 | 0.133 |
| COL | *CD53* | 0.121 | 0.807 |
| ESO | *POLR3H* | 0.431 | 0.479 |
| ESO | *PPM1M* | 0.474 | -0.130 |
| ESO | *CNNM2* | 0.431 | -0.442 |
| HEA | *ECHDC3* | 0.431 | 0.135 |
| HEA | *WDR81* | 0.431 | -0.446 |
| HEA | *SNORD46* | 0.431 | -0.452 |
| LIV | *JARID2* | 0.431 | -0.193 |
| LIV | *KANSL2* | 0.314 | -0.496 |
| LIV | *SKI* | 0.474 | 0.119 |
| LUN | *PTPN21* | 0.474 | -0.291 |
| LUN | *TXNDC17* | 0.365 | -0.156 |
| LUN | *CST6* | 0.431 | -0.256 |
| NER | *PSMD2* | 0.431 | -0.153 |
| NER | *PDIA6* | 0.386 | 0.430 |
| NER | *HSP90B1* | 0.345 | 0.346 |
| PIT | *DAPK3* | 0.474 | 0.260 |
| PIT | *ARID1B* | 0.431 | 0.283 |
| PIT | *INTS5* | 0.345 | -0.553 |
| THY | *ADAMTSL3* | 0.474 | 0.067 |
| THY | *RTN3* | 0.386 | 0.021 |
| THY | *PLXNB1* | 0.345 | -0.508 |

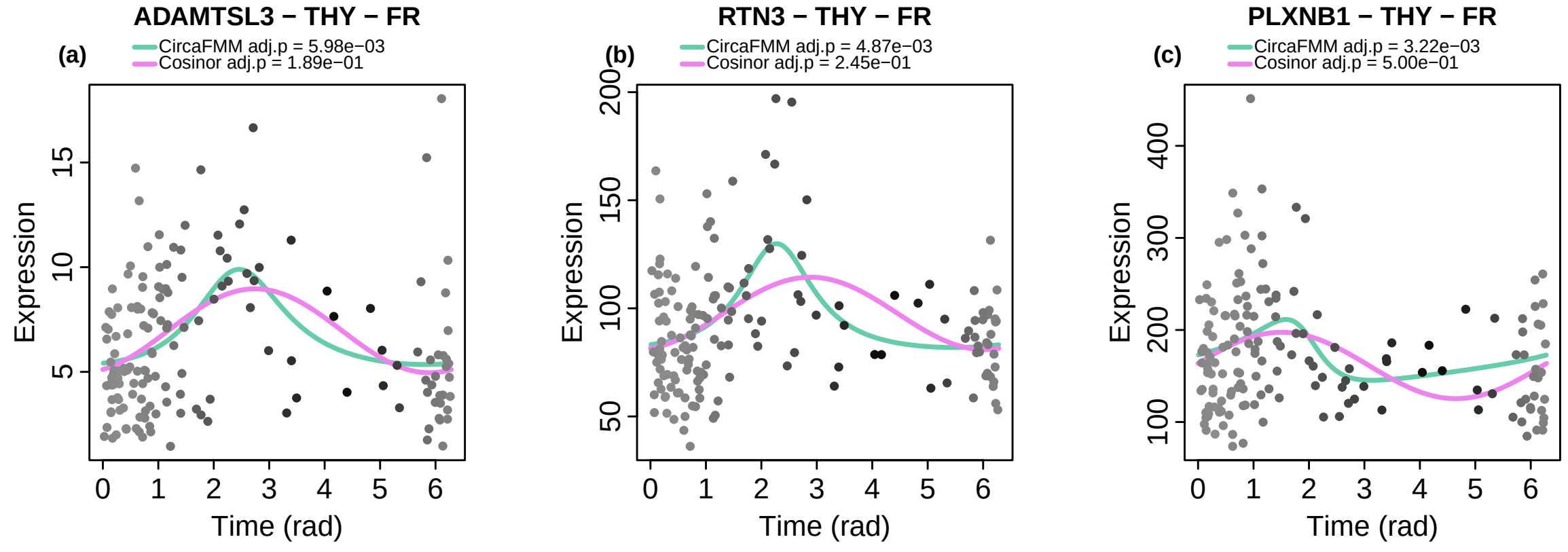


Figure S26: Representative FR genes selected to illustrate asymmetric CircaFMM profiles in THY. Genes were selected among FR detections with $\hat{\omega} < 0.5$. Points represent observed expression values, and grayscale intensity indicates observation weights $w_i$, with darker points corresponding to higher weights.

Table S10: Summary of $\log(1 + \text{SNR})$ values for FR genes across GTEx-13T tissues. SNR was computed gene-wise as the ratio between the variance explained by a cyclic spline trend and residual variance.

| Tissue | Q1 | Median | Q3 | Mean | SD |
|---|---|---|---|---|---|
| SUB | 0.098 | 0.148 | 0.211 | 0.160 | 0.083 |
| VIS | 0.106 | 0.159 | 0.220 | 0.169 | 0.085 |
| AOR | 0.105 | 0.150 | 0.204 | 0.159 | 0.072 |
| COR | 0.098 | 0.141 | 0.192 | 0.151 | 0.074 |
| TIB | 0.087 | 0.129 | 0.180 | 0.140 | 0.074 |
| COL | 0.090 | 0.136 | 0.190 | 0.147 | 0.076 |
| ESO | 0.087 | 0.154 | 0.230 | 0.163 | 0.095 |
| HEA | 0.109 | 0.167 | 0.228 | 0.177 | 0.096 |
| LIV | 0.110 | 0.158 | 0.208 | 0.161 | 0.065 |
| LUN | 0.095 | 0.148 | 0.207 | 0.155 | 0.077 |
| NER | 0.090 | 0.132 | 0.191 | 0.146 | 0.075 |
| PIT | 0.093 | 0.138 | 0.197 | 0.151 | 0.079 |
| THY | 0.100 | 0.153 | 0.212 | 0.162 | 0.084 |

Table S11: Representative FR genes selected from the lower quartile of $\log(1 + \text{SNR})$ in TIB, COL and NER. The table reports the gene-wise log-transformed variance-based SNR.

| Tissue | Gene | $\log(1 + \text{SNR})$ |
|---|---|---|
| TIB | *NUDT6* | 0.076 |
| TIB | *RAB11FIP5* | 0.030 |
| TIB | *CDK4* | 0.072 |
| COL | *SMARCD2* | 0.054 |
| COL | *ASH2L* | 0.044 |
| COL | *NBR2* | 0.069 |
| NER | *BAMBI* | 0.074 |
| NER | *MORC4* | 0.048 |
| NER | *LAMA2* | 0.084 |

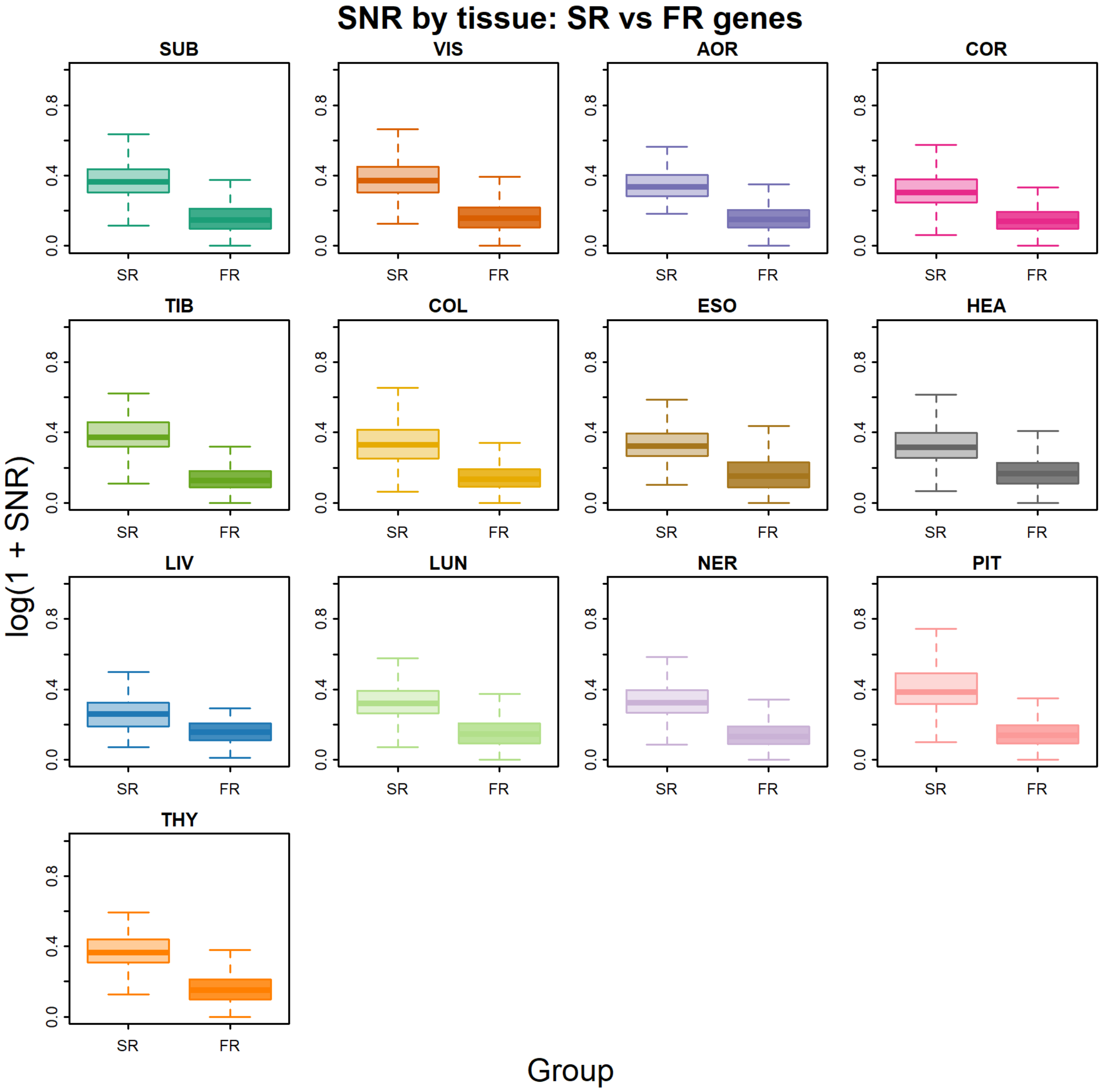


Figure S27: Distribution of log-transformed variance-based SNR values for SR and FR genes across GTEx-13T tissues. SNR was computed gene-wise as the ratio between the variance explained by a cyclic spline trend and the residual variance.

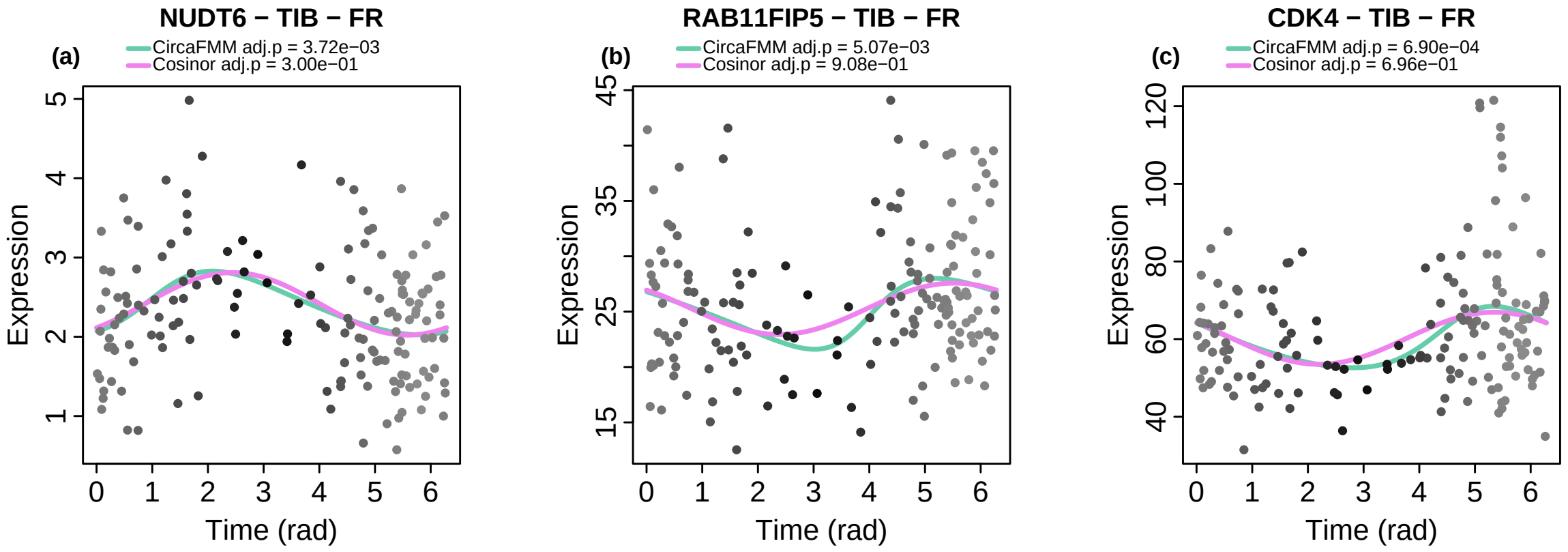


Figure S28: Randomly selected low-SNR FR genes in TIB. Genes were sampled among FR detections within the lower quartile of $\log(1 + \text{SNR})$. Grayscale intensity represents observation weights $w_i$, with darker points indicating higher weights.

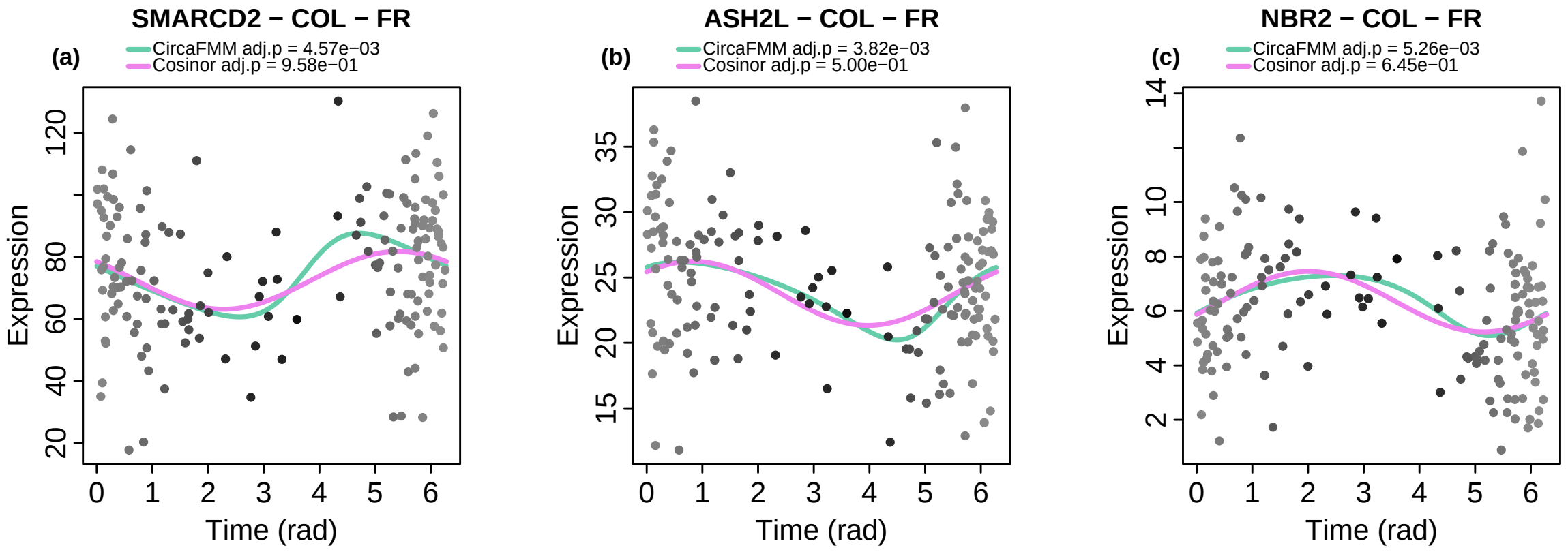


Figure S29: Randomly selected low-SNR FR genes in COL. Genes were sampled among FR detections within the lower quartile of $\log(1 + \text{SNR})$. Grayscale intensity represents observation weights $w_i$, with darker points indicating higher weights.

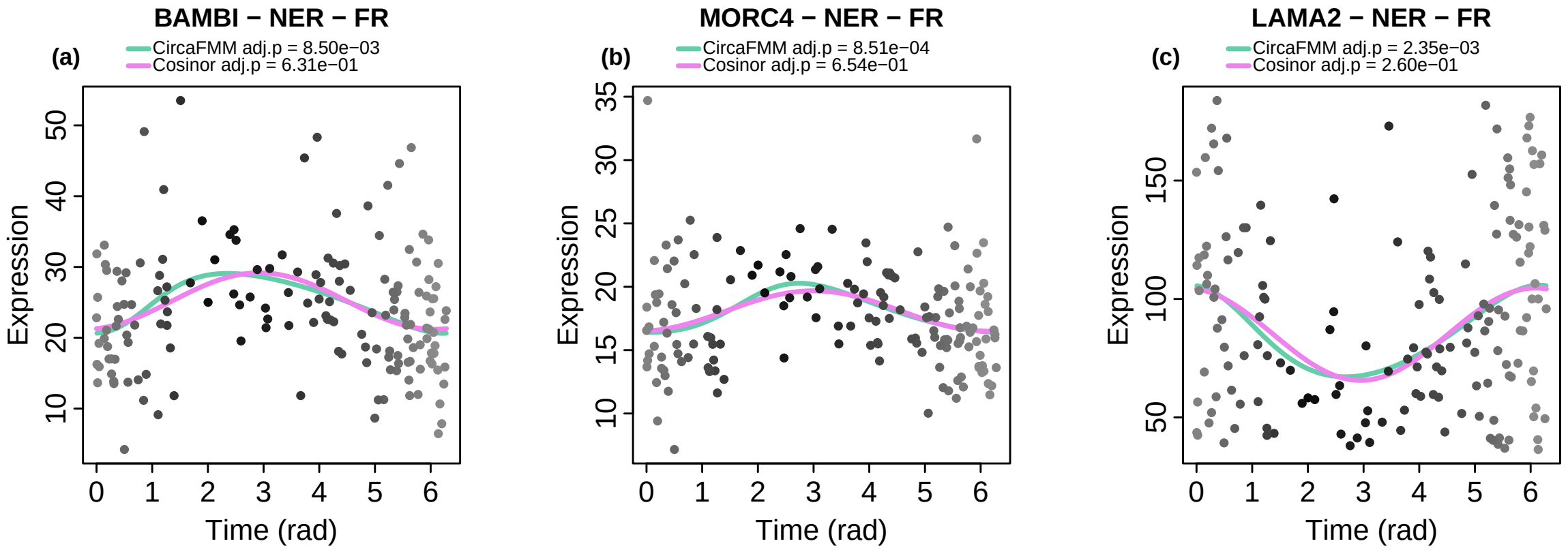


Figure S30: Randomly selected low-SNR FR genes in NER. Genes were sampled among FR detections within the lower quartile of $\log(1 + \text{SNR})$. Grayscale intensity represents observation weights $w_i$, with darker points indicating higher weights.

### S6.4.2 Cluster structures identified by Cosinor and CircaFMM

Representative cluster mean curves for Cosinor and CircaFMM detections are shown in Figures S31 and S32, respectively. In general, when only a small number of clusters was identified, both methods tended to show one dominant cluster. As the number of clusters increased, cluster sizes became more balanced.

Cosinor FCA consistently produced relatively simple clustering structures, typically ranging from two to five clusters and dominated by symmetric sinusoidal profiles primarily separated by peak-phase location (see Figure S31). The largest numbers of Cosinor clusters were observed in TIB and COL. However, in these tissues, representative Cosinor profiles showed limited structural differentiation, mainly differing by small peak-phase shifts rather than by substantial changes in waveform shape. This apparent increase in Cosinor cluster number may reflect the sensitivity of sinusoidal fits under low-SNR or irregular sampling conditions.

This apparent over-segmentation may reflect reduced robustness of sinusoidal fits under challenging data conditions. In TIB, this interpretation is supported by several indicators of low signal strength: the tissue was among those with the lowest FR SNR levels (Table S10), showed low tissue-level $RobSNR$ values (Table S6), and exhibited low cluster-level $Me$(SNR) values (Table S12).

In COL, the larger number of Cosinor clusters may additionally be related to its irregular sampling-time distribution. This tissue showed elevated $\%Max(Gaps)$ and $Cv(Gaps)$ values (Table S6), indicating large temporal gaps and uneven spacing of observations. Under these conditions, small differences in estimated phase may lead to additional Cosinor cluster separation without necessarily reflecting genuinely distinct waveform morphologies.

In contrast, CircaFMM FCA identified between four and seven clusters across tissues, reflecting a markedly higher degree of waveform heterogeneity (see Figure S32). This greater structural diversity remained evident in tissues with marked irregular temporal coverage, such as COR, and in low-SNR tissues such as NER (see Table S6). Moreover, CircaFMM simultaneously identified symmetric and strongly asymmetric rhythmic patterns within the same tissue, as observed in AOR, and separated waveform structures with similar peak phases but different temporal shapes, as observed in PIT.

Unlike Cosinor clusters, whose average peak and trough times were generally close to the $\pi$-separated configuration expected under symmetric sinusoidal oscillation, CircaFMM clusters often deviated from this configuration. This indicates the presence of asymmetric waveform structures within the same tissue.

Cluster-wise summaries are reported in Table S12. For CircaFMM clusters, the estimated sharpness parameter $\hat{\omega}$ and the median absolute duration asymmetry index $Me(|AD_{\mathrm{Dur}}|)$ further supported the presence of heterogeneous waveform shapes. Across tissues and clusters, $\overline{\omega}$ ranged from approximately 0.49 to 0.89, indicating substantial variability in waveform sharpness and departure from purely sinusoidal behavior. In some tissues, such as ESO, $\overline{\omega}$ exhibited particularly wide within-tissue variability, ranging from 0.528 to 0.784, suggesting the simultaneous presence of near-symmetric and more strongly asymmetric rhythmic profiles.

Similarly, $Me(|AD_{Dur}|)$ values ranged from approximately 0.05 to 0.34 across CircaFMM clusters, confirming the broad diversity of asymmetric waveform organizations captured by CircaFMM. Relatively large asymmetry values were especially evident in tissues such as ESO, consistent with the pronounced asymmetric signature observed for this tissue in the gene-level asymmetry analysis. These asymmetry-related quantities are intrinsically absent under the Cosinor framework, where waveform geometry is constrained to perfectly symmetric sinusoidal oscillations. In this sense, Cosinor representative profiles correspond to the limiting configuration $\omega = 1$ and $AD_{\mathrm{Dur}} = 0$.

To complete the characterization of waveform structures, Table S12 also reports values for $Me$(SNR) across clusters and tissues. This metric distinguished tissues and clusters with relatively higher rhythmic signal stability, such as COR, from low-SNR tissues such as NER, previously highlighted in the main text for their weak rhythmic signal structure. Nevertheless, even under these challenging conditions, CircaFMM consistently identified coherent asymmetric waveform organizations while maintaining lower median SNR levels than those observed in Cosinor clusters.

Table S12: Cluster-wise summaries for Cosinor and CircaFMM detections across GTEx-13T tissues. For each cluster, ($N_G$ denotes the number of genes, $\overline{t^U}$ and $\overline{t^L}$ denote the average of estimated peak and trough times, respectively, and $Me$(SNR) denotes the median log-transformed variance-based SNR. For CircaFMM clusters, the table also reports the mean selected sharpness parameter $\overline{\omega}$ and the median absolute duration asymmetry index $Me(|AD_{\mathrm{Dur}}|)$.

| Tissue | Method | Metric | $k=1$ | $k=2$ | $k=3$ | $k=4$ | $k=5$ | $k=6$ | $k=7$ |
|---|---|---|---|---|---|---|---|---|---|
| SUB | Cosinor | $N_G$ | 147 | 2 | | | | | |
| | | $\overline{t^U}$ | 3.730 | 1.148 | | | | | |
| | | $\overline{t^L}$ | 0.588 | 4.290 | | | | | |
| | | $Me(SNR)$ | 0.252 | 0.208 | | | | | |
| | CircaFMM | $N_G$ | 821 | 765 | 486 | 473 | 426 | 350 | 198 |
| | | $\overline{t^U}$ | 3.970 | 3.617 | 0.623 | 4.251 | 2.754 | 0.059 | 1.742 |

*Continued on next page*

| Tissue | Method | Metric | $k=1$ | $k=2$ | $k=3$ | $k=4$ | $k=5$ | $k=6$ | $k=7$ |
|---|---|---|---|---|---|---|---|---|---|
| | | $\overline{t^L}$ | 0.251 | 0.041 | 4.125 | 0.571 | 6.109 | 3.146 | 5.051 |
| | | $\overline{\omega}$ | 0.707 | 0.723 | 0.718 | 0.582 | 0.642 | 0.672 | 0.667 |
| | | $Me(\lvert AD_{\text{Dur}}\rvert)$ | 0.192 | 0.147 | 0.157 | 0.197 | 0.130 | 0.135 | 0.138 |
| | | $Me(SNR)$ | 0.165 | 0.148 | 0.151 | 0.149 | 0.149 | 0.152 | 0.142 |
| VIS | Cosinor | $N_G$ | 1629 | 82 | 59 | | | | |
| | | $\overline{t^U}$ | 2.557 | 5.485 | 5.059 | | | | |
| | | $\overline{t^L}$ | 5.699 | 2.344 | 1.918 | | | | |
| | | $Me(SNR)$ | 0.272 | 0.234 | 0.256 | | | | |
| | CircaFMM | $N_G$ | 5703 | 816 | 700 | 363 | | | |
| | | $\overline{t^U}$ | 2.454 | 5.263 | 1.874 | 5.948 | | | |
| | | $\overline{t^L}$ | 5.579 | 2.236 | 5.005 | 2.857 | | | |
| | | $\overline{\omega}$ | 0.819 | 0.665 | 0.798 | 0.753 | | | |
| | | $Me(\lvert AD_{\text{Dur}}\rvert)$ | 0.048 | 0.085 | 0.069 | 0.093 | | | |
| | | $Me(SNR)$ | 0.199 | 0.179 | 0.172 | 0.162 | | | |
| AOR | Cosinor | $N_G$ | 27 | 27 | 25 | 20 | 16 | | |
| | | $\overline{t^U}$ | 3.737 | 3.035 | 2.617 | 1.608 | 4.840 | | |
| | | $\overline{t^L}$ | 0.595 | 6.177 | 5.759 | 4.749 | 1.699 | | |
| | | $Me(SNR)$ | 0.266 | 0.227 | 0.244 | 0.255 | 0.260 | | |
| | CircaFMM | $N_G$ | 966 | 905 | 846 | 764 | 495 | | |
| | | $\overline{t^U}$ | 0.526 | 3.303 | 1.500 | 2.334 | 4.598 | | |
| | | $\overline{t^L}$ | 3.677 | 0.201 | 4.666 | 5.723 | 1.349 | | |
| | | $\overline{\omega}$ | 0.744 | 0.758 | 0.775 | 0.753 | 0.710 | | |
| | | $Me(\lvert AD_{\text{Dur}}\rvert)$ | 0.116 | 0.085 | 0.097 | 0.120 | 0.130 | | |
| | | $Me(SNR)$ | 0.149 | 0.154 | 0.155 | 0.151 | 0.151 | | |
| COR | Cosinor | $N_G$ | 909 | 232 | 213 | 62 | | | |
| | | $\overline{t^U}$ | 3.508 | 4.067 | 4.594 | 0.974 | | | |
| | | $\overline{t^L}$ | 0.366 | 0.925 | 1.452 | 4.116 | | | |
| | | $Me(SNR)$ | 0.222 | 0.223 | 0.200 | 0.209 | | | |
| | CircaFMM | $N_G$ | 1255 | 1132 | 903 | 666 | 535 | 488 | |
| | | $\overline{t^U}$ | 2.954 | 2.698 | 3.788 | 0.689 | 5.402 | 1.640 | |
| | | $\overline{t^L}$ | 0.896 | 0.121 | 0.877 | 3.284 | 1.494 | 5.439 | |
| | | $\overline{\omega}$ | 0.539 | 0.640 | 0.535 | 0.492 | 0.616 | 0.614 | |
| | | $Me(\lvert AD_{\text{Dur}}\rvert)$ | 0.337 | 0.195 | 0.204 | 0.276 | 0.259 | 0.230 | |
| | | $Me(SNR)$ | 0.187 | 0.168 | 0.182 | 0.150 | 0.167 | 0.147 | |
| TIB | Cosinor | $N_G$ | 35 | 30 | 23 | 19 | 12 | 9 | |
| | | $\overline{t^U}$ | 0.526 | 2.606 | 2.991 | 3.804 | 1.970 | 6.014 | |
| | | $\overline{t^L}$ | 3.667 | 5.748 | 6.133 | 0.662 | 5.111 | 2.872 | |
| | | $Me(SNR)$ | 0.256 | 0.269 | 0.266 | 0.165 | 0.267 | 0.284 | |
| | CircaFMM | $N_G$ | 1565 | 1470 | 1266 | 1148 | | | |
| | | $\overline{t^U}$ | 2.647 | 0.360 | 5.384 | 1.165 | | | |
| | | $\overline{t^L}$ | 5.861 | 3.448 | 2.268 | 4.888 | | | |
| | | $\overline{\omega}$ | 0.765 | 0.650 | 0.699 | 0.656 | | | |
| | | $Me(\lvert AD_{\text{Dur}}\rvert)$ | 0.078 | 0.130 | 0.106 | 0.198 | | | |
| | | $Me(SNR)$ | 0.132 | 0.133 | 0.127 | 0.129 | | | |
| COL | Cosinor | $N_G$ | 83 | 54 | 37 | 35 | 35 | 19 | |
| | | $\overline{t^U}$ | 4.007 | 4.556 | 2.872 | 2.438 | 1.869 | 3.395 | |
| | | $\overline{t^L}$ | 0.866 | 1.415 | 6.014 | 5.580 | 5.010 | 0.253 | |
| | | $Me(SNR)$ | 0.180 | 0.199 | 0.281 | 0.313 | 0.228 | 0.157 | |
| | CircaFMM | $N_G$ | 1285 | 1042 | 979 | 899 | | | |
| | | $\overline{t^U}$ | 2.003 | 3.383 | 4.547 | 5.909 | | | |
| | | $\overline{t^L}$ | 5.264 | 6.140 | 1.221 | 2.934 | | | |
| | | $\overline{\omega}$ | 0.734 | 0.677 | 0.670 | 0.712 | | | |
| | | $Me(\lvert AD_{\text{Dur}}\rvert)$ | 0.119 | 0.130 | 0.140 | 0.118 | | | |
| | | $Me(SNR)$ | 0.141 | 0.138 | 0.146 | 0.138 | | | |
| ESO | Cosinor | $N_G$ | 630 | 36 | | | | | |

*Continued on next page*

| Tissue | Method | Metric | $k=1$ | $k=2$ | $k=3$ | $k=4$ | $k=5$ | $k=6$ | $k=7$ |
|---|---|---|---|---|---|---|---|---|---|
| | | $\overline{t^U}$ | 4.555 | 5.483 | | | | | |
| | | $\overline{t^L}$ | 1.414 | 2.342 | | | | | |
| | | $Me(SNR)$ | 0.234 | 0.206 | | | | | |
| | CircaFMM | $N_G$ | 698 | 457 | 269 | 185 | 170 | 133 | 96 |
| | | $\overline{t^U}$ | 4.962 | 4.256 | 0.524 | 1.012 | 6.224 | 5.754 | 3.440 |
| | | $\overline{t^L}$ | 1.314 | 1.232 | 3.766 | 4.291 | 3.257 | 2.215 | 0.805 |
| | | $\overline{\omega}$ | 0.634 | 0.784 | 0.756 | 0.735 | 0.740 | 0.528 | 0.700 |
| | | $Me(\|AD_{\text{Dur}}\|)$ | 0.169 | 0.076 | 0.127 | 0.095 | 0.097 | 0.230 | 0.188 |
| | | $Me(SNR)$ | 0.217 | 0.216 | 0.174 | 0.164 | 0.149 | 0.177 | 0.198 |
| HEA | Cosinor | $N_G$ | 560 | 83 | 82 | 73 | 63 | | |
| | | $\overline{t^U}$ | 3.065 | 2.465 | 4.075 | 4.948 | 4.525 | | |
| | | $\overline{t^L}$ | 6.206 | 5.606 | 0.933 | 1.807 | 1.383 | | |
| | | $Me(SNR)$ | 0.232 | 0.271 | 0.257 | 0.262 | 0.305 | | |
| | CircaFMM | $N_G$ | 2134 | 831 | 557 | 416 | 379 | 311 | |
| | | $\overline{t^U}$ | 2.769 | 3.215 | 2.116 | 4.078 | 5.341 | 5.941 | |
| | | $\overline{t^L}$ | 5.759 | 0.359 | 4.770 | 1.004 | 1.308 | 2.502 | |
| | | $\overline{\omega}$ | 0.624 | 0.703 | 0.636 | 0.605 | 0.599 | 0.495 | |
| | | $Me(\|AD_{\text{Dur}}\|)$ | 0.126 | 0.128 | 0.231 | 0.140 | 0.308 | 0.265 | |
| | | $Me(SNR)$ | 0.184 | 0.193 | 0.176 | 0.192 | 0.203 | 0.178 | |
| LIV | Cosinor | $N_G$ | 162 | 17 | | | | | |
| | | $\overline{t^U}$ | 3.462 | 2.472 | | | | | |
| | | $\overline{t^L}$ | 0.320 | 5.613 | | | | | |
| | | $Me(SNR)$ | 0.178 | 0.172 | | | | | |
| | CircaFMM | $N_G$ | 73 | 55 | 47 | 33 | | | |
| | | $\overline{t^U}$ | 3.025 | 2.578 | 3.560 | 1.897 | | | |
| | | $\overline{t^L}$ | 6.075 | 5.824 | 0.326 | 4.916 | | | |
| | | $\overline{\omega}$ | 0.773 | 0.740 | 0.661 | 0.648 | | | |
| | | $Me(\|AD_{\text{Dur}}\|)$ | 0.090 | 0.119 | 0.069 | 0.118 | | | |
| | | $Me(SNR)$ | 0.183 | 0.176 | 0.172 | 0.158 | | | |
| LUN | Cosinor | $N_G$ | 105 | 51 | 7 | | | | |
| | | $\overline{t^U}$ | 2.322 | 1.752 | 5.026 | | | | |
| | | $\overline{t^L}$ | 5.463 | 4.894 | 1.884 | | | | |
| | | $Me(SNR)$ | 0.177 | 0.250 | 0.227 | | | | |
| | CircaFMM | $N_G$ | 264 | 236 | 214 | 210 | 161 | 99 | |
| | | $\overline{t^U}$ | 2.265 | 5.123 | 1.366 | 5.355 | 2.237 | 5.901 | |
| | | $\overline{t^L}$ | 5.302 | 1.474 | 4.803 | 2.243 | 5.447 | 2.808 | |
| | | $\overline{\omega}$ | 0.596 | 0.639 | 0.710 | 0.708 | 0.756 | 0.741 | |
| | | $Me(\|AD_{\text{Dur}}\|)$ | 0.093 | 0.211 | 0.170 | 0.059 | 0.048 | 0.083 | |
| | | $Me(SNR)$ | 0.159 | 0.163 | 0.162 | 0.139 | 0.168 | 0.148 | |
| NER | Cosinor | $N_G$ | 475 | 116 | | | | | |
| | | $\overline{t^U}$ | 2.658 | 5.861 | | | | | |
| | | $\overline{t^L}$ | 5.800 | 2.719 | | | | | |
| | | $Me(SNR)$ | 0.239 | 0.225 | | | | | |
| | CircaFMM | $N_G$ | 2514 | 812 | 762 | 732 | | | |
| | | $\overline{t^U}$ | 2.862 | 5.676 | 6.040 | 2.368 | | | |
| | | $\overline{t^L}$ | 5.893 | 2.492 | 3.122 | 5.668 | | | |
| | | $\overline{\omega}$ | 0.782 | 0.596 | 0.635 | 0.719 | | | |
| | | $Me(\|AD_{\text{Dur}}\|)$ | 0.068 | 0.086 | 0.116 | 0.081 | | | |
| | | $Me(SNR)$ | 0.148 | 0.138 | 0.141 | 0.144 | | | |
| PIT | Cosinor | $N_G$ | 43 | 38 | 37 | 36 | | | |
| | | $\overline{t^U}$ | 1.934 | 4.584 | 4.183 | 2.766 | | | |
| | | $\overline{t^L}$ | 5.076 | 1.443 | 1.042 | 5.907 | | | |
| | | $Me(SNR)$ | 0.283 | 0.229 | 0.281 | 0.226 | | | |
| | CircaFMM | $N_G$ | 816 | 775 | 504 | 417 | 396 | 281 | |
| | | $\overline{t^U}$ | 1.710 | 2.292 | 3.081 | 1.153 | 4.502 | 5.366 | |
| | | $\overline{t^L}$ | 5.138 | 5.510 | 6.022 | 4.166 | 1.046 | 2.324 | |

*Continued on next page*

| Tissue | Method | Metric | $k=1$ | $k=2$ | $k=3$ | $k=4$ | $k=5$ | $k=6$ | $k=7$ |
|---|---|---|---|---|---|---|---|---|---|
| | | $\overline{\omega}$ | 0.711 | 0.707 | 0.687 | 0.662 | 0.686 | 0.640 | |
| | | $Me(\lvert AD_{\text{Dur}}\rvert)$ | 0.152 | 0.128 | 0.111 | 0.148 | 0.139 | 0.133 | |
| | | $Me(SNR)$ | 0.142 | 0.142 | 0.147 | 0.136 | 0.156 | 0.142 | |
| THY | Cosinor | $N_G$ | 49 | 49 | 36 | 33 | | | |
| | | $\overline{t^U}$ | 4.089 | 2.642 | 1.639 | 3.180 | | | |
| | | $\overline{t^L}$ | 0.947 | 5.784 | 4.780 | 0.038 | | | |
| | | $Me(SNR)$ | 0.241 | 0.272 | 0.219 | 0.260 | | | |
| | CircaFMM | $N_G$ | 1694 | 1127 | 870 | 701 | 412 | 389 | |
| | | $\overline{t^U}$ | 2.437 | 1.853 | 1.730 | 1.049 | 3.640 | 0.007 | |
| | | $\overline{t^L}$ | 5.760 | 4.927 | 4.302 | 4.026 | 0.280 | 3.326 | |
| | | $\overline{\omega}$ | 0.661 | 0.643 | 0.686 | 0.719 | 0.725 | 0.671 | |
| | | $Me(\lvert AD_{\text{Dur}}\rvert)$ | 0.128 | 0.191 | 0.195 | 0.103 | 0.133 | 0.177 | |
| | | $Me(SNR)$ | 0.157 | 0.155 | 0.157 | 0.153 | 0.167 | 0.157 | |

### S6.4.3 Cluster correspondence between Cosinor and CircaFMM

Table S13 provides the cross-classification between Cosinor and CircaFMM waveform clusters after reordering clusters according to prevalence within each tissue. Values correspond to column-wise percentages, so each column shows how genes assigned to a given Cosinor cluster were redistributed across CircaFMM clusters. This representation enables direct assessment of cluster concordance and structural mismatching between both rhythmicity frameworks.

Overall, Cosinor clusters frequently mapped onto multiple CircaFMM waveform structures rather than exhibiting a one-to-one correspondence. This indicates that genes grouped together under the sinusoidal Cosinor representation can display heterogeneous waveform organizations when asymmetric rhythmic patterns are allowed. This phenomenon was particularly evident in tissues with large discrepancies in the number of detected clusters between methods, such as SUB and ESO, where approximately half of the CircaFMM clusters displayed overlap rates lower than 10% with Cosinor detections.

## S6.5 Circadian peak analysis

This section provides supplementary details for the circadian phase analyses reported in the main text. We first define the circular metrics used to compare CircaFMM and Cosinor peak estimates for SR genes, and then describe the bootstrap procedure used to quantify peak and trough uncertainty for FR detections.

### S6.5.1 Circular agreement metrics for SR detections

For SR genes, agreement between CircaFMM and Cosinor peak estimates was assessed using three complementary circular statistics. Let $\theta_g^{FMM}$ denote the CircaFMM peak time and $\theta_g^{Cos}$ the Cosinor peak time for gene $g$, both expressed in radians on $[0, 2\pi)$. The metrics are defined as follows:

- Circular correlation ($\rho$): global agreement between CircaFMM and Cosinor peak timings was quantified using the circular correlation coefficient of Fisher and Lee (1983), computed from $\theta_g^{FMM}$ and $\theta_g^{Cos}$. This statistic measures the association between two sets of angular estimates while accounting for their circular nature. High values indicate that both methods preserve a similar relative phase organization across genes.
- Gene-wise peak disagreement ($Me(|\Delta|)$): gene-specific discrepancies between methods were summarized through the signed circular phase difference

$$\Delta_g = \operatorname{atan2}\left\{\sin(\theta_g^{FMM} - \theta_g^{Cos}), \cos(\theta_g^{FMM} - \theta_g^{Cos})\right\},$$

  which maps phase differences to the interval $(-\pi, \pi]$ and therefore represents the shortest angular displacement between both estimates. The corresponding absolute angular error was expressed in hours as

$$|\Delta_g|_{\text{hours}} = |\Delta_g|\frac{24}{2\pi}.$$

  For each tissue, we reported the median absolute angular error, $Me(|\Delta|)$, as a robust summary of gene-wise peak disagreement.
- Peak concentration ($\overline{R}$): concentration of peak timings within each method was quantified using the mean resultant length

$$\overline{R} = \left[\left(\frac{1}{G}\sum_{g=1}^{G}\cos\theta_g\right)^2 + \left(\frac{1}{G}\sum_{g=1}^{G}\sin\theta_g\right)^2\right]^{1/2}.$$

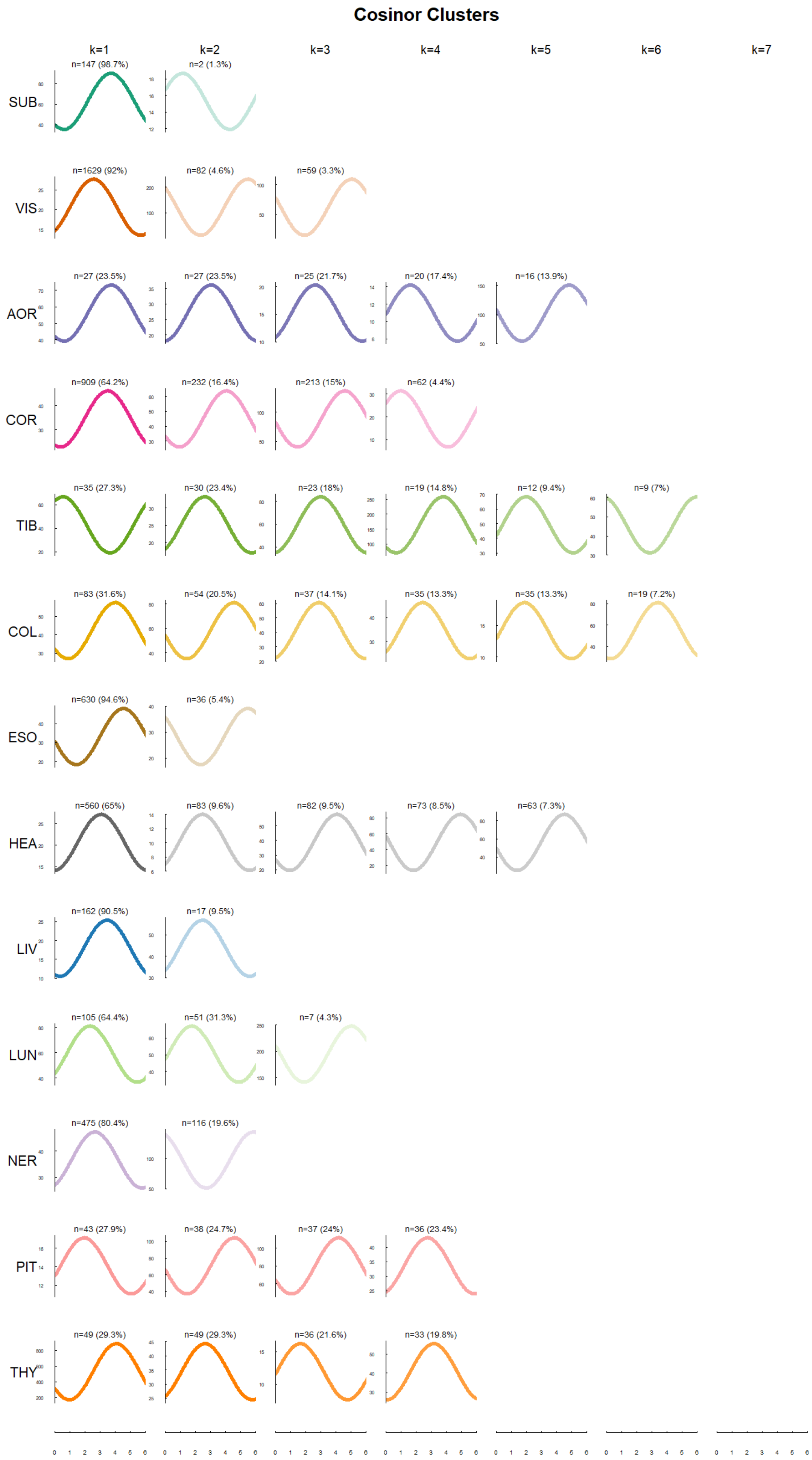


Figure S31: Representative cluster mean curves for Cosinor detections across tissues. Rows correspond to tissues and columns to clusters, ordered by cluster prevalence within each tissue from most to least represented. Curves represent the average fitted Cosinor waveform for each cluster. Colour intensity is proportional to the percentage of genes assigned to the corresponding cluster, with darker shades indicating higher prevalence.

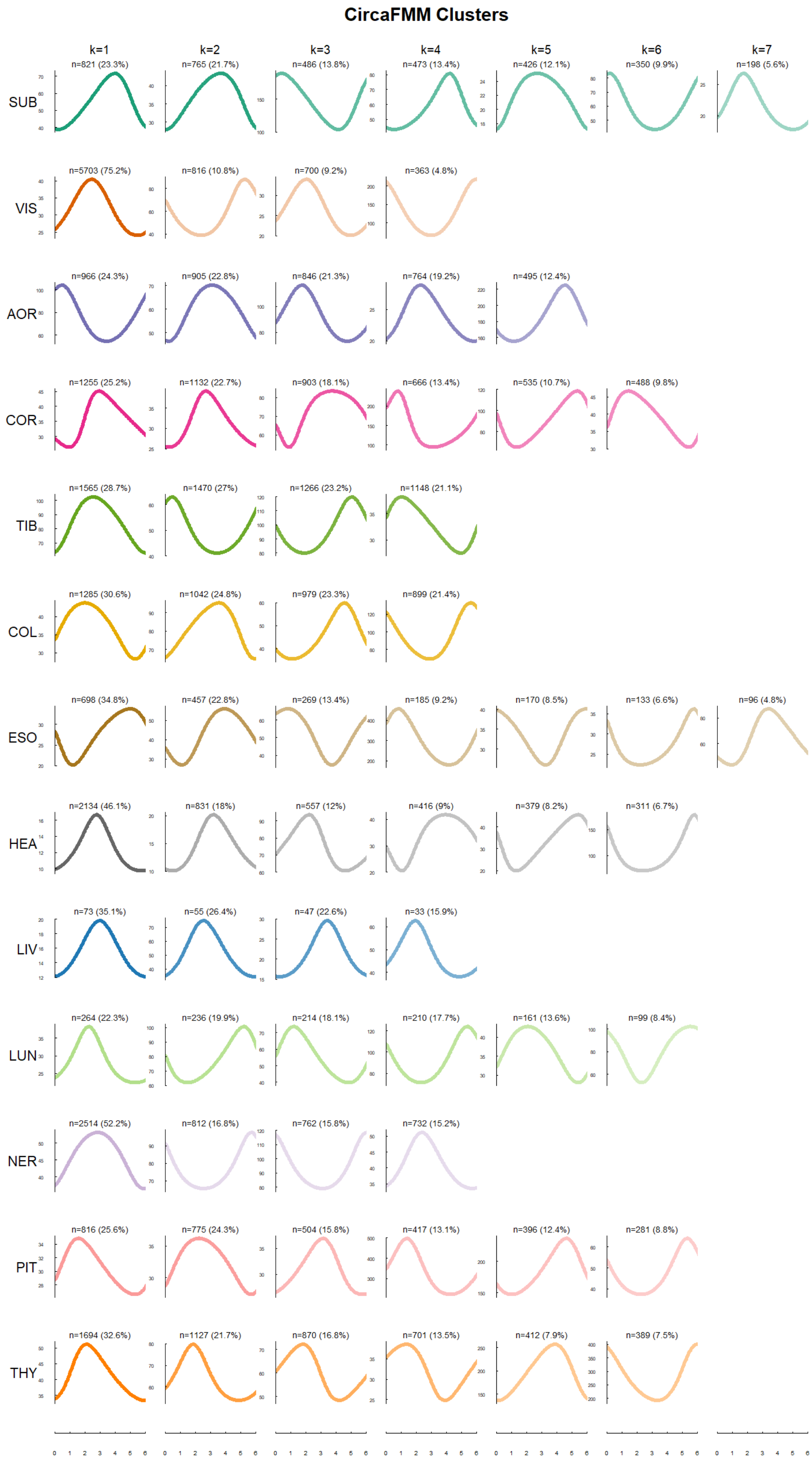


Figure S32: Representative cluster mean curves for CircaFMM detections across tissues. Rows correspond to tissues and columns to clusters, ordered by cluster prevalence within each tissue from most to least represented. Curves represent the average fitted CircaFMM waveform for each cluster. Colour intensity is proportional to the percentage of genes assigned to the corresponding cluster, with darker shades indicating higher prevalence.

**SUB**

| CircaFMM | Cosinor k=1 | k=2 |
|---|---|---|
| k=1 | 46.575 | 0.000 |
| k=2 | 27.397 | 0.000 |
| k=3 | 0.000 | 100.000 |
| k=4 | 23.973 | 0.000 |
| k=5 | 2.055 | 0.000 |
| k=6 | 0.000 | 0.000 |
| k=7 | 0.000 | 0.000 |

**VIS**

| CircaFMM | Cosinor k=1 | k=2 | k=3 |
|---|---|---|---|
| k=1 | 96.681 | 0.000 | 0.000 |
| k=2 | 0.000 | 80.488 | 100.000 |
| k=3 | 3.319 | 0.000 | 0.000 |
| k=4 | 0.000 | 19.512 | 0.000 |

**AOR**

| CircaFMM | Cosinor k=1 | k=2 | k=3 | k=4 | k=5 |
|---|---|---|---|---|---|
| k=1 | 0.000 | 0.000 | 0.000 | 15.000 | 6.250 |
| k=2 | 85.185 | 55.556 | 0.000 | 0.000 | 0.000 |
| k=3 | 0.000 | 0.000 | 0.000 | 80.000 | 0.000 |
| k=4 | 0.000 | 44.444 | 100.000 | 5.000 | 0.000 |
| k=5 | 14.815 | 0.000 | 0.000 | 0.000 | 93.750 |

**COR**

| CircaFMM | Cosinor k=1 | k=2 | k=3 | k=4 |
|---|---|---|---|---|
| k=1 | 54.181 | 5.479 | 0.000 | 0.000 |
| k=2 | 30.506 | 0.000 | 0.000 | 0.000 |
| k=3 | 14.959 | 89.041 | 15.429 | 0.000 |
| k=4 | 0.000 | 0.000 | 0.571 | 77.049 |
| k=5 | 0.000 | 5.479 | 84.000 | 0.000 |
| k=6 | 0.353 | 0.000 | 0.000 | 22.951 |

**TIB**

| CircaFMM | Cosinor k=1 | k=2 | k=3 | k=4 | k=5 | k=6 |
|---|---|---|---|---|---|---|
| k=1 | 0.000 | 100.000 | 100.000 | 83.333 | 41.667 | 0.000 |
| k=2 | 94.286 | 0.000 | 0.000 | 0.000 | 0.000 | 44.444 |
| k=3 | 0.000 | 0.000 | 0.000 | 16.667 | 0.000 | 55.556 |
| k=4 | 5.714 | 0.000 | 0.000 | 0.000 | 58.333 | 0.000 |

**COL**

| CircaFMM | Cosinor k=1 | k=2 | k=3 | k=4 | k=5 | k=6 |
|---|---|---|---|---|---|---|
| k=1 | 0.000 | 0.000 | 37.838 | 100.000 | 100.000 | 7.692 |
| k=2 | 16.438 | 0.000 | 62.162 | 0.000 | 0.000 | 92.308 |
| k=3 | 83.562 | 100.000 | 0.000 | 0.000 | 0.000 | 0.000 |
| k=4 | 0.000 | 0.000 | 0.000 | 0.000 | 0.000 | 0.000 |

**ESO**

| CircaFMM | Cosinor k=1 | k=2 |
|---|---|---|
| k=1 | 49.393 | 26.087 |
| k=2 | 44.130 | 0.000 |
| k=3 | 0.000 | 4.348 |
| k=4 | 0.000 | 8.696 |
| k=5 | 0.000 | 4.348 |
| k=6 | 0.607 | 56.522 |
| k=7 | 5.870 | 0.000 |

**HEA**

| CircaFMM | Cosinor k=1 | k=2 | k=3 | k=4 | k=5 |
|---|---|---|---|---|---|
| k=1 | 59.607 | 79.221 | 0.000 | 0.000 | 0.000 |
| k=2 | 39.083 | 0.000 | 8.750 | 0.000 | 0.000 |
| k=3 | 0.000 | 20.779 | 0.000 | 0.000 | 0.000 |
| k=4 | 1.310 | 0.000 | 83.750 | 0.000 | 14.516 |
| k=5 | 0.000 | 0.000 | 7.500 | 61.972 | 85.484 |
| k=6 | 0.000 | 0.000 | 0.000 | 38.028 | 0.000 |

**LIV**

| CircaFMM | Cosinor k=1 | k=2 |
|---|---|---|
| k=1 | 45.238 | 0.000 |
| k=2 | 28.571 | 33.333 |
| k=3 | 26.190 | 0.000 |
| k=4 | 0.000 | 66.667 |

**LUN**

| CircaFMM | Cosinor k=1 | k=2 | k=3 |
|---|---|---|---|
| k=1 | 57.447 | 4.545 | 0.000 |
| k=2 | 0.000 | 0.000 | 57.143 |
| k=3 | 2.128 | 63.636 | 0.000 |
| k=4 | 0.000 | 0.000 | 42.857 |
| k=5 | 40.426 | 31.818 | 0.000 |
| k=6 | 0.000 | 0.000 | 0.000 |

**NER**

| CircaFMM | Cosinor k=1 | k=2 |
|---|---|---|
| k=1 | 77.215 | 0.000 |
| k=2 | 0.000 | 44.828 |
| k=3 | 0.000 | 55.172 |
| k=4 | 22.785 | 0.000 |

**PIT**

| CircaFMM | Cosinor k=1 | k=2 | k=3 | k=4 |
|---|---|---|---|---|
| k=1 | 73.810 | 0.000 | 0.000 | 0.000 |
| k=2 | 21.429 | 0.000 | 0.000 | 36.364 |
| k=3 | 0.000 | 0.000 | 0.000 | 63.636 |
| k=4 | 4.762 | 0.000 | 0.000 | 0.000 |
| k=5 | 0.000 | 86.842 | 100.000 | 0.000 |
| k=6 | 0.000 | 13.158 | 0.000 | 0.000 |

**THY**

| CircaFMM | Cosinor k=1 | k=2 | k=3 | k=4 |
|---|---|---|---|---|
| k=1 | 0.000 | 85.417 | 0.000 | 84.375 |
| k=2 | 0.000 | 14.583 | 22.857 | 0.000 |
| k=3 | 0.000 | 0.000 | 45.714 | 0.000 |
| k=4 | 0.000 | 0.000 | 31.429 | 0.000 |
| k=5 | 90.909 | 0.000 | 0.000 | 15.625 |
| k=6 | 9.091 | 0.000 | 0.000 | 0.000 |

Table S13: Cross-classification of Cosinor SR detections across CircaFMM cluster structures. Columns represent Cosinor clusters and rows represent CircaFMM clusters after reordering clusters by prevalence within each tissue. Values are column-wise percentages and indicate how genes assigned to each Cosinor cluster are redistributed across CircaFMM clusters.

This quantity was computed separately for CircaFMM ($\overline{R}_{FMM}$) and Cosinor ($\overline{R}_{Cosinor}$) peak estimates. Values close to one indicate that peak times are highly concentrated around a common phase, whereas values close to indicate a more dispersed phase distribution.

Table 3 of the main text revealed a general agreement between CircaFMM and Cosinor peak estimates across tissues. Notably, some tissue-specific differences were also observed. COR and THY combined high circular correlations with larger median absolute angular errors, suggesting systematic phase shifts or wrapping effects between methods. In contrast, NER showed a lower circular correlation despite a very small median absolute angular error, consistent with broader dispersion of peak timings, as reflected by reduced concentration measures ($\overline{R}$). While VIS and AOR exhibited similarly small gene-wise peak discrepancies, their circular correlations differed markedly, indicating that overall phase organization across genes was more consistent between methods in AOR than in VIS.

### S6.5.2 Bootstrap uncertainty analysis for FR peak and trough estimation

The analysis of FR peak phases revealed distributions spanning the entire circular domain and generally exhibiting lower phase concentration than SR genes (Figure S33; compare with Figure 9 in the main text). For comparison, Figure S33 also displays the corresponding Cosinor peak estimates for the same FR genes. The 22 recurrent FR genes detected in at least 10 tissues showed marked tissue-specific variation in peak timing, with visual patterns ranging from relatively concentrated peaks to broader or multimodal phase distributions (Figure S34).

For example, *ZNF236*, *NOXA1*, and *SERPINA3* showed peak phases concentrated within a dominant temporal window of approximately 4–6 hours. In contrast, other genes displayed more complex phase architectures: *C5orf45* (*CIART*) and *BCL6* showed dispersed phase patterns, with tissue-specific peaks distributed across opposite regions of the circadian cycle, including clusters near phases 0 and $\pi$. Similarly, *ALG3*, *C12orf5*, and *ZFAND5* showed two separated phase clusters that were not always arranged in antiphase. These patterns suggest distinct tissue-specific temporal programs among CircaFMM-specific rhythmic candidates.

These findings further support the heterogeneous and tissue-specific nature of FR rhythmic patterns, motivating an uncertainty analysis of CircaFMM peak ($t^U$) and trough ($t^L$) estimates. The stability of these circadian landmarks was assessed using the following parametric bootstrap procedure.

For each FR gene $g$ within each tissue, let $\hat{X}_g(t_i)$ denote the fitted CircaFMM signal at observation time $t_i$, $i = 1, \dots, n$. The residual variance was estimated as

$$\hat{\sigma}_g^2 = \frac{\mathrm{RSS}_g}{n - p}$$

where $\mathrm{RSS}_g$ is the residual sum of squares for gene $g$, $n$ is the number of observations, and $p$ is the number of estimated model parameters.

Bootstrap samples were generated according to the parametric bootstrap scheme

$$X_g^{*(b)}(t_i) = \hat{X}_g(t_i) + \varepsilon_g^{*(b)}(t_i), \qquad \varepsilon_g^{(*b)}(t_i) \sim N(0, \hat{\sigma}_g^2),$$

for $b = 1, \dots, B$, with $B = 250$ bootstrap replicates per gene. Each bootstrap sample was subsequently refitted using the CircaFMM estimation procedure, obtaining bootstrap parameter estimates ($\hat{\alpha}^*, \hat{\beta}^*, \hat{\omega}^*$) and their associated peak ($t^{*U}$) and trough ($t^{*L}$) timings.

Because peak and trough timings are circular quantities defined on $[0, 2\pi)$, confidence intervals were computed using a circular percentile procedure. First, for a generic set of bootstrap phase estimates $\theta_1, \dots, \theta_B$, the circular mean was computed as

$$\mu = \mathrm{atan2}\left( \frac{1}{B} \sum_{b=1}^{B} \sin\theta_b, \frac{1}{B} \sum_{b=1}^{B} \cos\theta_b \right).$$

Bootstrap phase estimates were then recentered around $\mu$ through

$$\delta_b = \mathrm{atan2}\left[\sin(\theta_b - \mu), \cos(\theta_b - \mu)\right],$$

thereby mapping angular deviations to the interval $(-\pi, \pi]$ and representing the shortest circular displacement relative to the bootstrap mean phase.

Percentile confidence intervals were computed from the empirical quantiles of $\delta_b$, and the interval bounds were transformed back to the original circular domain ($[0, 2\pi)$ by adding the bootstrap circular mean $\mu$.

Finally, phase uncertainty was quantified through bootstrap standard errors, denoted by $SE_{\mathrm{boot}}$, computed from the recentered angular deviations and expressed in hours as

$$SE_{\mathrm{boot}} = SE_{\mathrm{rad}} \frac{24}{2\pi}.$$

The proportion of FR genes with overlapping peak and trough confidence intervals was computed as an additional indicator of phase-estimation robustness. Minimal overlap between $t^U$ and $t^L$ intervals supports that CircaFMM detections correspond to stable and biologically interpretable asymmetric waveforms, rather than to highly peaked or excessively asymmetric profiles that could artificially drive rhythmic detections. Table 4 in the

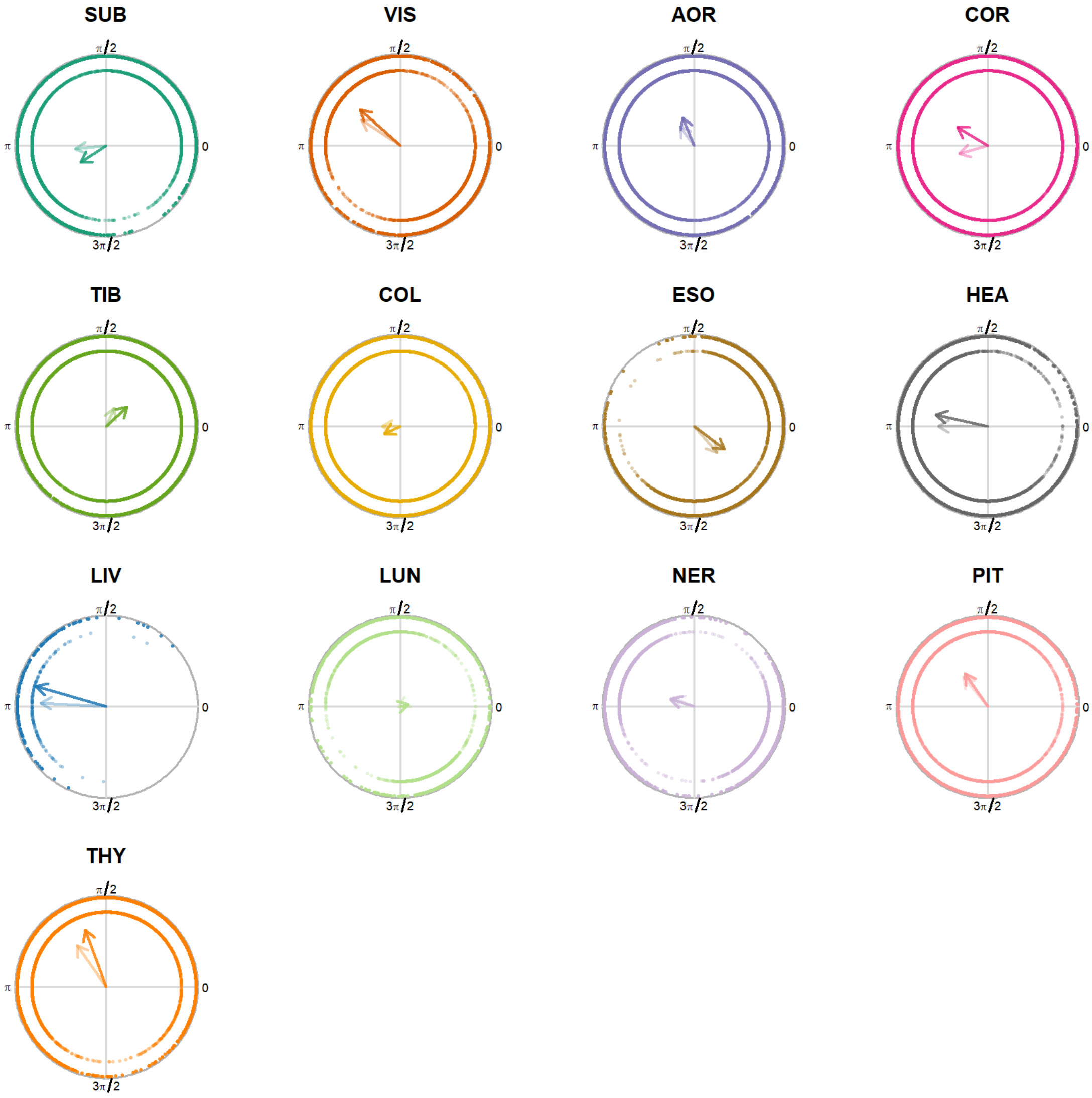


Figure S33: Circular distribution of peak expression times ($t^U$) for FR genes across tissues. Points on the outer ring represent gene-specific peak estimates obtained with CircaFMM, whereas points on the inner ring represent the corresponding peak estimates obtained with Cosinor for the same genes. Arrows denote the mean resultant vector for each method, with direction indicating the circular mean peak time and length reflecting the concentration of peak estimates ($\overline{R}$), scaled relative to the corresponding ring radius.

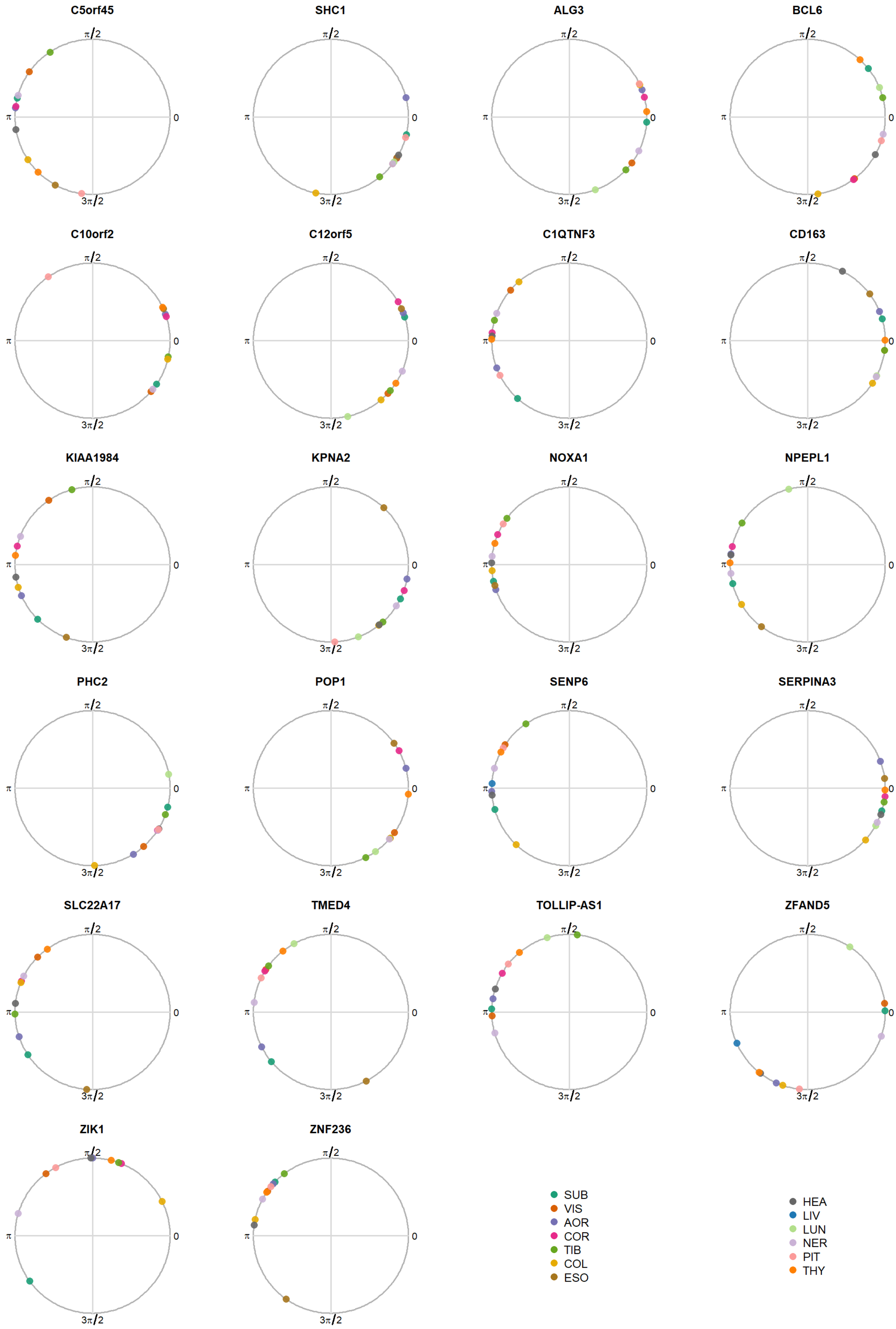


Figure S34: Peak time estimates ($t^U$) for the 22 recurrent FR genes detected in at least 10 tissues. Each panel corresponds to one gene, and each point represents the estimated peak in a specific tissue.

main text summarizes the stability of circadian landmarks estimation in terms of bootstrap standard errors and confidence-interval overlap proportions. The complete set of bootstrap uncertainty estimates, together with the corresponding CircaFMM phase estimates for both SR and FR genes, is available in Supplementary Data 1.



\end{document}